\documentclass[twocolumn, superscriptaddress]{revtex4-2}
\usepackage{amsmath}

\usepackage[utf8]{inputenc}
\usepackage{graphicx}
\usepackage{soul}
\usepackage[colorlinks=true, allcolors=blue]{hyperref}
\usepackage[capitalise]{cleveref}
\usepackage{wasysym}
\usepackage[table]{xcolor}
\usepackage{braket}
\usepackage{siunitx}
\usepackage{units}

\usepackage{adjustbox,pgfplots,pgfplotstable,xfp}
\usetikzlibrary{external}
\usepgfplotslibrary{colorbrewer,groupplots}
\pgfplotsset{compat=1.16}
\tikzsetexternalprefix{fig/}
\newcommand{\comment}[1]{}

\newcommand{\MHz}[1]{\SI{#1}{\MHz}}
\newcommand{\GHz}[1]{\SI{#1}{\GHz}}
\newcommand{\us}[1]{\SI{#1}{\us}}

\newcommand{\puribe}{\begin{eqnarray}}
\newcommand{\puriee}{\end{eqnarray}}

\newcommand{\puribra}[1]{\langle{#1}|}
\newcommand{\puriket}[1]{| #1 \rangle}
\newcommand{\purihta}{{a}}

\newcommand{\puricatpm}{\puriket{\mathcal{C}^\pm_\alpha}}
\newcommand{\puricatmp}{\puriket{\mathcal{C}^\mp_\alpha}}
\newcommand{\puricatp}{\puriket{\mathcal{C}^+_\alpha}}
\newcommand{\puricatpb}{\puribra{\mathcal{C}^+_\alpha}}
\newcommand{\puricatm}{\puriket{\mathcal{C}^-_\alpha}}
\newcommand{\puricatmb}{\puribra{\mathcal{C}^-_\alpha}}
\newcommand{\puricatpmb}{\puribra{\mathcal{C}^\pm_\alpha}}
\usepackage{longtable}

\newcommand{\dg}{^\dagger}
\usepackage{xcolor}

\newcommand{\CX}{\mathrm{CX}}
\newcommand{\CZ}{\mathrm{CZ}}
\newcommand{\CD}{\mathrm{CD}}

\crefname{section}{Sec.}{Secs.}
\Crefname{section}{Section}{Sections}

\begin{document}

\title{Practical quantum error correction with the XZZX code and Kerr-cat qubits}

\author{Andrew S. Darmawan}
\affiliation{Yukawa Institute for Theoretical Physics (YITP), Kyoto University, Kitashirakawa Oiwakecho, Sakyo-ku, Kyoto 606-8502, Japan}
\email[]{andrew.darmawan@yukawa.kyoto-u.ac.jp}
\affiliation{JST, PRESTO, 4-1-8 Honcho, Kawaguchi, Saitama 332-0012, Japan}
\author{Benjamin J. Brown}
\author{Arne L. Grimsmo}
\author{David K. Tuckett}
\affiliation{Centre for Engineered Quantum Systems, School of Physics, University of Sydney, Sydney, New South Wales 2006, Australia}
\author{Shruti Puri}
\affiliation{Department of Applied Physics, Yale University, New Haven, Connecticut 06511, USA}
\email[]{shruti.puri@yale.edu}
\affiliation{Yale Quantum Institute, Yale University, New Haven, Connecticut 06511, USA}
\date{\today}

%New abstract
 \begin{abstract}

    The development of robust architectures capable of large-scale fault-tolerant quantum computation should consider both their quantum error-correcting codes, and the underlying physical qubits upon which they are built, in tandem. Following this design principle we demonstrate remarkable error-correction performance by concatenating the XZZX surface code with Kerr-cat qubits.
We contrast several variants of fault-tolerant systems undergoing different circuit-noise models that reflect the physics of Kerr-cat qubits. 
Our simulations show that our system is scalable below a threshold gate infidelity of $p_{\CX} \sim 6.5\%$ within a physically reasonable parameter regime, where $p_{\CX}$ is the infidelity of the noisiest gate of our system; the controlled-not gate.
This threshold can be reached in a superconducting-circuit architecture with a Kerr nonlinearity of $10$MHz, a 
$\sim 6.25$ photon cat qubit, single-photon lifetime of $\gtrsim 64\mu$s, and thermal photon population $\lesssim 8\%$.
Such parameters are routinely achieved in superconducting circuits.

 \end{abstract}

\maketitle

\section{Introduction}

A scalable quantum computer will require a large number of almost perfect qubits. To this end, impressive progress has been made in engineering high-quality  quantum systems in the laboratory that can perform some requisite set of operations that are needed for quantum computation~\cite{Reed12, Barends14, Nigg14, kelly2015state,Corcoles15,Takita16,  Ofek16, Hu:2019aa, Fluhmann:2019aa, Campagne2020, Grimm2020,Lescanne2020, Sun2020, chen_exponential_2021}. While we cannot expect to reduce qubit noise arbitrarily with laboratory engineering alone, the theory of quantum error correction~\cite{Shor96, Kitaev03, Dennis02, Terhal15, Brown16, Campbell17} has shown that we can produce logical qubits that experience next-to-no noise by using a redundancy of high-quality physical qubits, provided the noise strength on the physical qubits as they perform computational operations is below some threshold value~\cite{AharonovBen-Or97}. Nevertheless, it still remains a significant experimental challenge to produce scalable quantum error-correcting codes using available technology. 
In order to overcome this challenge we should co-design the code in unison with the underlying qubit architecture. This way we can build better error-correcting systems that specifically target the dominant sources of errors.

A promising approach to realize a high-quality physical qubit is to encode it in a bosonic mode~\cite{Chuang97, cochrane99, gottesman01, mirrahimi2014dynamically, Ofek16, Hu:2019aa, Fluhmann:2019aa, Campagne2020, Grimm2020,Lescanne2020, Sun2020, gertler2021protecting, Vuillot2018, Noh:2019aa, Guillaud2019, Grimsmo2020, chamberland2020building, noh2021low, Terhal2020, joshi2021quantum}. Broadly speaking, encoding a qubit in the larger Hilbert space of an oscillator gives more room to protect the encoded information from noise, either through active error correction or autonomous stabilization of the codespace.
One of the simplest incarnations of this idea is the Kerr-cat qubit, where orthogonal superpositions of two coherent states $\ket{\pm\alpha}$ define a physical qubit~\cite{cochrane99,Goto2016,Puri2017,Puri2020}. The Kerr-cat qubit is realized by parametric pumping of a Kerr-nonlinear oscillator, as recently demonstrated in a superconducting circuit platform using a
Superconducting Nonlinear Asymmetric Inductive eLement (SNAIL)~\cite{Grimm2020}.
This qubit is stabilized against bit flips. Specifically, bit flips are exponentially suppressed with the photon number $\sim |\alpha|^2$, leading to a highly biased noise channel for large $\alpha$~\cite{Puri2020}.
Surprisingly, it has recently been proposed that the controlled-not gate ($\CX$) can be executed with Kerr-cat qubits without changing this noise profile~\cite{Puri2020}. This enables us to go beyond schemes where it is assumed that only diagonal entangling gates can be used without introducing bit-flip errors at a significant rate~\cite{aliferis_fault-tolerant_2008, aliferis_fault-tolerant_2009}.

A number of error-correction protocols have been proposed recently that are adapted specifically to biased noise~\cite{aliferis_fault-tolerant_2008, aliferis_fault-tolerant_2009,Stephens13bias, Tuckett18,  Tuckett19,Xu19,Li2019, Tuckett20, Huang20, guillaud2021error, hanggli_enhanced_2020, higgott_subsystem_2020, chamberland2020building}. 
In particular, the XZZX surface code~\cite{Wen03, BonillaAtaides20}, a surface code variant~\cite{Kitaev03, Bravyi1998, Dennis02}, has recently been shown to have a very high threshold under biased noise. This is because it mimics the behavior of the classical repetition code in the limit that a particular type of Pauli error is dominant. As such, minimum-weight perfect-matching decoders~\cite{Edmonds65, Dennis02, Kolmogorov09} have been developed by exploiting the symmetries of this code~\cite{Kitaev03, Dennis02, Brown20, Tuckett20}. These practical decoders demonstrate exceptionally high fault-tolerant thresholds when the dominant source of noise is dephasing~\cite{BonillaAtaides20}. 

The error correction capabilities of the XZZX code complement the biased noise of the Kerr-cat qubits. This motivates the present work where we demonstrate a very robust architecture for scalable fault-tolerant quantum computing by combining Kerr-cat qubits and the XZZX code. To this end, we perform elaborate simulations for a concatenated scheme undergoing a circuit error model that reflects noise sources that have been identified in recent experimental demonstrations of Kerr-cat qubits~\cite{Grimm2020}. Simulating these details enables us to express the threshold error rates we obtain in terms of real laboratory parameters to interrogate the experimental viability of our proposal directly.

\begin{figure}
    \centering
    \includegraphics[width=0.5\textwidth]{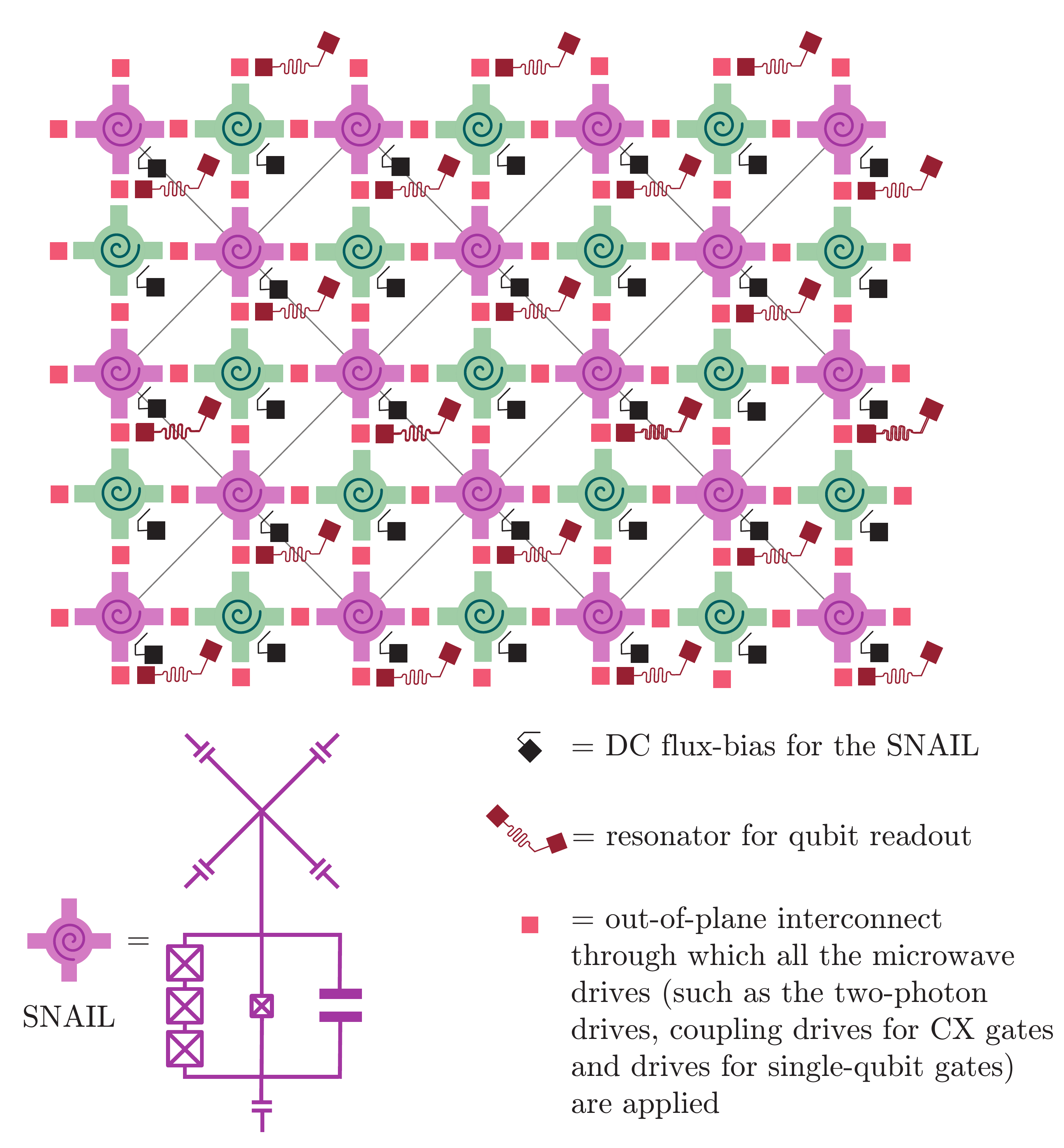}
    \caption{Schematic for a possible realization of the XZZX surface code using a 2D lattice of nearest-neighbor capacitively coupled SNAIL-based Kerr-cat qubits. Data qubits are represented in magenta and ancilla qubits in green. A DC flux line is present to bias the SNAIL at the point of desired nonlinearities. The lattice arrangement is such that neighboring SNAILs resonate at different frequencies to minimize cross-talk. Microwave drives applied at appropriate frequencies through the out-of-plane interconnects parametrically activate gate operations~\cite{Puri2020,Grimm2020}. Importantly, no extra nonlinear control/coupling elements are required unlike other biased noise bosonic qubits. For direct readout of ancilla and data qubits, an additional readout resonator is available at each site. This readout resonator also provides the ability to parametrically introduce two-photon dissipation in the SNAILs which could be used for autonomous leakage correction.  }
    \label{fig:snail_cat}
\end{figure}

\Cref{fig:snail_cat} shows what a Kerr-cat XZZX chip may look like if implemented with SNAILs. The dominant physical error channels for Kerr-cat qubits are photon loss and gain, quantified by a single-photon loss rate~$\kappa$ and the oscillator's thermal population $n_\mathrm{th}$~\cite{Grimm2020}. The noise model of the Kerr-cat qubit is a function of~$\kappa$, $n_\mathrm{th}$, as well as the ``cat size''~$|\alpha|^2$ and the Kerr-nonlinearity of the oscillator~$K$. With $|\alpha|^2=6.25$ and a thermal population of $8\%$, we find a threshold of $\kappa/K = 2.5 \times 10^{-4}$. This corresponds to a total infidelity of $ \sim 6.5\%$ for the noisiest logical operation in the error correction circuit, $\CX$, for which pure Pauli Z errors occur $\sim 351$ times more frequently than other Pauli errors during the gate.
%whose average bias is $\sim 351$ where we define bias as the ratio of the probability of Pauli $Z$ errors and the probability of any other Pauli error occurring during the gate. 
This noise threshold can be reached, for example, using a superconducting Kerr cat realization with $K/(2\pi) = \unit[10]{MHz}$, a parametric pump strength of $\unit[62.5]{MHz}$, and a single photon lifetime $\kappa^{-1} = \unit[63.6]{\mu s}$.
We remark that the experimental demonstration of the Kerr-cat qubit in Ref.~\onlinecite{Grimm2020} showed that frequency fluctuations are heavily suppressed when the SNAIL is pumped into the cat-subspace, such that requirements on flux noise (characterized by the $T_2$ time of the un-pumped SNAIL) are greatly relaxed compared to frequency tunable transmons.
These parameters are routinely achieved in superconducting circuits~\cite{jurcevic2021demonstration,kjaergaard2020superconducting}, suggesting that Kerr cats well below threshold may be realized in the near future, making the proposed architecture a promising approach to large-scale fault-tolerant quantum computing.

While our protocol achieves exceptional performance using Kerr-cat qubits, we also show that similar improvements may be attainable in other biased-noise architectures. 
Using the XZZX code with a generic biased circuit noise model we observe roughly a factor of two improvement in the threshold error rate over the standard CSS version of the surface code. Surprisingly, we still find that the XZZX code has a $50\%$ higher threshold error rate than the CSS surface code when using standard $\CX$ gates rather than bias-preserving $\CX$ gates. 

We also perform exact simulations on small codes that represent near-term experiments. These simulations allow us to study non-Pauli noise models with optimal decoders and reveal that the high performance of our scheme at both high and low biases is relatively insensitive to suboptimal implementation of gates and decoding compared to other schemes.

This paper is structured as follows. In \cref{s:sc_definitions} we review the definition of the XZZX surface code and describe how it can be implemented in a Kerr-cat architecture. In \cref{s:cat_definitions} we review Kerr-cat qubits and describe the bias-preserving gates available to them. In \cref{s:cat_sim} we describe the important noise processes affecting Kerr-cat qubits and the methods we have used to simulate noisy gate execution. In \cref{s:threshold} we present results of numerical simulations, demonstrating that very high thresholds can be achieved using XZZX codes and Kerr-cat qubits. 
The performance of low-distance codes under biased noise is presented in \cref{s:low_distance}. A discussion of the results along with potential future research directions are provided in \cref{s:discussion}. Details of the surface code simulation methods and some advantages of Kerr-cat qubits over purely dissipative cat qubits are included in \cref{s:simulation,s:kerr_diss2} respectively.

\section{The XZZX surface code}
\label{s:sc_definitions}

\begin{figure}
    \centering
    \includegraphics[width=0.45\textwidth]{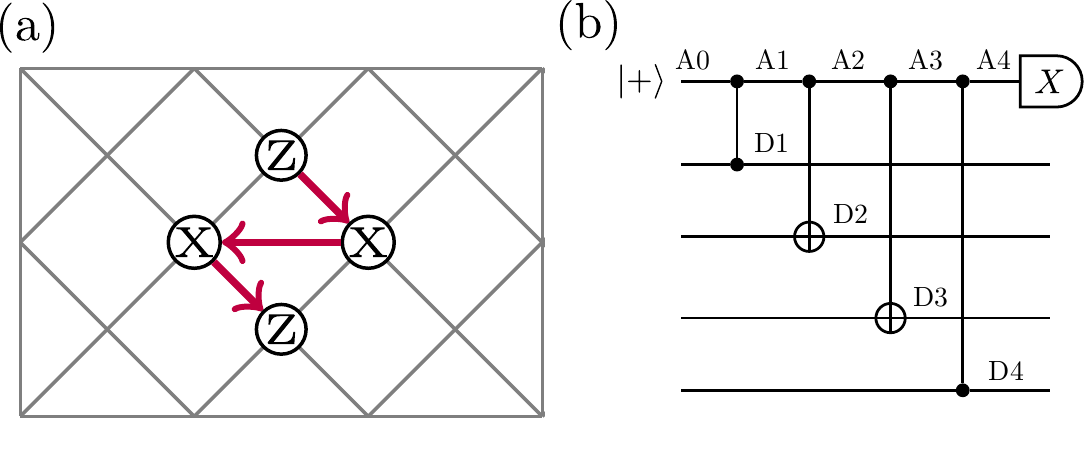}
    \caption{(a) An XZZX surface code check as a product of two $X$ and two $Z$ Pauli operators on qubits arranged on the vertices of a square lattice. Every face corresponds to a measured check operator and all checks (except for the three-qubit checks on the boundary) have the same form. 
    The order in which data qubits of a check are coupled to the ancilla (at the center of each face) is specified by the arrows. We have used the same order for every check.
    (b) Check measurement circuit for  the XZZX surface code. The four data qubits are coupled to an ancilla qubit prepared in the $\ket{+}$ state before measuring the ancilla in the $X$ basis.
    Distinct error locations on the ancilla qubits are labelled A0-4, and on data qubits are labelled D1-4. The error probabilities at these locations determine the weights of the syndrome graph for decoding.} 
    \label{fig:xzzx_check}
\end{figure}

We describe the XZZX code using the stabilizer formalism, i.e., a list of commuting Pauli operators whose common $+1$ eigenspace defines the codespace of the code. We lay out the code with a single qubit on each vertex of the square lattice. Up to its boundaries, all of its stabilizers are the product of four Pauli terms; $S_f =X\otimes Z\otimes Z\otimes X$ acting on the four qubits at the corners of each of the faces $f$ of the square lattice, see \cref{fig:xzzx_check}(a).

The XZZX code is locally equivalent~\cite{Nussinov09, Brown11} to the conventional CSS surface code~\cite{Kitaev03, Dennis02}. As such it inherits its code properties. We choose a lattice of size $\sim d_\mathrm{x}\times d_\mathrm{z} $ using $n = O(d_\mathrm{x} d_\mathrm{z})$ qubits, where $d_\mathrm{x}$ and $d_\mathrm{z}$ correspond to the lengths of the smallest $X$ and $Z$ logical operators, which have a string-like support and traverse homologically non-trivial paths between opposite boundaries of the code. With open boundary conditions, the XZZX code encodes one logical qubit.

\subsection{Stabilizer measurements}

We measure stabilizer operators to identify the locations of errors that occur on the qubits of the code. As the codespace is defined as the $+1$ eigenspace of the stabilizers, measuring a stabilizer in the ${-}1$ state indicates an error has occurred that takes the system out of its codespace. Our stabilizer readout circuit is based on that given in Ref.~\cite{Fowler12a}, up to the Hadamard gates that rotate the standard surface code onto the XZZX code. Here we briefly summarize the circuit we use.

We measure each stabilizer generator using the circuit shown in \cref{fig:xzzx_check}(b). All of the stabilizer generators can be measured in parallel. We place a single ancilla in the $|+\rangle$ state on each face. We subsequently apply two-qubit entangling gates between the ancilla and its four nearest-neighbor data qubits such that the ancilla collects the parity information for the stabilizer~$S_f$ on its respective face~$f$. We finally measure the ancilla in the Pauli-$X$ basis to learn the value of the check. 

In \cref{fig:xzzx_check}(a), we indicate the order in which qubits should be coupled to the ancilla. With the exception of qubits on the boundary, every data qubit is coupled with an ancilla qubit at each time step. The same ordering is used for checks on the boundary, except when there is no neighbouring ancilla or data qubit to couple with, at which time the qubit is left idle. This ordering enables parallel readout of all the stabilizer operators~\cite{Fowler12a}. The exception to this ordering is for the small $n=9$ codes in \cref{s:low_distance} where we use the rotated surface code layout. Here, on every second face the order of application of the two $\CX$ gates is swapped. For that particular layout, this order was shown in Ref.~\onlinecite{tomita_low-distance_2014} to mitigate the effect of hook errors, in which a single error on an ancilla can propagate to two data qubits. %, also help to mitigate the spread of the error.

In a realistic device, errors may occur during gate execution, state preparation, measurements and on idle qubits.
Repeated measurement of stabilizer generators provides data, called the syndrome, that allows us to correct for all of these types of errors, provided we adopt an appropriate decoding strategy.

 \subsection{Decoding}

A decoder is a classical algorithm that uses the error syndrome as an input to estimate the best choice of correction that can be applied to recover the encoded state.  We use a minimum-weight perfect-matching (MWPM) decoding algorithm to correct for all types of errors~\cite{Dennis02}. Note that correctly identifying the error, up to stabilizer equivalence, is equivalent to finding a correction for that error. We format the error syndrome as a list of defects, where we say that a defect appears at face $f$ and at time $t$ if $S_f(t-1) S_f(t) = -1$, where $S_f(t)$ denotes the outcome of the stabilizer measurement $S_f$ made at time $t$. Describing the error syndrome as a list of defects means that errors can be interpreted as strings where defects lie at their endpoints~\cite{Dennis02}. This interpretation enables us to employ MWPM to estimate the error that has occurred.

The MWPM algorithm takes a graph with weighted edges and returns a matching such that the sum of the weights of the matching is minimal. Given the error model and the syndrome, the graph input to MWPM is constructed such that each node corresponds to a defect, and the weight of an edge connecting two defects is $-\log p$, where $p$ is the probability of the most likely error that can give rise to that pair of defects. This choice of weights implies that the edges returned by the MWPM algorithm correspond globally to the most likely error consistent with the observed syndrome. 

This practical decoder is particularly well suited for decoding biased noise using the XZZX code. For the XZZX code, a $Z$ error on any data qubit will produce a pair of defects displaced in the same direction, independent of the qubit, and this is perpendicular to the displacement produced by an $X$ error.
Thus, when the noise is strongly biased towards $Z$ errors, the edge weights in the syndrome graph have much higher weight in one direction. This allows errors to be identified less ambiguously using MWPM, compared to when edges in both directions are more equally weighted, and results in a high threshold under biased noise.  
This was demonstrated using a phenomenological noise model in Ref.~\onlinecite{BonillaAtaides20}.

%We optimize the stabilizer readout circuit and our decoder to account for the correlated errors that occur when we consider the errors that individual components of stabilizer readout circuits might introduce. 
%We optimize the decoder to account for error correlations that components of stabilizer readout circuits introduce. 
%In more realistic noise models where we account for the behavior of stabilizer readout circuits, single error events can give rise to pairs of defects separated `diagonally' i.e. in both space and time as well as hook errors. We give details on how the decoder accounts for these correlations in \cref{s:circuit_decoder}\asd{maybe supplementary material}. One type of error, that we call a hook error, occurs when the ancilla qubit experiences a low rate error halfway through the stabilizer readout circuit [an $X$ error at A2 in \cref{fig:xzzx_check}(b)]. This error is copied onto two of the data qubits by the entangling gates.  However, as the hook error is caused by a low rate event, this effect of these errors is relatively benign for biased noise. 
In realistic circuit level noise, single error events can give rise to pairs of defects separated `diagonally' i.e. in both space and time as well as hook errors, which occur when the ancilla qubit experiences a $X$ error halfway through the stabilizer readout circuit [an $X$ error at A2 in \cref{fig:xzzx_check}(b)]. This error is copied onto two of the data qubits by the entangling gates.  However, as the hook error is caused by a low rate event, the effect of these errors is relatively benign for biased noise. Our decoder takes into account all of these types of errors by appropriately assigning edge weights in the construction of the syndrome graph.

\section{Cat qubits and bias preserving gates}
\label{s:cat_definitions}

\begin{figure}
 \includegraphics[width=\columnwidth]{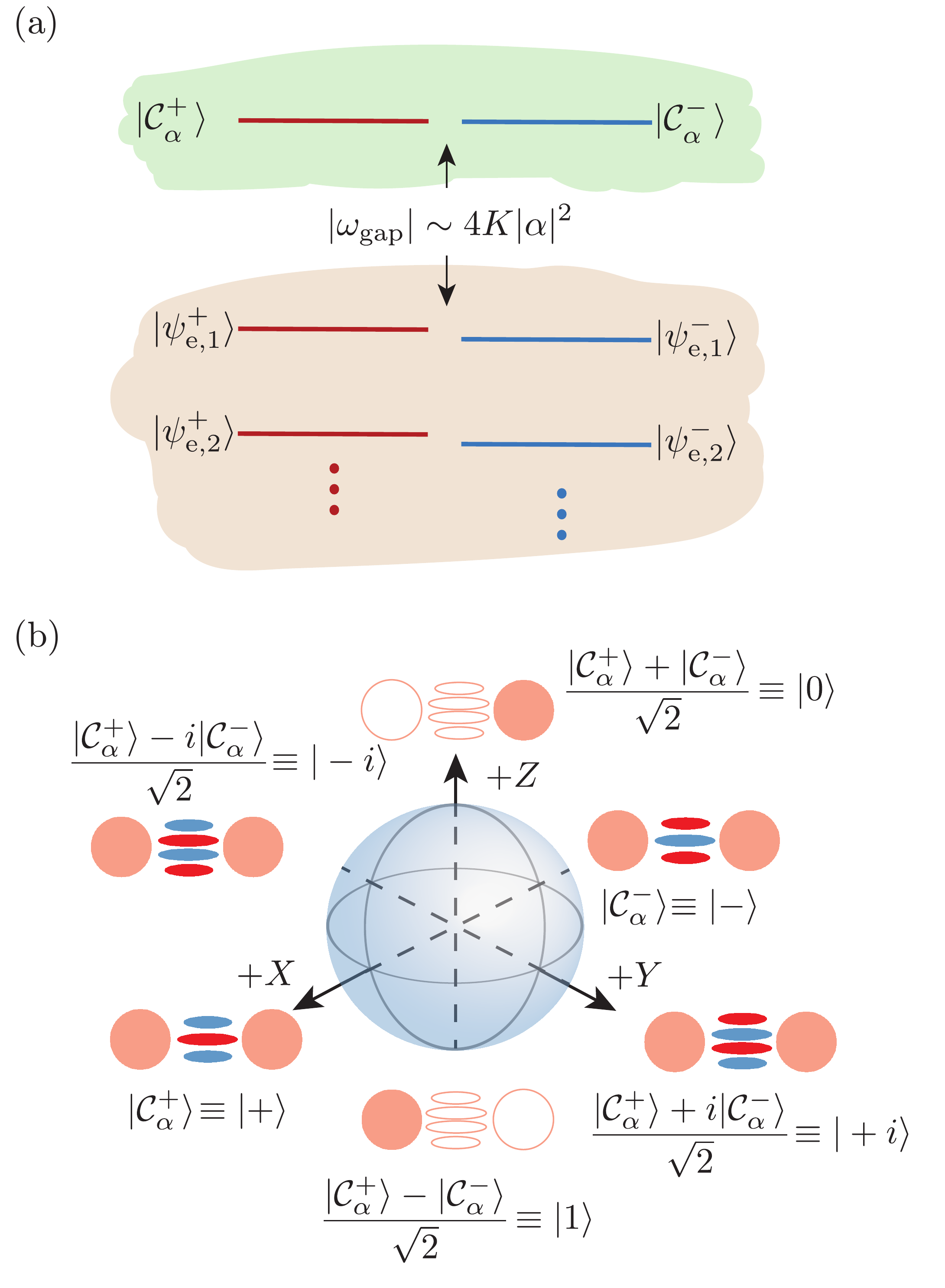}
 \caption{ (a) Eigenspectrum of the two-photon driven nonlinear oscillator in the rotating frame.
 The cat states $\puricatpm$ are exactly degenerate. The eigenspectrum can be divided into an even and an odd parity manifold. The cat subspace, highlighted in green, is separated from the first excited state by an energy gap $\omega_\mathrm{gap}|
 \sim4K\alpha^2$. (b) Bloch sphere of the cat qubit. The figure also shows cartoons of the Wigner functions corresponding to the eigenstates of Pauli-$X$,$Y$,$Z$ operators.  }
 \label{fig:bloch}
 \end{figure}

Here we describe the Kerr cat~\cite{Puri2017,Grimm2020} which, in our proposal, forms the elementary qubit from which the XZZX code is built. The Kerr-cat qubit is realized in a two-photon driven Kerr nonlinear oscillator. The Hamiltonian of such an oscillator in a frame rotating at the oscillator frequency $\omega_\mathrm{t}$ is given by
\begin{align}
{H}_\mathrm{cat}=-K\purihta^{\dag 2}\purihta^2+K\alpha^2(\purihta^{\dag 2}+\purihta^2).
\label{puri_H0_2}
\end{align}
Here, $K>0$ is the strength of the nonlinearity while $K\alpha^2$ is the amplitude of the two-photon drive. 
The orthogonal even- and odd-parity cat states $\puricatpm=\mathcal{N}_\pm (\puriket{\alpha}\pm\puriket{{-}\alpha})$ are degenerate eigenstates of this Hamiltonian. 
As illustrated in \cref{fig:bloch}, the degenerate cat-subspace $\mathcal{C} = \text{span}\{\puricatpm\}$ is separated from the rest of the Hilbert space $\mathcal{C}_\perp$ by an energy gap, which in the limit of large $\alpha$ is well approximated by $|\omega_\mathrm{gap}|\sim 4K\alpha^2$. We will work in the basis where the cat states $\puricatp \equiv\puriket{+}$ and $\puricatm\equiv \puriket{-}$ are aligned along the $X$-axis of the Bloch sphere, and their superpositions $(\puricatp+\puricatm)/\sqrt{2}\equiv  \puriket{0} {\simeq \ket\alpha}$ and $(\puricatp-\puricatm)/\sqrt{2}\equiv \puriket{1} {\simeq \ket{-\alpha}}$ are aligned along the $Z$-axis (where the approximations hold for large $|\alpha|$).

 The Kerr-cat qubit has been recently realized experimentally with a capacitively shunted SNAIL~\cite{Grimm2020}. The SNAIL is a flux-biased nonlinear device comprising of four Josephson junctions (\cref{fig:snail_cat}). Both third- and fourth-order (Kerr) nonlinearities exist in the SNAIL at an appropriate value of the external magnetic flux. Due to the third-order non-linearity, a single microwave drive applied at twice the resonance frequency generates the two-photon term in the Hamiltonian in \cref{puri_H0_2}. 
 
 It has been shown that the error channel of the Kerr-cat qubit is dominated by phase-flip errors while $X$ and $Y$ errors are exponentially suppressed with the average photon number $\braket{a\dg a} \sim |\alpha|^2$~\cite{Puri2020}. Some noise sources, such as thermal excitation, frequency-fluctuations, etc., may introduce leakage errors. Leakage can, to some extent, be autonomously corrected or ``cooled" by introducing any additional photon dissipation channel without adversely affecting the bias~\cite{Puri2020}. A two photon dissipation source is preferred for such autonomous leakage correction.
 
\begin{figure}
    \centering
    \includegraphics[width=0.5\textwidth]{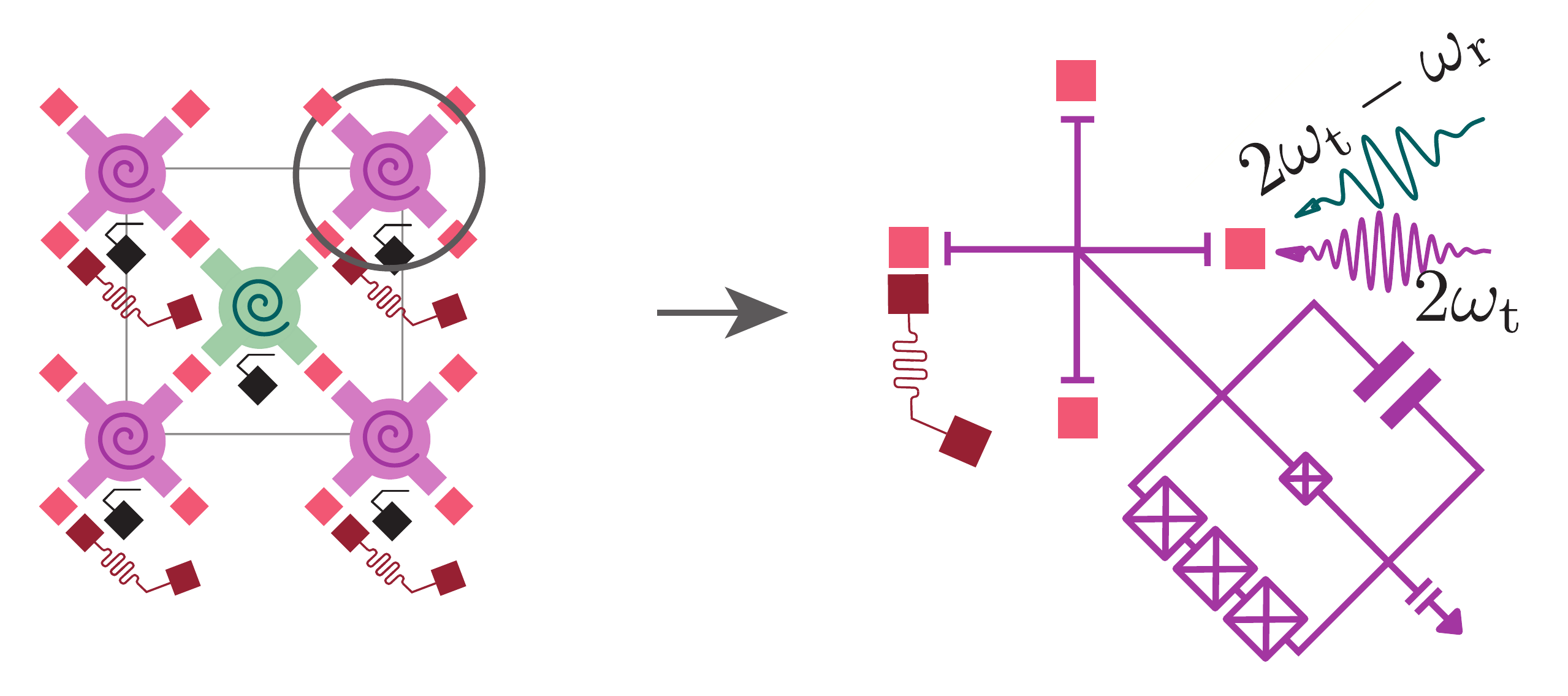}
    \caption{ A drive at $2\omega_\mathrm{t}$ applied to the SNAIL implements the two-photon drive required to realize the Kerr-cat Hamiltonian in \cref{puri_H0_2}. Here $\omega_\mathrm{t}$ is the resonance frequency of the SNAIL. It is also possible to activate the two-photon dissipation for autonomous leakage correction by applying a microwave drive at $2\omega_\mathrm{t}-\omega_\mathrm{r}$. Here $\omega_\mathrm{r}$ is the resonance frequency of the readout resonator coupled to the SNAIL. }
    \label{fig:snail1}
\end{figure}

 The inherent nonlinearity in the Kerr cat enables a straight-forward realization of two-photon dissipation for autonomous leakage correction. For example, consider the setup in \cref{fig:snail1} which is the building block of the surface code architecture. Suppose the resonance frequency of one of the data qubit SNAILs is $\omega_\mathrm{t}$. Each SNAIL is capacitively coupled to a readout resonator of frequency  $\omega_\mathrm{r}$. A microwave drive applied to the SNAIL at $2\omega_\mathrm{t}-\omega_\mathrm{r}$ can induce the two-photon dissipation  as follows.  Due to the Kerr-nonlinearity in the SNAIL, the drive photons are consumed to convert two photons in the SNAIL to one photon in the readout mode. The photon in the low-Q readout is rapidly lost to the environment, thus effectively inducing two-photon dissipation in the SNAIL. Depending on the drive amplitude, mixing between the readout and SNAIL mode, and lifetime of the readout, the strength of the two-photon dissipation achieved can be easily engineered to be $K/100-K/10$.

 Single-photon loss and thermal noise were observed to be the two main sources of errors in the experiment in Ref.~\onlinecite{Grimm2020}. Consequently,  we also consider these noise sources for the simulations in the present work. Importantly, the experiment found that non-phase-flip errors decreased with increasing cat size, confirming the theoretically predicted biased-noise property of the qubit.

 \subsection{Kerr cats and dissipative cats}
 \label{s:kerr_diss}
In our work, the choice of the Kerr cat as the biased-noise system used for the numerical analysis is motivated by several characteristics that makes it an excellent candidate for large-scale quantum computation. Firstly, the large nonlinearities inherently present in superconducting circuits ensures a large energy gap $|\omega_\mathrm{gap}|$, typically in the range of $10-300$ MHz. As further described in the next section, this large gap allows fast gates ($10-100$ ns). The typical lifetimes in these circuits, $10-100\mu$s~\cite{jurcevic2021demonstration,kjaergaard2020superconducting,sete2021parametric,kelly2015state}, are $\sim100-1000$ times longer than the gate times ensuring high-fidelity gate operations~\cite{Puri2020,Grimm2020}. Consequently, it becomes much easier to reach the threshold parameters estimated in our work with current experimental techniques. The inherent nonlinearity can also be exploited to implement all the  single- and multi-qubit gate operations and measurements using microwave pulses, thus eliminating any need for external coupling elements and providing a hardware efficient  solution for scaling up~\cite{Puri2020,Grimm2020}. 
\\
\indent 
We now contrast the Kerr-cat qubit with the purely dissipative cat qubit which also has a strongly biased noise channel~\cite{mirrahimi2014dynamically}. The dissipative cat is realized by engineering a two-photon drive and two-photon dissipation in a linear resonator via external nonlinear modes, such as a transmon~\cite{leghtas2015confining,Lescanne2020,chamberland2020building}. So far, the strength of the two-photon dissipation induced this way has been limited to $10-100$ KHz, leading to slow gate operations ($\gtrsim 1 \mu s$)~\cite{touzard2018coherent}.

In addition, the lack of self-Kerr nonlinearity in purely dissipative cats makes implementation of gate operations, preparation, and measurements more challenging. This is evident, for example, from the scheme for $X$-basis measurement. The Kerr-cat readout (see \cref{s:cat_sim} for more details) proceeds by effectively rotating the qubit around the $Y$ axis, which maps the $X$ basis states to the $Z$ basis. The $Z$ basis states are subsequently mapped onto a linear readout resonator and a homodyne measurement of its field reveals the state of the Kerr-cat qubit. A useful feature of this scheme is that the strength of the homodyne field, and hence measurement fidelity, increases with the size of the Kerr-cat. The $Y$ axis rotation is possible by freely evolving the qubit under the self-Kerr nonlinearity in the Kerr-cat mode. This rotation is  not possible in purely dissipative cats and hence this simple measurement scheme cannot be employed. 

In the case of dissipative cats, $X$-basis measurement proceeds by adiabatic deflation of the even and odd parity cat states to vacuum and single-photon Fock states respectively. The Fock states are subsequently read out via coupling to other nonlinear modes like a transmon or asymmetrically threaded SQUID ~\cite{Lescanne2020,chamberland2020building}. Compared to the Kerr-cat readout scheme, this scheme is slow and its fidelity is effectively independent of cat-size.

Another challenge with dissipative cats is that additional nonlinear coupling elements required to implement quantum operations may open up new sources of dissipation in the system which can reduce the noise bias~\cite{leghtas2015confining,touzard2018coherent}. Finally, during gate operations, dissipative cats may suffer from non-adiabatic phase-flip errors (see \cref{s:kerr_diss2})~\cite{chamberland2020building}. Because this extra source of phase-flips is absent in Kerr cats, it is easier to achieve the threshold parameters in practice.

Future theoretical and experimental research in optimizing the hardware and control architecture will certainly improve both biased noise cat platforms and may even lead to development of new biased-noise qubits~\cite{hanggli_enhanced_2020}. But because of the reasons outlined above and the current state of technology, this paper focuses on the XZZX code implementation with Kerr-cat qubits.

\subsection{Logical Cat Operations and Noise Channels}
\label{s:cat_sim}
We assume that at the beginning of the computation, each SNAIL (data and ancilla) has been pumped into the cat-qubit basis by the adiabatic pumping technique described in Refs.~\cite{Puri2017,Grimm2020}. Once in the logical cat subspace, the set of bias-preserving operations in the Kerr-cat platform, relevant for the XZZX code implementation, are the projective $X$-basis measurement, $\mathcal{M}_X$, and the gates $\{ S,S^\dag,\CZ,\CX \}$. We use the projective measurement $\mathcal{M}_X$ for state preparation $\mathcal{P}_{\puriket{+}}$. 
We will now describe how these operations can be implemented.

\subsubsection{$S,S^\dag$ gates}

Consider the Hamiltonian of a Kerr-cat qubit when, in addition to the two-photon drive, an additional resonant microwave drive of amplitude $\varepsilon$ is applied $H_Z =H_\mathrm{cat}+ (a \varepsilon^* + a\dg \varepsilon)$. Projecting the interaction onto the logical subspace gives $P_\mathcal C H_Z P_\mathcal C = \alpha (\varepsilon+\varepsilon^*) Z + \mathcal {O}^*(e^{-2\alpha^2})$, where $\mathcal P_C = \ket 0\bra 0 + \ket 1\bra 1 = \puricatp\puricatpb + \puricatm\puricatmb$ and $Z$ is Pauli-$Z$ in the logical cat basis (we use $\mathcal O^*[f(x)]$ notation to suppress polynomial factors in $x$~\cite{woeginger2008}). $S$ and $S\dg$ gates are implemented by applying the microwave drive, in phase $(\varepsilon=\varepsilon^*>0)$ and $\pi$  out-of-phase $(\varepsilon=\varepsilon^*<0)$ with the two-photon drive respectively for time $T_S=\pi/8\alpha\varepsilon$, see~\cref{fig:CX_circuit}(a). Clearly, by increasing the cat size and strength of the microwave drive $|\varepsilon|$ it is possible to decrease $T_S$, thereby decreasing the probability of  errors during the gate operation. Note that the drive can cause virtual transitions out of the computational cat subspace. Consequently, we must have $|\varepsilon|\ll  |\omega_\mathrm{gap}|$ so that these virtual transitions and leakage outside the qubit manifold $\sim |\varepsilon/\omega_\mathrm{gap}|^2$ remain negligible.

\subsubsection{$\CZ$ gate}

The implementation of a $\CZ$ gate between two Kerr cats requires a Hamiltonian interaction of the form $H_{ZZ} =H_\mathrm{cat,c}+H_\mathrm{cat,t}- J (a_\mathrm{c}\dg a_\mathrm{t} + a_\mathrm{c} a_\mathrm{t}\dg)+J(a_\mathrm{c}+a^\dag_\mathrm{c})$, see~\cref{fig:CX_circuit}(b) for the corresponding drive scheme. Here the subscripts $\mathrm{c},\mathrm{t}$ refer to the control and target qubits respectively. Evolution under this Hamiltonian for time $T_\CZ=\pi/8J\alpha^2$ realizes the $\CZ$ gate. Clearly, larger $\alpha$ and $J$ lead to a faster gate. To minimize virtual  transitions out of the computational cat subspace we must have $J\alpha \ll |\omega_\mathrm{gap}|$~\cite{Puri2017}.
The term $\propto (a_\mathrm{c}\dg a_\mathrm{t} + a_\mathrm{c} a_\mathrm{t}\dg)$ is a beamsplitter interaction between the two qubits and is realized by applying a drive to either of the two qubits at their frequency difference. The third-order nonlinearity in the SNAIL consumes the drive photons to convert photons in one SNAIL to photons in the other, effective realizing the beamsplitter Hamiltonian. The term {$\propto (a_\mathrm{c}+a^\dag_\mathrm{c})$} is realized by applying a resonant microwave drive to the control SNAIL.

\subsubsection{$\CX$ gate}
\begin{figure}
 \includegraphics[width=\columnwidth]{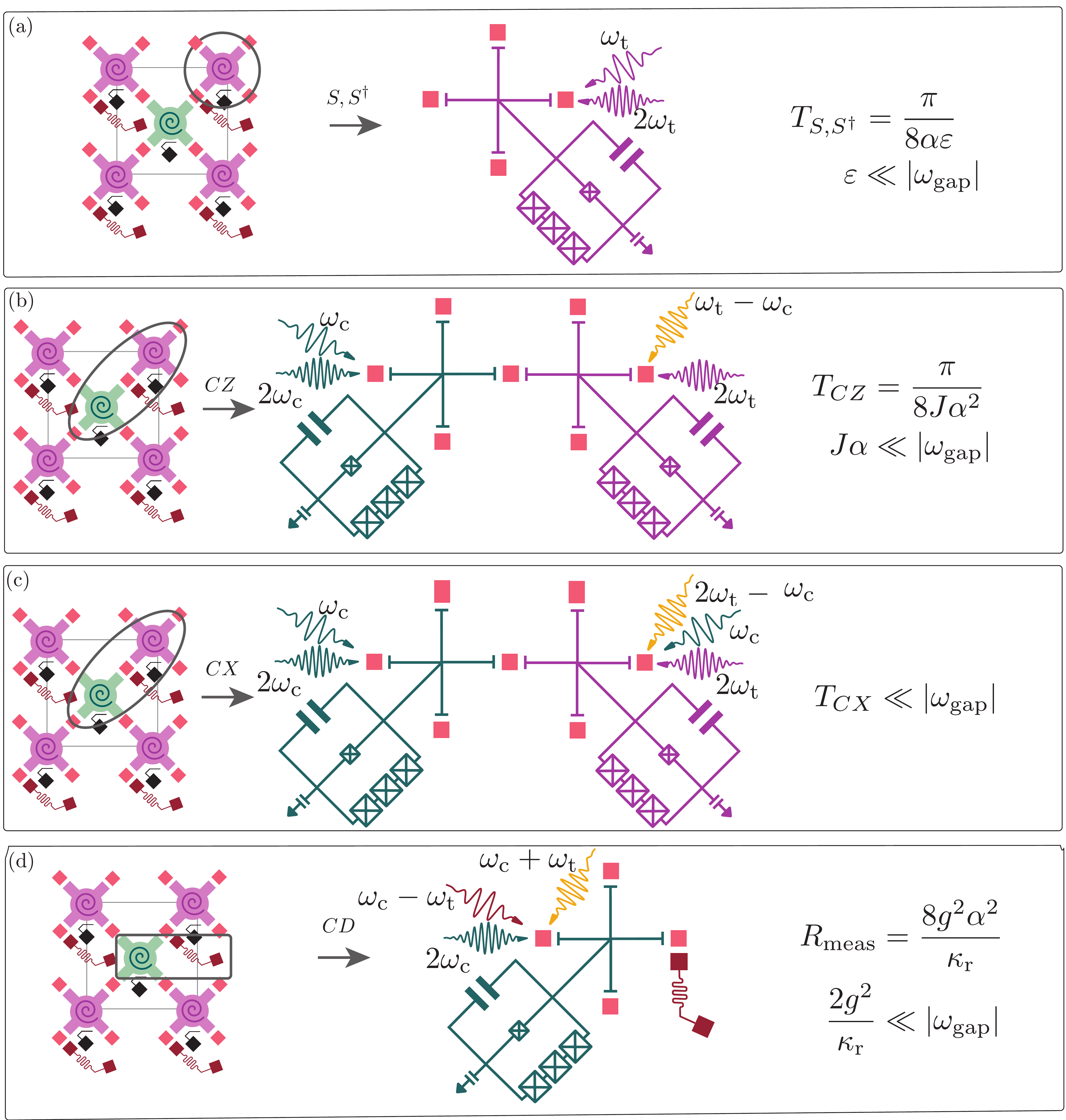}
 \caption{Schematic showing parametrically driven (a) $S,S^\dag$, (b) $\CZ$, (c) $\CX$, and (d) $\mathrm{CD}$ operations required to realize syndrome measurement circuits for the surface codes. The $X(\pi/4)$ gate required for ancilla readout is not shown here. It simply requires turning off the two-photon drive on the ancilla Kerr cat for time $\pi/2K$. The speed of the homodyne measurement after the $\mathrm{CD}$ operation is characterized by the rate $R_\mathrm{meas}=2\kappa_\mathrm{r} |\beta|^2$, where $|\beta|=2g\alpha/\kappa_r$ is the maximum of  readout-resonator displacement caused by the  Kerr-cat qubit~\cite{gambetta2006qubit}. }
 \label{fig:CX_circuit}
 \end{figure}

The diagonal $S$, $S\dg$, and $\CZ$ gates are trivially bias preserving. A more challenging gate to realize, however, is the bias-preserving $\CX$ gate. Such a gate can be achieved by exploiting the continuous variable nature of the cat encoding, and temporarily leaving the cat-code subspace during the gate. The intuition for the $\CX$ gate is best understood by first considering a bias preserving $X$ (or NOT) gate. An $X$ gate can be executed by adiabatically changing the phase of the two coherent states $\ket{\pm \alpha} \to \ket{\pm \alpha e^{i\phi(t)}}$ such that $\phi(0)=0$ and $\phi(T) = \pi$. At time $T$ we have thus mapped $\ket{\pm\alpha} \to \ket{\mp\alpha}$ and implemented an $X$ gate (in the large $\alpha$ limit where $\ket{0/1} \simeq \ket{\pm\alpha}$). Similarly, a $\CX$ gate can be executed if we can condition the phase $\phi(T)$ on the state of the control qubit. As shown in Ref.~\onlinecite{Puri2020} this can be achieved using a total Hamiltonian
\begin{equation}
    \begin{aligned}
    H_{CX} ={}& - K(a_\mathrm{c}^{\dag 2} - \beta^2)(a_\mathrm{c}^2 - \beta^2)\\
    &- K\left[a_\mathrm{t}^{\dag 2} - \alpha^2 e^{-i2\phi(t)}\left( \frac{\beta - a_\mathrm{c}\dg}{2\beta} \right) - \alpha^2 \left( \frac{\beta + a_\mathrm{c}\dg}{2\beta} \right)\right]\\
    &\times \left[a_\mathrm{t}^{2} - \alpha^2 e^{i2\phi(t)}\left( \frac{\beta - a_\mathrm{c}}{2\beta} \right) - \alpha^2 \left( \frac{\beta + a_\mathrm{c}}{2\beta} \right)\right] \\
    & - \frac{\dot\phi(t)}{4\beta} a_\mathrm{t}\dg a_\mathrm{t} ( 2\beta - a_\mathrm{c}\dg - a_\mathrm{c}).
    \end{aligned}
    \label{eq:cx}
\end{equation}
The interactions needed for the $\CX$ gate can be activated parametrically as shown in \cref{fig:CX_circuit}(c) (adapted from Ref.~\onlinecite{Puri2020}). The $\CX$ gate time is limited by the adiabaticity requirement $\dot{\phi}(t)\ll |\omega_\mathrm{gap}|$. 

\subsubsection{$\mathcal{M}_X, \mathcal{P}_{\ket{+}}$}
\begin{figure}
 \includegraphics[width=\columnwidth]{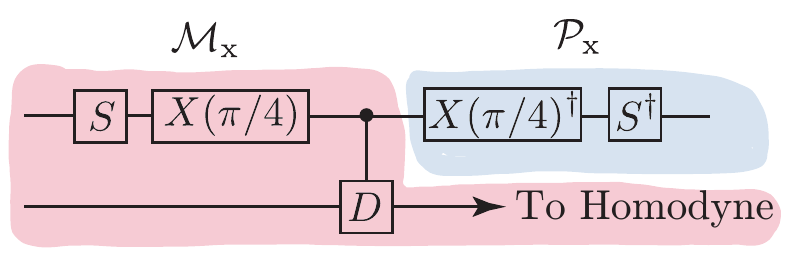}
 \caption{Circuit for projective measurement and subsequent preparation of basis states in the $X$ basis. The top line is the data qubit and the botton is the readout resonator. The controlled-displacement or CD gate maps the state of the data qubit to the displacement of the readout mode. A homodyne measurement of the signal from the readout resonator then reveals the state of the qubit.}
 \label{fig:meas_prep}
 \end{figure}

The remaining components required for the surface code implementation are the projective $X$-basis measurement, $\mathcal M_X$, and preparation $\mathcal{P}_{\ket{+}}$. The projective measurement proceeds in two steps. The first step comprises of rotating the $\ket{\pm}$ ancilla states to the computational basis  states $\ket{0,1}$ respectively, using an $S$ gate followed by $X(\pi/4)=(1+iX)/\sqrt{2}$ operation (see \cref{fig:meas_prep}).

The $X(\pi/4)$ gate is a non-bias preserving gate but since this gate is being applied to the ancilla during measurement the effect  of a non-dephasing  error is the same as a dephasing error~\cite{aliferis_fault-tolerant_2008}. It is implemented by turning  off the parametric drive and evolving the cats freely under the self-Kerr nonlinearity for time $\pi/2K$~\cite{puri2019stabilized,Grimm2020}. 

Once the $\ket{\pm}$ states are rotated  to $\ket{0,1}$, a controlled-displacement ($\CD$) operation between the Kerr-cat qubit and a linear readout resonator can be used to read out in the computational basis. The  required (rotating frame) interaction Hamiltonian is $H_\mathrm{readout}=H_\mathrm{cat}+g(a_1^\dag+a_1)( a_\mathrm{r}^\dag +a_\mathrm{r})/2$, where $a_1$, $a_\mathrm{r}$ are the annihilation operators for the  Kerr-cat  mode and the readout mode respectively and $g$ 
is the beam splitter coupling between the two modes~\cite{puri2019stabilized,Grimm2020}.  Projecting onto the qubit subspace gives $P_\mathcal{C}H_\mathrm{readout}P_\mathcal{C}=g\alpha Z_1(a_\mathrm{r}^\dag+a_\mathrm{r})$. From this projected Hamiltonian it is clear that the readout resonator experiences a displacement conditioned on the state of the qubit. 

If the single-photon decay rate of the readout mode is $\kappa_\mathrm{r}$, then the qubit-dependent displacement amplitude is $\sim\mp (2gi\alpha/\kappa_\mathrm{r})(1-e^{-\kappa_\mathrm{r}t/2})$. This  displacement amplitude can be subsequently  measured  via a typical homodyne measurement and the outcome of  the homodyne measurement can be used to infer the state of the ancilla qubit. Since the displacement of the readout-resonator increases with $\alpha$, the subsequent homodyne measurement of its field becomes more efficient as the cat-size is increased. Consequently, higher fidelity measurements are possible by making the cat-size larger.

The interaction Hamiltonian, $H_\mathrm{readout}$, can  be achieved by applying two microwave drives to the SNAIL: one at the frequency difference of the readout resonator and the SNAIL and the other at the sum of these two frequencies [see \cref{fig:CX_circuit}(d)]. The first drive gives rise to a beam-splitter type coupling $(a_1^\dag a_\mathrm{r}+a_\mathrm{r}^\dag a_1)$, while the second drive gives the squeezing type coupling $(a_1^\dag a^\dag_\mathrm{r}+a_\mathrm{r} a_1)$. The amplitudes of the two drives can be adjusted so that the strength of the beam-splitter and squeezing interactions are  the same leading to the desired Hamiltonian $H_\mathrm{readout}$. In the limit of large photon number cat qubits, only one drive at the difference frequency is sufficient. In this limit, the resulting beam-splitter interaction projected in the cat-basis is $P_\mathcal{C}(a_1^\dag a_\mathrm{r}+a_\mathrm{r}^\dag a_1)P_\mathcal{C}\sim g\alpha Z_1(a_\mathrm{r}^\dag+a_\mathrm{r})$ to a very good approximation [$\mathcal{O}^*(e^{-2\alpha^2})$]. This readout scheme, called {\it quadrature readout} has been demonstrated experimentally~\cite{Grimm2020}. After the measurement is complete, the ancilla is projected onto one of the computational basis states conditioned on the measurement  outcome. In order to re-initialize the ancilla in $\ket{+}$ or $\ket{-}$ for the next round of syndrome extraction, the states can be rotated from the computational basis by applying  $X(\pi/4)^\dag=X(-\pi/4)$ followed by $S^\dag$.

\section{Thresholds}
\label{s:threshold}
A crucial figure of merit for a quantum error-correcting code is the threshold: the maximum noise strength below which it becomes possible to efficiently reduce the noise strength on the logical level arbitrarily by increasing the number of physical qubits and gates. We perform numerical simulations to estimate the threshold noise strength using both a generic biased circuit level noise model in \cref{s:threshold_toy}, and a realistic Kerr-cat qubit noise model in \cref{s:threshold_cat}. The former demonstrates the generic performance increase of the XZZX code when using bias-preserving gates and operations. The latter provides a target noise strength in a Kerr-cat architecture based on a realistic noise model. 

The threshold is calculated by numerically computing logical failure rates at different sizes, and determining the maximum physical noise strength below which the logical error rate decreases exponentially with system size. In each simulated run, the circuit implementing $d_\mathrm{z}$ rounds of noisy syndrome extraction is applied followed by a single round of noiseless syndrome extraction. The final round can be thought of as a readout, in which each data qubit is measured individually, and without loss of generality can be considered noiseless as any measurement error in this round can be equally regarded as having occurred in the previous (noisy) round. The decoder is run on the observed syndrome and a logical failure occurs if the physical error and the correction, computed by the decoder, are homologically distinct. Further details of the Pauli simulation method are provided in \cref{s:pauli_simulation}.

As is standard in most numerical studies of quantum error correcting codes, a Pauli noise model is used in these threshold calculations, enabling efficient simulation of large system sizes. Although the noise model for Kerr-cat qubits obtained through numerically solving the Lindblad master equation is not a Pauli channel, we have demonstrated with exact simulations of smaller systems, described in \cref{s:pauli_approximation}, that a Pauli approximation to the noise does not appear to significantly affect calculated logical error rates.

The results we present in this section for the XZZX code are obtained with the conventional open-boundary layout of the surface code, which has three-qubit checks on the boundary~\cite{Bravyi1998}. Alternative boundary conditions (including periodic in both in space and time) were also tested, and similar threshold values were obtained in all cases (despite large differences in logical error rates). 

As was shown in Ref.~\onlinecite{BonillaAtaides20}, with the standard layout and open boundary conditions, the aspect ratio of the code can be tuned to the bias. By increasing the length of the logical $Z$ operator compared to the logical $X$ operator, an overall lower logical error rate can be obtained for a given number of qubits. Unless otherwise specified, we determine the aspect ratio using small systems by varying the length of the code relative to its width until the logical $X$ error rate is comparable to the logical $Z$ error rate near threshold. This aspect ratio is then fixed when scaling up the code. To make more system sizes computationally accessible, we restrict to aspect ratios such that the length of the code is a multiple of its width. As with boundary conditions, varying the aspect ratio does not appear to appreciably change the value of the threshold, provided the aspect ratio is fixed as the system is scaled up.

We also perform simulations of the standard CSS version of the surface code and an alternative variant of the surface code proposed for biased noise which we call the tailored surface code (TSC). The TSC is simply the standard surface code, but with $Z$ checks replaced with $Y$ checks, such that a single phase flip error on a data qubit creates four adjacent defects. This version of the surface code was studied in Refs.~\onlinecite{Tuckett18, Tuckett19}, where it was found to have strong resilience to phase-flip noise. However, we found that the thresholds obtained with the TSC with circuit level noise and a modified MWPM decoder were somewhat lower than those of the XZZX code, so we have not presented them here. Nevertheless, the TSC may have some advantages for finite-sized systems, which we discuss in \cref{s:low_distance}. 

\subsection{Generic biased circuit noise thresholds}
\label{s:threshold_toy}
First, we estimate the threshold of the XZZX surface code using a generic biased circuit noise model. We compare  this to the threshold of the standard CSS version of the surface code under the same noise model and find that the XZZX code has a significantly higher threshold. Surprisingly, we also find that the XZZX code maintains a large advantage over the CSS surface code even if a standard $\CX$ gate for two-level qubits, rather than a bias-preserving $\CX$ gate, is used. 

\begin{figure*}[!tbhp]
    \centering
      \includegraphics[width=\linewidth]{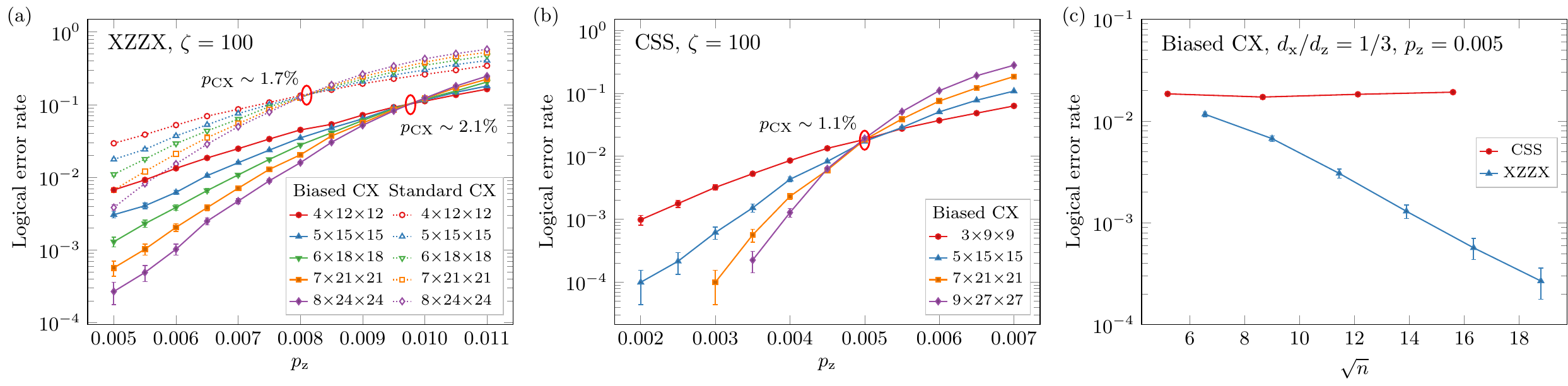}
    \caption{Error thresholds of surface code variants with a generic circuit noise model and bias $\zeta=100$.
    In (a), we compare the threshold of the XZZX code using bias-preserving CX gates (solid lines) to its threshold using standard non-bias-preserving CX gates (dotted lines).
    We find that the availability of bias-preserving CX gates leads to a $\sim 25\%$ improvement in the threshold.
    This result highlights the fact that along with total gate error, structure of noise in the gate plays an important role in determining the performance of the surface code variant.
    In (b), the threshold of the CSS code using bias-preserving CX gates is shown for comparison and is seen to be approximately half that of the XZZX code using the same bias-preserving gates.
    In this model, both types of CX gate have error rates approximately equal to $2p_\mathrm{z}$ (see text for details of the noise model).
    The crossing points are circled, and the total $\CX$ error probability $p_{\CX}$ at these points is shown.
    In (c), we plot the logical error rate of both code variants against $\sqrt{n}$ near the threshold of the CSS code, found in (b), where $n$ is the number of data qubits. 
    In all plots, error bars indicate one standard deviation.
    }
    \label{fig:cartoon}
\end{figure*}

The generic noise model has two parameters $p_\mathrm{z}$ and the noise bias $\zeta$. In the following we assume, without loss of generality, that the Pauli errors are applied after the gate. On the $\rm CZ$ gates, the errors $Z_\mathrm{a}\otimes I_\mathrm{d}$ and $I_\mathrm{a}\otimes Z_\mathrm{d}$, are each assigned the probability $p_\mathrm{z}$ and all other two-qubit Pauli errors are assigned the probability $p_\mathrm{z}/\zeta$. We remark  that in this simplistic toy model we are assuming probability of correlated $Z_\mathrm{a}\otimes Z_\mathrm{d}$ error to be the same as the probability of non-dephasing errors $p_\mathrm{z}/\zeta$. 
This gives a total error probability (or infidelity) of $p_\mathrm{z}(2+13/\zeta)$ per $\rm CZ$ gate. Here  the subscripts $\mathrm{d},\mathrm{a}$ refer to the data and ancilla qubits respectively. Single qubit noise including idle noise on data qubits and noise during ancilla preparation have a single qubit Pauli error model with $X$, $Y$ and $Z$ error probabilities given by $p_\mathrm{z}/\zeta$, $p_\mathrm{z}/\zeta$ and $p_\mathrm{z}$ respectively. Faulty measurements are modelled by flipping the  ancilla measurement outcomes with probability $p_\mathrm{z}+p_\mathrm{z}/\zeta$.

 In the case of ordinary discrete two-level qubits, evolution under an interaction Hamiltonian of the form $\chi\left[(1+Z_\mathrm{a})I_\mathrm{d}/2+(1-Z_\mathrm{a})X_\mathrm{d}/2\right]$ for time $\pi/2\chi$ would be required for the $\CX$ operation. It is easy to show that if the probability of a $Z$ error during the gate is $p_\mathrm{z}$, then the resulting error channel is (writing only the Pauli-terms for brevity) $p_\mathrm{z} Z_\mathrm{a}I_\mathrm{d} \rho Z_\mathrm{a}I_\mathrm{d}+p_\mathrm{z}'I_\mathrm{a}Z_\mathrm{d} \rho I_\mathrm{a}Z_\mathrm{d}
+p_\mathrm{z}'Z_\mathrm{a}Z_\mathrm{d} \rho Z_\mathrm{a}Z_\mathrm{d}+
p_\mathrm{z}''I_\mathrm{a}Y_\mathrm{d} \rho I_\mathrm{a}Y_\mathrm{d}+
p_\mathrm{z}''Z_\mathrm{a}Y_\mathrm{d} \rho Z_\mathrm{a}Y_\mathrm{d}
$ with $p_\mathrm{z}'=(2p_\mathrm{z}/\pi)\int_0^{\pi/2}\cos^4(\phi)d\phi\sim 0.375p_\mathrm{z} $ and $p_\mathrm{z}''=(2p_\mathrm{z}/\pi)\int_0^{\pi/2}\cos^2(\phi)\sin^2(\phi)d\phi\sim 0.125p_\mathrm{z} $. Consequently, we use this equation for probability assignment for the $I_\mathrm{a}Z_\mathrm{d},Z_\mathrm{a}I_\mathrm{d},Z_\mathrm{a}Z_\mathrm{d},I_\mathrm{a}Y_\mathrm{d}$, and $Z_\mathrm{a}Y_\mathrm{d}$ errors, while all other two-qubit Pauli errors are assigned the probability $p_\mathrm{z}/\zeta$. The total infidelity of the $\CX$ gate in this case is $p_\mathrm{z}(2+10/\zeta)$.

For the bias-preserving $\CX$ gate in the generic noise model, unlike its ordinary counterpart described above, phase-flip errors on the data qubits may convert to $Z_\mathrm{a}\otimes Z_\mathrm{d}$ but they do not convert to $Z_\mathrm{a}\otimes Y_\mathrm{d}$ or $I_\mathrm{a}\otimes Y_\mathrm{d}$ errors. Consequently, in this case $I_\mathrm{a}\otimes Z_\mathrm{d}$ and $Z_\mathrm{a}\otimes Z_\mathrm{d}$ errors are assigned the probability $p_\mathrm{z}/2$, $Z_\mathrm{a}\otimes I_\mathrm{d}$ is assigned the probability $p_\mathrm{z}$ and all other two-qubit Pauli errors are assigned the probability $p_\mathrm{z}/\zeta$. Note that total infidelity of the CX gate in this case is larger than the infidelity of the generic non-bias-preserving $\CX$ gate by $2p_\mathrm{z}/\zeta$.
For large $\zeta$, for example $\zeta=100$ used here, this difference becomes negligibly small.  The resulting logical error probability vs $p_\mathrm{z}$ is plotted for the surface code with both bias preserving and non-bias preserving $\CX$ gates in \cref{fig:cartoon}(a). 

For comparison, we also perform simulations of the conventional CSS surface code using the bias-preserving $\CX$ gate and the same generic biased error model. 
In this case, we use the `rotated' layout of Ref.~\onlinecite{bombin_optimal_2007}  with open boundary conditions and 2-qubit checks on the boundary, which has a code distance that is a factor of $\sqrt{2}$ larger than the XZZX code on the unrotated geometry using an equal number of physical qubits. 
The resulting logical error rates are shown in \cref{fig:cartoon}(b). We do not use the rotated layout for the XZZX code, since the all $Z$ logical operator in this case is oriented at $45^{\circ}$ to the lattice boundaries, and we therefore cannot improve the resilience of the code against phase flips by varying the code's aspect ratio (unlike with the standard layout).
We also present the logical failure rate as a function of $\sqrt{n}$ for both the XZZX code and the CSS surface code at physical error rate $p = 0.005$ in \cref{fig:cartoon}(c).

The procedure for choosing the aspect ratio for the XZZX and CSS codes using bias-preserving $\CX$ gates is as described above, which yields the same aspect ratio for both codes near threshold of around $3:1$. To simplify comparison, we use the same aspect ratio for the XZZX code with standard CX gates, however, this may not be optimal.

As can be seen, the XZZX code using a bias-preserving gate achieves the highest threshold $p_\mathrm{z}=0.98\pm0.05\%$, corresponding to a two-qubit error rate of close to 2\%. This is approximately double the threshold of the CSS surface code with the same noise model, which achieves a threshold of $p_\mathrm{z}=0.50\pm0.05\%$. The XZZX code with a standard non-bias-preserving CX gate performs worse than the XZZX code with a bias-preserving gate, achieving a threshold of $0.80\pm 0.05\%$.  However this is clearly much higher than the CSS surface code threshold, suggesting that the XZZX code may be a good choice in biased-noise architectures, even when a bias-preserving $\CX$ gate is not available.

\subsection{Kerr-cat noise thresholds}
\label{s:threshold_cat}
To obtain the noise models used for threshold estimates of the XZZX Kerr-cat code, we first perform a master equation simulation of the operations needed for the syndrome extraction circuit (\cref{fig:xzzx_check}): the $\CX$ and $\CZ$ gates, as well as measurement and preparation schemes described above. The simulations are performed for two cat sizes, $\alpha^2=1$ and $\alpha^2=6.25$. 

The noise channel of the larger cat is more biased than that of the smaller one. Based on the observations in the recent experimental work~\cite{Grimm2020}, we subject the qubit to a white thermal noise channel represented by the Lindbladian $\kappa_1(1+n_\mathrm{th})\mathcal{D}[a]\rho+\kappa_1 n_\mathrm{th}\mathcal{D}[a^\dag]\rho$. Thermal noise introduces leakage and in order to autonomously correct some of the leakage we add an additional two-photon dissipation to the Lindbladian $\kappa_2\mathcal{D}[a^2]\rho$, where the rate of two-photon dissipation $\kappa_2$ is fixed to $\kappa_2=K/10$.

When the two-photon dissipation is added care must be taken to adjust the phase of the two-photon drive to preserve the cat-qubit subspace~\cite{Puri2020}. In our simulations, the two-photon dissipation is added \emph{while} the gates are being implemented for all but the $\CX$ gate. The $\CX$ gate requires a time-dependent two-photon dissipation which becomes hard to numerically simulate. In practice, it also increases the number of drives that must be carefully phase-tuned in an experiment.

In order to avoid these complications, we simulate the $\CX$ gate with the following  steps. First the master equation with the Hamiltonian in \cref{eq:cx} and dissipative terms $\mathcal{L}[\rho]=\sum_{i=c,t}\kappa_1(1+n_\mathrm{th})\mathcal{D}[a_i]\rho+\kappa_1 n_\mathrm{th}\mathcal{D}[a^\dag_i]\rho$ is simulated with $\phi(t)=\pi t/T$ for time $T$. Next, the interactions between the control and target Kerr cats are removed and the Hamiltonian of the system is set to that of two uncoupled Kerr cats. Finally, the master equation with the two uncoupled Kerr cats is simulated with time-independent two-photon dissipation, $\sum_{i=c,t} \kappa_2\mathcal{D}[a^2_i]\rho$, in addition to $\mathcal{L}[\rho]$ for time $T$~\cite{Puri2020}.

Of course, more phase-flip errors are introduced in the second step of the simulation. The resulting error probabilities in this two-step $\CX$ gate, which takes time $T_{\CX}=2T$, will be roughly twice as high as that in the $\CX$ gate simulated with time-dependent two-photon dissipation in one step, which takes time $T_{\CX}=T$. 

For estimating measurement errors, we performed a master equation simulation of the $X(\pm \pi/4)$ and $S^{(\dagger)}$ gates. 
Subsequent steps of CD and finite efficiency homodyne measurements are simulated by stochastic master equation simulations. We assume a homodyne measurement efficiency of $60\%$~\cite{touzard2019gated}. We have found quantitative agreement between the analytic expression for homodyne measurement fidelity in Ref.~\onlinecite{didier2015fast} and numerical results from stochastic master equation simulations. For $\alpha^2=6.25$ cat we use the beam-splitter Hamiltonian $H_\mathrm{readout}=H_\mathrm{cat}+g(a_1^\dag a_\mathrm{r}+a_\mathrm{r}^\dag a_1)$, while for $\alpha^2=1$ we use $H_\mathrm{readout}=H_\mathrm{cat}+g(a_1^\dag+a_1)( a_\mathrm{r}^\dag +a_\mathrm{r})/2$. 
Note that, if the probability of bit-flip errors during the $\mathrm{CD}$ gate is small (a condition which is satisfied in the biased noise cat-qubit) this step in the measurement protocol is equivalent to the well studied longitudinal readout scheme~\cite{didier2015fast}. Consequently, known analytic expressions for longitudinal readout error may be used.

Observe from \cref{fig:CX_circuit} that all gate operations [other than $X(\pi/4)$] become faster with increasing cat-size. For a given $\alpha$ we choose the gate speeds so that the probability of leakage outside the cat-manifold is at least an order of magnitude smaller than total infidelity. This choice is justified as the goal, when designing pulses for experimental implementation of gates, is to not limit the gate fidelity by the amount of leakage. While no pulse shaping is being used in our simulations, such techniques combined with DRAG can be used to  further reduce leakage without sacrificing the gate fidelity~\cite{werninghaus2021leakage,chen2016measuring}. In our simulations, the leakage probability  per qubit, per stabilizer round is between $10^{-3}$ and $10^{-4}$, which is consistent with recent experiments~\cite{mcewen2021removing}. Additionally, the two-photon dissipation channel applied during or after the gates also checks the leakage growth. For the cat sizes considered here, the two-photon dissipation after every $\CX$  gate reduces the leakage by a factor of $\sim 10$.

In this way, in our simulations of the qubit operations, we suppress leakage to a reasonable amount at the physical level but then neglect the residual leakage in the further surface code simulations. While, this is the dominant practice in literature for studying the performance of a code, further investigation must be carried out to model, quantify, and counteract the effect of leakage in the Kerr-cat-surface code architecture. Like in other platforms, further suppression of leakage may require leakage reduction units (LRUs). These units reduce leakage errors to errors in the computational basis~\cite{aliferis2005fault,suchara2015leakage}. In the case of Kerr cats, leakage in one qubit is unlikely to propagate to others during two-qubit gates. Moreover, leakage is primarily confined to the lower energy subspace so that the computational-basis errors induced after LRUs remain biased. Of course, additional circuit complexity of the LRUs may effect thresholds. Assessing the performance of the surface code with LRUs optimized for Kerr cats is beyond the scope of the current work, but will be the topic of future study.

We simulate the XZZX surface code using the noise channel obtained numerically as described above. \Cref{table2,table1} list the average fidelities, bias, and leakage in different operations obtained from numerical simulations for cat qubits with $\alpha^2=1.0$ and $\alpha^2=6.25$ respectively.

\begin{table}[htbp]
\footnotesize
\centering
\begin{tabular}[c]{c c c}
%\toprule
\multicolumn{3}{c}{ $\alpha^2=1.0$, $\kappa_2=K/10$, $n_\mathrm{th}=8\%$}\\
\multicolumn{3}{c}{$\CX$ gate, $T_\CX=112.6/K$}\\
\hline
Single-photon loss rate & Infidelity   & Average bias  \\
% loss rate  & probability  & bias \\
   & Biased-$\CX$   & Biased-$\CX$  \\
\hline
%\midrule
& &  \\
$0.67\times 10^{-4}$  & 0.02 &  16.3  \\%& $4.2\times 10^{-5}$\\ 
$0.83\times 10^{-4}$  & 0.024 &  17.2 \\%& $5.2\times 10^{-5}$\\ 
$1.11\times 10^{-4}$  & 0.031 &  16.4 \\%& $7.0\times 10^{-5}$\\ 
$1.45\times 10^{-4}$  & 0.041 &  16.5 \\%& $9.1\times 10^{-5}$\\ 
$1.67\times 10^{-4}$  & 0.046 &  16.8 \\%& $1\times 10^{-4}$\\ 
\hline
\\
\multicolumn{3}{c}{CZ gate, $J\sim 0.013K$, $T_\CZ\sim 30.8/K$}\\
\hline
Single-photon loss rate & Infidelity   & Average bias \\%& Leakage \\
% loss rate  & probability  & bias \\
\hline
%\midrule
& &  \\
$0.67\times 10^{-4}$  & 0.006 & 9.6  \\%& $1.2\times 10^{-4}$\\ 
$0.83\times 10^{-4}$  & 0.008 & 10.6  \\%& $1.3\times 10^{-4}$\\ 
$1.11\times 10^{-4}$  & 0.01 & 12.5 \\%& $1.5\times 10^{-4}$\\ 
$1.45\times 10^{-4}$  & 0.013 & 13.2  \\%& $1.6\times 10^{-4}$\\ 
$1.67\times 10^{-4}$  & 0.014 & 14.5  \\%& $1.7\times 10^{-4}$\\ 
\hline
\\
\end{tabular}
\\
\begin{tabular}[c]{c c}
\multicolumn{2}{c}{Preparation, $\varepsilon= 0.026K$, $T_\mathrm{P}\sim 16.5/K$}\\
\hline
{Single-photon loss rate}  & Infidelity \\%& Leakage \\
%{loss rate}  &    \\
\hline
%\midrule
&   \\
{$0.67\times 10^{-4}$}  & 0.0015  \\%& $9.1\times 10^{-5}$\\ 
{$0.83\times 10^{-4}$}  &  0.0019 \\%& $9.5\times 10^{-5}$\\ 
{$1.11\times 10^{-4}$}  & 0.0027   \\%& $1.0\times 10^{-4}$\\
{$1.45\times 10^{-4}$}  &  0.0032 \\%& $1.1\times 10^{-4}$\\ 
{$1.67\times 10^{-4}$}  & 0.004  \\%& $1.2\times 10^{-4}$\\ 
\hline
\\
\end{tabular}
\begin{tabular}[c]{c c}
\multicolumn{2}{c}{Measurement, $\kappa_\mathrm{r}=0.1K$, $g\sim 0.06K$}\\
\multicolumn{2}{c}{$T=T_\mathrm{P}+T_\mathrm{CD}, T_\mathrm{CD}=42/K$}\\
\multicolumn{2}{c}{Measurement efficiency$=60\%$}\\
\hline
Single-photon loss rate  & Total error  \\
\hline
%\midrule
$0.67\times 10^{-4}$  & 0.017  \\ 
$0.83\times 10^{-4}$  & 0.017  \\ 
$1.11\times 10^{-4}$  &  0.018 \\
$1.45\times 10^{-4}$  & 0.019 \\ 
$1.67\times 10^{-4}$  & 0.020 \\ 
\hline
\end{tabular}
\caption{Main parameters characterizing the operations used in syndrome extraction for $1$-photon Kerr-cat qubit. The average gate bias is defined as the ratio of the sum of the probabilities of $I\otimes Z$, $Z\otimes I$, and $Z\otimes Z$ errors and the sum of the remaining two-qubit Pauli errors. The effective Pauli noise channel of preparation and measurement operations are naturally biased. The total error in preparation is simply the sum of the probability of a $Z$ and a $Y$ error. Like for $\alpha^2=6.25$, the $\CX$ is  carried  out in two steps so  that $T_{\CX}=2T$  with $T$  the time of each step. Leakage in all the operations is $\sim 10^{-4}-10^{-5}$. }
\label{table2}
\end{table}

\begin{table}[htbp]
\footnotesize
\centering
\begin{tabular}[c]{c c c}
%\toprule
\multicolumn{3}{c}{ $\alpha^2=6.25$, $\kappa_2=K/10$, $n_\mathrm{th}=8\%$}\\
\multicolumn{3}{c}{$\CX$ gate, $T_{\CX}=18.47/K$ (bias-preserving)}\\
\multicolumn{3}{c}{$T_{\CX}=15.0/K$ ($R\hspace{0.1cm}\CZ\hspace{0.1cm}R^\dag$)}\\
\hline
Single-photon loss rate & Infidelity   & Average  bias \\
 %loss rate  & probability  & bias \\
   & Biased-$\CX$   & Biased-$\CX$  \\
    & ($R\hspace{0.1cm}\CZ\hspace{0.1cm}R^\dag$)  &  ($R\hspace{0.1cm}\CZ\hspace{0.1cm}R^\dag$) \\
\hline
%\midrule
& &  \\
$1.67\times 10^{-4}$  & 0.047 (0.035) &  319 (11.7)\\% & $2.6\times 10^{-5}$\\ 
$2.08\times 10^{-4}$  & 0.054 (0.043) &  337 (12.0)\\%& $2.7\times 10^{-5}$\\ 
$2.77\times 10^{-4}$  & 0.071 (0.057) &  357 (12.4)\\%& $2.8\times 10^{-5}$\\ 
$4.16\times 10^{-4}$  & 0.104 (0.084) &  375 (12.7)\\%& $4\times 10^{-5}$\\ 
\hline
\\
\multicolumn{3}{c}{CZ gate, $J\sim 0.08K$, $T_\CZ\sim 0.8/K$}\\
\hline
Single-photon loss rate & Infidelity   & Average bias \\
%   & probability  & bias \\
\hline
%\midrule
& &  \\
$1.67\times 10^{-4}$  & 0.0029 &  $1.7\times 10^3$ \\%& $1.9\times 10^{-4}$\\ 
$2.08\times 10^{-4}$  & 0.0036 &  $1.9\times 10^3$ \\%& $1.9\times 10^{-4}$\\ 
$2.77\times 10^{-4}$  & 0.0039 &  $2.3\times 10^3$  \\%& $2.0\times 10^{-4}$\\
$4.16\times 10^{-4}$  & 0.0056 &  $3.1\times 10^3$ \\%& $2.1\times 10^{-4}$\\ 
\hline
\\
\end{tabular}
\\
\begin{tabular}[c]{c c}
\multicolumn{2}{c}{Preparation, $\varepsilon\sim 0.15K$, $T_\mathrm{P}=2.6/K$}\\
\hline
Single-photon loss rate & Infidelity \\
%{}  &    \\
\hline
%\midrule
&   \\
{$1.67\times 10^{-4}$}  & 0.003  \\%& $2.8\times 10^{-4}$\\ 
{$2.08\times 10^{-4}$}  & 0.004  \\%& $3.5\times 10^{-4}$\\ 
{$2.77\times 10^{-4}$}  & 0.005   \\%& $4.4\times 10^{-4}$\\
{$4.16\times 10^{-4}$}  & 0.007  \\%& $6.8\times 10^{-4}$\\ 
\hline
\\
\end{tabular}
\begin{tabular}[c]{c c}
\multicolumn{2}{c}{Measurement, $\kappa_\mathrm{r}=1.0K$, $g\sim 0.5K$}\\
\multicolumn{2}{c}{$T_\mathrm{M}=T_\mathrm{P}+T_\mathrm{CD}, T_{\mathrm{CD}}=2.5/K$}\\
\multicolumn{2}{c}{Measurement efficiency$=60\%$}\\
\hline
Single-photon loss rate  & Infidelity  \\
\hline
%\midrule
$1.67\times 10^{-4}$  & 0.008  \\ 
$2.08\times 10^{-4}$  & 0.009  \\ 
$2.77\times 10^{-4}$  & 0.011  \\
$4.16\times 10^{-4}$  & 0.012 \\ 
\hline
\end{tabular}
\caption{Main parameters characterizing the operations used in syndrome extraction for $6.25$-photon Kerr-cat qubit. The average gate bias is defined as the ratio of the sum of the probabilities of $I\otimes Z$, $Z\otimes I$, and $Z\otimes Z$ errors and the sum of the remaining two-qubit Pauli errors. Leakage in all the operations is $\sim 10^{-4}-10^{-5}$. The effective Pauli noise channel of preparation and measurement operations are naturally biased. The total error in preparation is simply the sum of the probability of a $Z$ and a $Y$ error. The bias-preserving $\CX$ gate simulation is carried out in two steps, first by simulating the Hamiltonian in \cref{eq:cx} with the Lindbladian $\mathcal{L}[\rho]=\sum_{i=c,t}\kappa_1(1+n_\mathrm{th})\mathcal{D}[a_i]\rho+\kappa_1 n_\mathrm{th}\mathcal{D}[a^\dag_i]\rho$ is simulated for time $T$ and in the second step the master equation of two uncoupled Kerr cats with Lindbladian $\sum_{i=c,t} \kappa_2\mathcal{D}[a^2_i]\rho+\mathcal{L}[\rho]$ for time $T$, so that $T_\mathrm{CX}=2T$. For fair-comparison, a two-photon dissipation channel after the $\CX_R=R \CZ R^\dag$ gate is also applied, for time $\sim  9/K$ so  that the residual leakage after the $\CX$ and $\CX_R$ gates are the same. The  time of $R$ gate is $2.6/K$, $\CZ=0.8/K$ giving a total time of $15/K$.   }
\label{table1}
\end{table}

\begin{figure}[!tbhp]
  \centering
 \includegraphics[width=\linewidth]{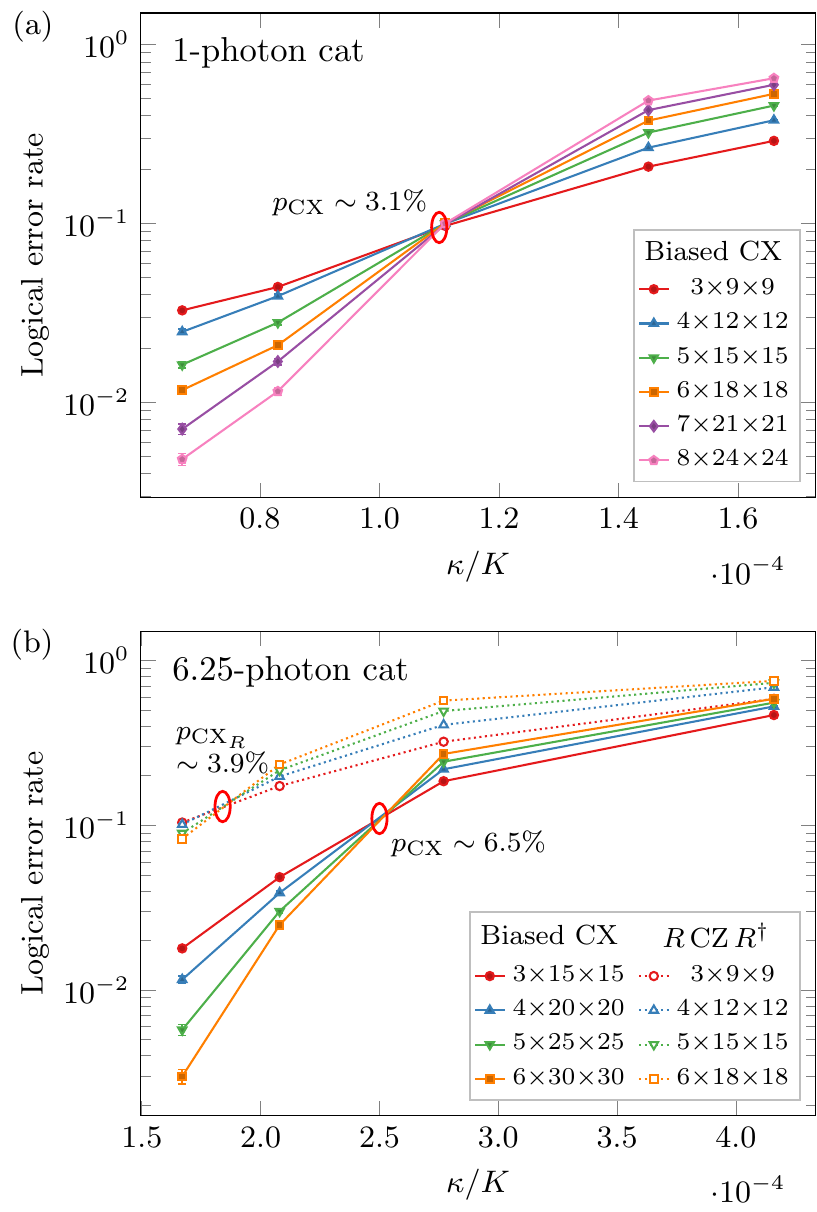}
 \caption{Error thresholds of the XZZX surface code with the circuit noise model for Kerr-cat qubits with a cat size (a) $\alpha^2=1$ photon and (b) $\alpha^2=6.25$ photons.
 The bottom axis shows the ratio of the single-photon loss rate $\kappa$ or equivalently the inverse $T_1$ time of the SNAIL used to realize the cat and the Kerr-nonlinear constant.
 The simulations also take into account a constant $8\%$ thermal noise and two-photon dissipation $\kappa_2=K/10$ added for autonomous leakage correction.
 The noise channel in (b) is significantly more biased than in (a).
 We find the threshold $\kappa/K$ of $1.1\times 10^{-4}$ and $2.5\times 10^{-4}$ in (a) and (b) respectively which corresponds to a $\CX$ error threshold of $\sim3.1\%$ and $\sim 6.5\%$ respectively. For comparison, the threshold with $\alpha^2=6.25$ photon cat and non-bias-preserving gate $\CX_R= R\,\CZ\,R^\dag$ with $R=\exp(i\pi Z_\mathrm{t}/4)\exp(i\pi X_\mathrm{t}/4)$ is also shown in (b). Even though the gate infidelity of $\CX_R$ is lower than that of the bias-preserving $\CX$ in \cref{eq:cx}, the threshold is lower $\sim 1.84\times 10^{-4}$ because the bias is smaller. 
 }
 \label{cat_1}
 \end{figure}

\Cref{cat_1} shows the logical failure rate of the XZZX surface code as a function of the ratio of the single-photon loss rate and Kerr-nonlinearity $\kappa/K$. The total gate error scales as $\propto \kappa/K$ and so this is the important metric for experimentalists aiming to design the system.

The threshold values obtained for $\kappa/K$ are given by $(\kappa/K)_\mathrm{thres}\sim 1.1\times 10^{-4}$ for $\alpha^2=1.0$ and $(\kappa/K)_\mathrm{thres}\sim 2.5\times 10^{-4}$ for $\alpha^2=6.25$. To illustrate the physical implication of these thresholds, consider a SNAIL-based Kerr cat with a Kerr-nonlinearity of  $K/2\pi=10$ MHz. Then $(\kappa/K)_\mathrm{thres}\sim 1.1\times 10^{-4}$ for $\alpha^2=1.0$ implies that the lifetime of the SNAIL ($1/\kappa$) should be longer than $144.7$ $\mu s$ for error correction with the XZZX surface code to be successful. For $\alpha^2=6.25$, $(\kappa/K)_\mathrm{thres}\sim 2.5\times 10^{-4}$ implies that the lifetime of the SNAIL ($1/\kappa$) should be longer than $63.6$ $\mu s$ for error correction with the XZZX surface code to be successful. 

The most noisy operation in the system is the $\CX $ gate and we see that, with the $\alpha^2=1.0$ cat, the $\CX$  gate infidelity must be lower than $3.1\%$. On the other hand, with the $\alpha^2=6.25$ cat the $\CX$  gate infidelity must be lower than $\sim 6.5\%$. The average $\CX$ gate bias at threshold for the $\alpha^2=6.25$ cat is $\sim 351$ while that for $\alpha^2=1$ cat is $\sim 16$. Clearly, as expected, we find that the XZZX surface code is able to tolerate more errors in the gates when the cat size is large because the noise bias increases.

The SNAIL lifetime in the first experimental demonstration of the Kerr cat was $\sim 15.5$ $\mu$s, which shows that an improvement in the lifetime is required to operate below threshold with Kerr-cat qubits. Lifetimes of $50-100$ $\mu$s are routinely achieved in superconducting Josephson-junction circuits~\cite{jurcevic2021demonstration,kjaergaard2020superconducting,sete2021parametric,kelly2015state}, so that reaching a lifetime above the $T_1$ threshold appears very reasonable using state of the art superconducting circuit devices and fabrication. Optimizing the gate implementations and incorporating a one-step $\CX$ will further improve the threshold requirements.  Finally, we note that the first Kerr-cat experiment~\cite{Grimm2020} showed a strong suppression of frequency fluctuations due to $1/f$ noise for the pumped cat, compared to when operating the SNAILmon as a conventional transmon qubit without pumping. In practice, frequency fluctuations of the SNAIL were near negligible when pumped to the cat subspace. This suggest that, albeit the Kerr cat is a flux-tunable device, recent improvements in coherence times for fixed frequency qubits may also carry over to Kerr-cat qubits~\cite{place2021new}.

\subsubsection{Comparison with non-bias-preserving $\CX$ gate}

Note that with Kerr-cat qubits it is possible to implement a non-bias-preserving $\CX$ gate by implementing the gate sequence $\CX_R = R\,\CZ \,R^\dag$ with $R=\exp(i\pi Z_\mathrm{t}/4)\exp(i\pi X_\mathrm{t}/4)$. The subscript $\mathrm{t}$ denotes that the single qubit gate $R$ acts on the target of the $\CZ$. We have already discussed how $\exp(i\pi Z_\mathrm{t}/4)=S^\dag$, $\exp(i\pi X_\mathrm{t}/4)=X(\pi/4)$, and  $\CZ$ gates are simulated. Unlike the bias-preserving $\CX$ gate  implementation  given in \cref{eq:cx}, $\CX_R$ does not preserve the bias in noise. However, the gate can be implemented faster resulting in higher total fidelity. \Cref{table1} also lists the gate fidelity and bias for the $\CX_R$ gate for the $\alpha^2=6.25$ photon cat. For fair-comparison, a two-photon dissipation channel after the $\CX_R$ gate is also applied so that the residual leakage after the $\CX$ and $\CX_R$ gates are the same for a given $\kappa_1$, $n_\mathrm{th}$, and $\kappa_2$.

To compare the two different gates,  we repeat the simulation of the XZZX surface code but this time using the noise channel of the $\CX_R$ gate. As shown in \cref{cat_1}(b), the threshold in this case is lowered by  $\sim 40\%$  compared to the threshold with the bias-preserving $\CX$. 
Moreover, the below-threshold logical error rate for the XZZX code is significantly higher with the $\CX_R$ gate then the bias-preserving $\CX$ gate, despite the $\CX_R$ gate having a lower total infidelity.

\subsubsection{Comparison with the CSS surface code}

We find that the CSS surface code with Kerr-cat noise in this parameter region does not perform as well as the XZZX code. 
At $\alpha=2.5$, over the range of $\kappa/K$ shown in \cref{table1}, we observe a logical error rate of close to $50\%$, which does not decrease with system size, indicating that the CSS code is far above threshold. At lower bias $\alpha=1$, we also do not observe a threshold in the overall error rate over the range of $\kappa/K$ used in our XZZX code simulations. 

We also consider the scenario of Ref.~\onlinecite{chamberland2020building} where the width of the CSS code is fixed $d_\mathrm{x}=3$, and only the length of the code $d_\mathrm{z}$ is scaled up. The same above threshold behavior is observed for $\alpha=2.5$, even when $X$ logical errors are completely neglected.  For $\alpha=1$, if we neglect logical $X$ logical errors, we observe a threshold in the $Z$ logical error rate comparable to the threshold in the overall error rate of the XZZX code at $\kappa/K=1.1\times 10^{-4}$. Given, however, that $X$ errors are non-negligible at this bias, the the overall failure rate still appears above threshold, and so error correction with the CSS surface code does not appear to be functional in the parameter regions considered.

The high thresholds we obtain with the XZZX code are due to the fact that it appears very well suited to Kerr-cat qubit noise. One of the major advantages of this pairing is that a full round of syndrome measurements can be  performed in a substantially shorter time than with the standard CSS surface code %using cat qubits studied in Ref.~\onlinecite{chamberland2020building}, 
and therefore is far less prone to error. This is because, for large $\alpha$, the gate that takes the longest to execute is the $\CX$ gate, while the relative time required to perform a $\CZ$ gate compared to $\CX$ tends to zero as $\alpha$ increases. For the CSS surface code, readout of an $X$ check requires four data qubits to be coupled to the ancilla via $\CX$ gates, which takes at least time $4T_{\CX}$. In contrast, every check in the XZZX code requires two $\CX$ gates and two $\CZ$ gates, and given that these can be parallelized, the entangling operations for a full round of syndrome measurements can be performed in time $2(T_{\CX}+ T_{\CZ})$. In the limit of large $\alpha$, the total time taken to apply the gates for a round of syndrome measurements with the XZZX code is $\sim 2T_{\CX}$, which is roughly equal to that of a 1D repetition code and half that of the CSS surface code.

\section{Small codes}
\label{s:low_distance}

% Section intro
Here we perform exact simulations of different quantum error-correcting codes for systems of small size undergoing biased noise. In the previous section we were able to analyse the scalability of different quantum error-correcting codes by simulating systems of a large number of qubits. We obtained these results using an efficient decoder and by approximating quantum noise channels as a Pauli noise channel to make simulations tractable. Exact simulations enable us to investigate other details of various codes that are specialised to biased noise. 

% Simulations
We perform several analyses using small-scale simulations using the Kerr-cat qubit noise model described in~\cref{s:cat_sim} and an optimal decoding algorithm. By focusing on small codes we can perform optimal decoding without concern of the efficiency of the decoder. We describe the optimal decoder and the methods we use to conduct exact simulations in~\cref{s:simulation}.

% Codes used in analysis; XZZX, TSC, repetition code
The smallest realization of a surface code that can correct at least one $X$ error and one $Z$ error has $n=9$ data qubits arranged on a $3 \times 3 $ lattice and uses the `rotated' layout of Ref.~\onlinecite{bombin_optimal_2007}. 
In our small-scale simulations we compare the XZZX code on this layout to the tailored surface code (TSC)~\cite{Tuckett19} on the same layout and to the one-dimensional repetition code with $n=9$ qubits. The TSC is an alternative surface code variant where we measure weight-four Pauli-$X$ stabilizer checks and weight-four Pauli-$Y$ stabilizer checks. This code has been demonstrated to have favourable properties to correct for biased noise~\cite{Tuckett19, Tuckett20}. We use open boundary conditions for all simulations of the $n=9$ codes with two-qubit checks on the boundaries of the surface code lattices. 

% Features of the different codes
While in \cref{s:threshold} we have shown that the XZZX code has impressive fault-tolerant thresholds with finite noise biases, the TSC with open boundary conditions has the advantage that it can tolerate up to $(n-1)/2$ dephasing errors. In contrast, a logical failure can be introduced to the $3\times 3$ XZZX code with as few as two Pauli-$Z$ errors~\cite{Tuckett18}. 

% Features of the repetition code
Like the TSC, the $n=9$ repetition code can also correct up to four dephasing errors with certainty. Moreover, it has the additional advantage that it requires only two entangling gates, rather than four, in the circuits used for stabilizer parity measurements. As such, the stabilizer readout circuits for the repetition code are easier to implement and less error prone than surface code stabilizers. We might therefore expect that the repetition code is more accessible to realisation using near-term experiments. 

This convenience of the repetition code comes at the expense that it has no protection against bit-flip errors.  Indeed, in Ref.~\onlinecite{chamberland2020building}, numerical simulations showed that the repetition code does not increase its suppression of logical failures for a realistic biased noise model when increasing the system size beyond $n\sim 10$ qubits. Nevertheless, for a small number of qubits and a very large noise bias such that the bit-flip errors are very rare, repetition codes may outperform surface codes. We can therefore expect variations in the performance of these codes in different parameter regimes.

Our exact simulation results highlight different features of the TSC, XZZX and repetition code when using using Kerr-cat qubits. The XZZX code in particular exhibits high performance in both high and low bias regimes, it is relatively insensitive to decoder miscalibration, and it does not suffer from significant increases in logical error rate when standard $\CX$ gates are used instead of bias-preserving $\CX$. 
In contrast, the repetition code will perform better when the bias is high relative to the system size, and the TSC may have an advantage in architectures where $\CX$ gates are not so noisy (compared to other operations).

\begin{figure*}[!tbhp]
    \centering
  \includegraphics[width=\linewidth]{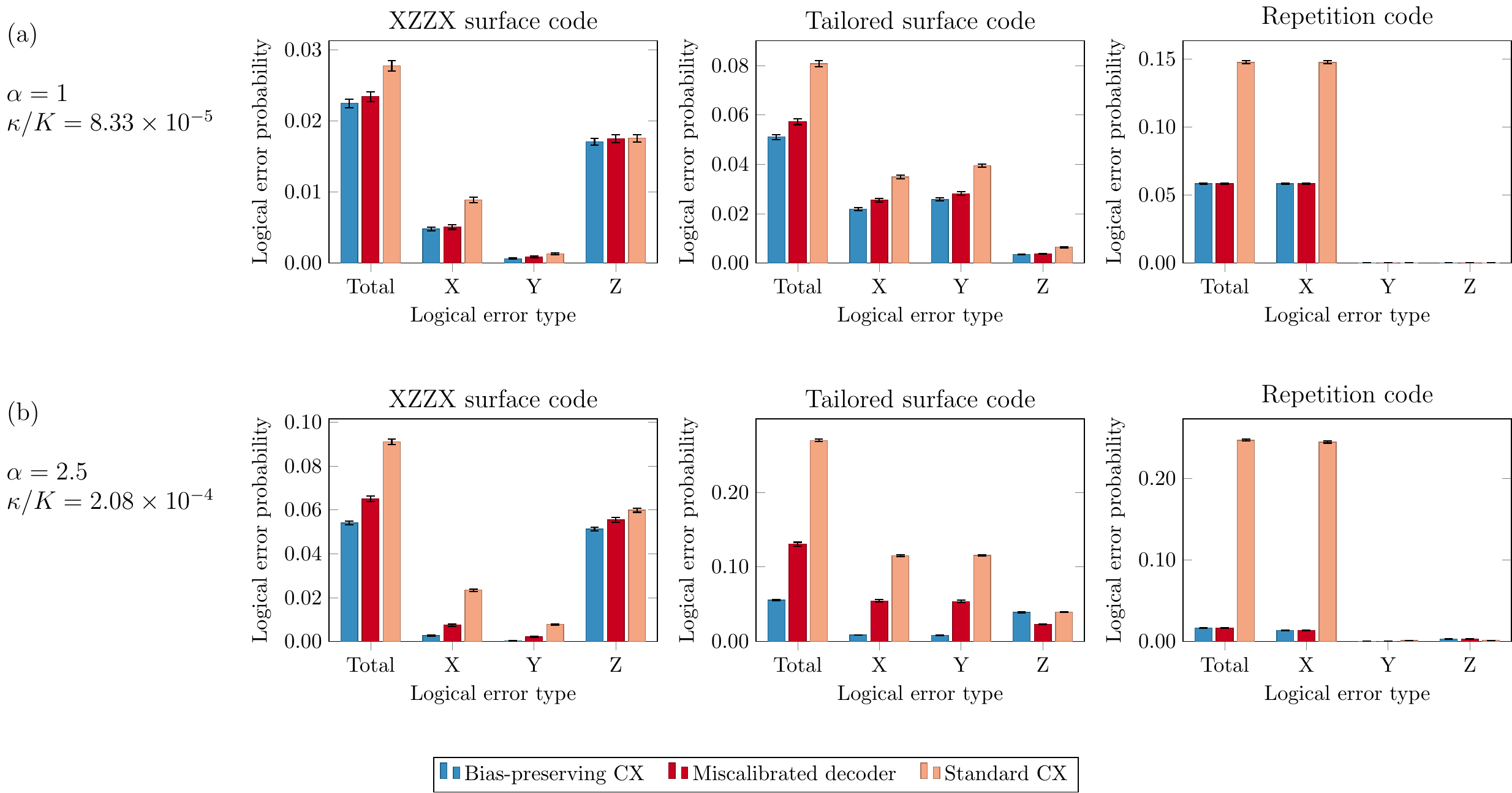}
    \caption{Logical error probabilities for low distance ($n=9$) codes obtained by exact simulation.
      In (a) a moderate bias, corresponding to $\alpha=1$, is used with $\kappa/K=8.33 \times 10^{-5}$.
      In (b) a high bias, corresponding to $\alpha=2.5$, is used with $\kappa/K=2.08 \times 10^{-4}$.
      In both cases, the overall noise strength $\kappa$ is chosen to be slightly below the threshold estimated in \cref{cat_1}.
      Optimal performance (blue) is compared to the performance obtained with a miscalibrated decoder (red), and non-bias preserving $\CX$ gate (orange) for the XZZX surface code, tailored surface code and repetition code.
      Each data point is evaluated over 12\,000 samples and error bars represent the standard error of the mean. Aside for the red bars, which are obtained using a miscalibrated decoder, optimal decoding is used in all of these simulations.
    }
    \label{fig:exact_error_rates_combined}
\end{figure*}

In \cref{fig:exact_error_rates_combined}, we have plotted logical error probabilities obtained using Kerr-cat qubits and bias-preserving operations alongside those obtained using non-bias preserving $\CX$ and a miscalibrated decoder (which we describe in \cref{s:miscalibrated_decoder}). The probabilities of $X$, $Y$ and $Z$ logical errors separately are shown as well as the total logical error probability (which is the sum of the former three). The noise parameters were chosen to be slightly below the threshold estimated in~\cref{cat_1}, which is a regime relevant to near-term experiments. 

Certain properties of these codes are reflected in these separate logical error probabilities. For instance, the $Z$ logical is the only one that would be affected in the presence of pure phase-flip noise (i.e. noise where the bit-flip probability is exactly zero). With the rotated layout on a $3\times3$ lattice, the weight of the smallest $Z$ logical of the XZZX code is 3, while for the TSC and repetition codes it is 9~\cite{Tuckett19}. The $Z$ logical is therefore much more protected than $X$ or $Y$ logicals for the TSC and repetition codes compared to the XZZX code, as seen in \cref{fig:exact_error_rates_combined}. We see that the repetition code's failures are almost exclusively due to bit-flip errors, rather than phase-flip errors since in \cref{fig:exact_error_rates_combined} the $Z$ logical errors are nearly undetectable.

We point out that, like in the Pauli simulations used to determine the threshold in \cref{s:threshold}, we have approximated the cat qubits in the exact surface code simulations as strict two-level systems, and have neglected additional levels into which the system may leak. While the leakage suppression with engineered two-photon dissipation (and noise caused by this process) is still accounted for in the qubit noise model, there remains a small amount of residual leakage that is ignored. While we believe this effect to be small (for the $\CX$ gate, the probability of leakage is an order of magnitude smaller than that of a non-dominant error), we leave a detailed study of this effect to future research. 

Under the following subheadings we discuss our small-scale analysis under several different criteria. 
We begin by examining the break-even point of different codes. 
This is a pseudo threshold that tells us the noise parameters below which a small quantum error-correcting code can outperform the constituent parts from which it is built. These numbers give us an indication of the quality of the qubits we need to demonstrate quantum error correction using small-scale experiments. In addition to this, we discuss the change in performance obtained by using bias-preserving gates as compared with a more conventional gate noise model. We also use exact simulations to compare the true noise model extracted from master equation simulations (but still neglecting leakage) to the Pauli-twirl approximation, which was used in~\cref{s:threshold_cat}.
Finally, we also use small-scale simulations to investigate the loss of performance due to a miscalibrated decoder.

\subsection{Break-even error rates}
An important milestone for experimental demonstrations of error correcting codes is reaching the psuedo-threshold or break-even point. We say that a code is below the pseudo-threshold if the probability of a logical error on the encoded qubit after a fixed number of rounds of error correction is less than the error probability of the noisiest element in the  syndrome circuit, which in this case is the $\CX$ gate. The number of rounds was set to three for all of the following simulations.

As shown in \cref{fig:exact_error_rates_combined}(b), at high bias $\alpha=2.5$ and $\kappa=2.08\times 10^{-4}$ with a bias-preserving $\CX$ and optimal decoding, the logical error rate for the XZZX code is 
%$0.0541\pm0.0009$ and TSC is $0.055\pm0.001$,
$5.41\pm0.09\%$ and TSC is $5.53\pm0.10\%$,
which for both codes is equal, within statistical error, to the error rate of the $\CX$ gate of 
%$0.0543$.
$5.43\%$.
This shows that the pseudo-threshold can be achieved with the XZZX and TSC at remarkably high $\CX$ gate infidelity. At this level of bias, where the probability of bit-flip errors is extremely low, the repetition code has an error rate of $1.62\pm0.02\%$, which is lower than the both surface code variants. This is due to its overall shorter circuit, having fewer qubits idle during $\CX$ gate execution and also its ability to correct up to four dephasing errors with certainty. 

In contrast, at a lower bias of $\alpha=1$ and with $\kappa=8.33\times 10^{-5}$ with a bias-preserving $\CX$ the XZZX code is the only code of the three to be below the pseudo-threshold. The advantage of the XZZX code over the repetition code here is that it can detect and correct bit-flip errors, which are non-negligible at this level of bias.

The drawback of the TSC in a Kerr cat architecture comes from the fact that it requires twice as many $\CX$ gates compared to the XZZX code (which are far noisier than $\CZ$ gates). While we do not present the results here, we have found that for simple biased noise models where $\CZ$ and $\CX$ gates have roughly equal error probabilities, the $3\times 3$ TSC with optimal decoding can outperform the XZZX code in terms of logical error rate for a wide range of biases. However, as efficient near-optimal decoding appears more complicated for the TSC than the XZZX code for larger systems, more work is still required before the TSC can be regarded as a practical or scalable alternative to the XZZX code under this kind of noise.

\subsection{The necessity of bias-preserving $\CX$ gates}
\label{s:pauli_approximation}

One striking difference between the different codes is the relative importance of using bias-preserving $\CX$ gates. To test this we compared the performance of each code using the bias-preserving $\CX$ gate to a non-bias preserving preserving $\CX$ gate. See \cref{s:threshold_toy} for a description of the latter $\CX$ gate. We defined this gate with the same infidelities as the bias-preserving $\CX$ gate.     

The blue and orange bars in \cref{fig:exact_error_rates_combined} represent the logical error probabilities obtained using bias-preserving $\CX$ gates and standard (non-bias preserving) $\CX$ gates respectively.  While both TSC and XZZX surface codes have a higher logical error rate when not using a bias preserving $\CX$ gate, the increase in error rate is much larger for the TSC. At high bias $\alpha=2.5$, for the XZZX code, the error rate increases by around $70\%$ when using non-bias preserving $\CX$, while the TSC sees a huge increase of nearly $400\%$ compared to when using a bias preserving $\CX$ gate. This is likely due to fact that the TSC uses twice as many $\CX$ gates as the XZZX code, and so using non-bias preserving $\CX$ gates reduces the overall noise bias less in the XZZX code than the TSC. This effect is also observed, although is less pronounced, at lower bias $\alpha=1$. 
Like the threshold results of \cref{s:threshold}, these results suggest that high performance may be achieved with Kerr cats and the XZZX code even if bias-preserving gates are not used (although the best performance is likely obtained with bias-preserving gates).

Unlike the surface code variants, the repetition code is entirely dependent on a bias-preserving $\CX$ gate. Since it is only designed to correct phase-flip errors, the additional bit-flip errors introduced with a non-bias preserving $\CX$ gate drastically reduce its performance. This can be seen in both $\alpha=1$ and $\alpha=2.5$ plots, where the logical error rate is roughly equal to the probability of an $X$ error occurring on a data qubit somewhere in the error correction circuit.

\subsection{Interrogating the Pauli-twirl approximation}

The threshold simulations in \cref{s:threshold} are performed with a Pauli noise model, which allows efficient simulation of large systems. This Pauli noise model is obtained by taking the Pauli twirl approximation of the full noise model, which has non-Pauli or coherent terms. The Pauli twirl approximation effectively sets non-Pauli terms in the noise to zero. We do not need to apply the Pauli twirl to the noise in the exact simulations (albeit we still neglect leakage to other states). However, our exact simulation data suggests that non-Pauli terms in the physical noise have a negligible contribution to the performance of the surface code variants.  

Using the Pauli twirl of the noise maps on the physical level in the exact simulation produces an average logical $\chi$ matrix with diagonal terms (which can be interpreted as logical Pauli error probabilities) within statistical error of those obtained using the full (non-Pauli) noise maps on the physical level. Furthermore, off-diagonal terms in the logical $\chi$ matrix are at least two orders of magnitude smaller than the diagonal terms for the averaged logical channel, suggesting that the averaged logical channel is close to a stochastic Pauli channel.  This provides some justification for the use of the Pauli twirl approximation in the threshold calculations in \cref{s:threshold}. 

In contrast, the repetition code preserves coherence on the logical level. The full noise-model results in slightly worse performance than its Pauli twirl, with about 10\% lower fidelity in the logical channel for $\alpha=1$ and $\kappa/K=8.33 \times 10^{-5}$. Furthermore, off-diagonal terms in the logical $\chi$ matrix are larger than the dominant diagonal terms. This is not particularly surprising, since the $X$-component of the noise is not transformed from the physical to the logical, e.g. a coherent $X$ rotation on the first physical qubit of the repetition code results in an undetectable coherent $X$ rotation by an equal angle of the logical qubit.

\subsection{Miscalibrated decoding algorithms}
\label{s:miscalibrated_decoder}

An exact density matrix simulator can also function as an optimal decoder. It can thus be used to probe the performance gains possible with improved decoding, or, conversely performance loss if the decoder is not optimally calibrated to the noise. To demonstrate the importance of calibrating the decoder to biased gate noise, we have compared the performance of a miscalibrated decoder to the optimal decoder. With the physical $\CX$ gate chosen to be bias preserving, the miscalibrated decoder is tuned to the noise model of the standard $\CX$ (rather than the bias-preserving $\CX$). Thus the miscalibrated decoder is not properly adapted to the noise on the $\CX$ gate, however it still correctly takes into account the bias on all other elements. The difference in performance observed between the miscalibrated decoder and optimal decoder tells us how important it is for the decoder to have an accurate characterisation of bias on the $\CX$ gate.

As shown in \cref{fig:exact_error_rates_combined} at high bias $\alpha=2.5$ the XZZX code is much less sensitive to decoder miscalibration than the TSC code at this bias. Using a miscalibrated decoder resulted in an increase of less than $20\%$ increase in logical error rate for the XZZX code, compared to over $100\%$ for the TSC. The insensitivity of the XZZX code to miscalibrated decoding suggests that near optimal performance may be achieved with this code using heuristic decoding methods (like minimum weight perfect matching) where a precise characterisation of the noise model does not need to be used. 

At lower bias $\alpha=1$ we see a smaller increase in logical error rate when the decoder is miscalibrated for both XZZX and TSC surface codes. This suggests that the importance of tailored decoding increases as the bias increases.
In contrast, the logical error rate of the repetition code, due to its relatively simple decoding requirements, appears almost completely unaffected by decoder miscalibration at both low and high bias.

\section{Discussion}
\label{s:discussion}

We propose a robust architecture for fault-tolerant quantum computing in which Kerr-cat qubits are used as the elementary qubits in an XZZX surface code. The resulting system demonstrates exceptional thresholds that can be met using a superconducting circuit architecture with parameters that appear fairly modest compared to the current state of the art. 
We make conservative assumptions about the Kerr-cat photon number and imperfections such as thermal noise and measurement inefficiency.
For the parameters used in our simulations, the Kerr cat has an anharmonicity, given by $\omega_\mathrm{gap}$, comparable to that of a transmon, leading to fast gates.
Note, however, that our simulations were performed without pulse optimization, which may yield significant improvements~\cite{Motzoi2009}.
Further optimization is also possible by improving the decoding algorithm, as our results were obtained using a standard minimum-weight perfect-matching decoder~\cite{Fowler12a}.

It remains to make a detailed analysis of the resource cost to use our error-correcting system to perform a computation below some target logical failure rate. This will be important to compare the performance of our proposal with others~\cite{aliferis_fault-tolerant_2008, aliferis_fault-tolerant_2009,Stephens13bias, Tuckett18,  Tuckett19,Xu19,Li2019, Tuckett20, Huang20, guillaud2021error, hanggli_enhanced_2020, higgott_subsystem_2020, chamberland2020building}. 
Resource estimates are highly sensitive to the details of noise in laboratory systems~\cite{Beverland18}, and can be improved dramatically with optimizations to the code (e.g. to the code layout and decoder) and hardware. 
Hence, we leave such calculations to future work. Nevertheless, given the remarkable threshold that our system exhibits, we expect we will be able to demonstrate a scalable system by realizing our proposal using modern technology.

Theoretical research should progress hand-in-hand with experimental development. To this end, details in our simulations have been motivated by the dominant sources of noise in Kerr-cat qubits that have been observed in recent experiments~\cite{Grimm2020}. We should expect that as the system is scaled up in terms of both the size of the cats, and the number of qubits on a chip, that other sources of errors such as crosstalk, drive-induced heating and leakage
%may become appreciable.
must be accounted for.
It will be important to develop better methods to characterize the noise in physical realizations of our error-correcting systems~\cite{Harper2020}, and to refine our quantum error-correcting codes accordingly.

The architecture motivating our present study is based on the Kerr-cat qubit realized with on-chip superconducting SNAILs. This is a  promising platform for realizing a 
large-scale surface-code architecture using systems presently available in the laboratory. The principles in the design and fabrication of a large-scale SNAIL chip are very similar to other circuit QED based approaches, for instance quantum processors based on tunable transmons~\cite{arute2019quantum}.
The scalability of our proposal with existing technologies suggests a practical solution to achieve fault-tolerant quantum computing in the near term.

\section*{Acknowledgements}
The authors would like to thank S. Bartlett, P. Bonilla Ataides, and S. Flammia for conversations about decoders and tailored codes. S.P would like to thank  N. Frattini for numerous discussions which were also helpful in the production of the conceptual figure for the Kerr-cat/SNAIL-surface code.
A.S.D. was supported by JST, PRESTO Grant No. JPMJPR1917, Japan. B.J.B., A.L.G. and D.K.T. are supported by the Australian Research Council via the Centre of Excellence in Engineered Quantum Systems (EQUS) project number CE170100009, and by the Army Research Office(ARO) under Grant Number: W911NF-21-1-0007. B.J.B. also received support from the University of Sydney Fellowship Programme, and A.L.G. is also supported by the Australian Research Council via a Discovery Early Career Research Award, project number DE190100380. S.P. is supported by the ARO under grant number W911NF-18-1-0212. 
Most of the numerical computation in this work was carried out at the Yukawa Institute Computer Facility.
Access to high-performance computing resources was also provided by the National Computational Infrastructure (NCI), which is supported by the Australian Government.
\appendix
\section{Surface-code simulation method}
\label{s:simulation}

We used two types of numerical simulation to evaluate the performance of cat-surface codes. For large systems, in order to determine error thresholds, we used circuit-level Pauli simulations and for small codes more accessible to near term implementation we used exact state-vector simulations. Using two simulations methods provides complementary insights into surface code performance: the Pauli simulations allow system size scaling analysis to be performed, while the exact simulations allow general types of error and to see the potential performance achievable with optimal decoding.

\subsection{Pauli simulation}
\label{s:pauli_simulation}
The Pauli simulations rely on the fact that the surface code is implemented with a Clifford circuit, which transforms Pauli operators to Pauli operators. One can therefore simulate a noisy circuit by sampling Pauli errors in the circuit (according to an error model of each circuit element) then, by commuting the Pauli errors through to the end of the circuit, determine which ancilla measurements are flipped (i.e. the syndrome) and the residual error on the data qubits. 

The Pauli simulation used for threshold calculation assumes the initial state is a noiseless surface code state, to which a number of rounds of noisy syndrome measurements are applied, followed by a single final round of noiseless syndrome measurements. The assumption that the final round of check measurements is noiseless is justified in practice since the final readout of a surface-code qubit is performed by single qubit measurements on the data qubits for which there is no distinction between a measurement error and a data qubit error. After sampling an error, the decoding algorithm is run on the corresponding syndrome to determine a correction. If the computed correction and the sampled error belong to the same logical equivalence class, we regard the error correction as successful. If not, a logical failure is said to have occurred. By repeating the simulation many times and counting the number of logical failures, the logical failure rate can be determined. To determine the error threshold, the simulation is performed at varying error rates as the code distance $d$ is increased. For code distance $d$ the number of rounds of noisy measurements was set to $d$. The threshold was determined to be the noise strength below which increasing $d$ led to an exponential decrease in logical failure rate.

\subsection{Exact simulation}
The exact simulation method involves storing the $N$ qubit density operator of the noisy code state $\rho$ as a $4^N$ component vector in memory and updating it as gates, noise and measurements are applied with matrix-vector multiplication. Given the exponential scaling in required memory, the method is only suitable for small system sizes. However, it has the advantage of allowing arbitrary noise models to be studied (rather than restricting to Pauli noise). Keeping track of the full density operator also means that the simulation algorithm may be used for optimal decoding. This type of decoding is similar to the approach described in Ref.~\onlinecite{darmawan17, Darmawan18}. By calculating how a linearly independent set of encoded states (represented as $2\times2$ matrices) transforms under error correction for a fixed syndrome $s$, it is possible to calculate the logical channel $\mathcal{E}_s$, represented as a $4\times 4$ $\chi$ matrix, conditioned on the syndrome. The optimal Pauli correction is then the one that minimises the distance of the logical channel to the identity i.e.
\begin{equation}
{\rm argmin}_{L\in I, X, Y, Z}||L\circ \mathcal{E}_s - I||\,.
\end{equation}
For Pauli noise, this is equivalent to maximum-likelihood decoding. One simplification made for the exact simulation was that the check measurements were performed sequentially, i.e. all the gates and the ancilla of a check were applied before measuring the next check. This is in contrast to the Pauli simulation, where all checks were measured simultaneously with an interleaved 2 qubit gate pattern (described in \cref{s:sc_definitions}). This simplification to the exact simulation was made to save computational time, since performing check measurements in parallel necessitates keeping all ancilla qubits in memory (in addition to the data qubits), rather than a single ancilla qubit at a time. We remark that there is a simulation method, described in Ref.~\onlinecite{huang_alibaba_2020}, based on novel tensor-network contraction techniques which can obviate the large memory cost of the simple exact simulation method we have used, however the overall computational costs in this method still remain daunting. 

There is some choice in where to put idle noise in the simulation when checks are measured sequentially. We have applied idle noise to data qubits and ancilla qubits such that the total strength of the idle noise on each qubit is the same as if the gates were interleaved to measure the checks in parallel. This idle noise includes the noise on qubits left idle during measurement and preparation of ancilla as well as noise on qubits left idle while $\CX$ gates are applied to other qubits. Noise on qubits left idle during rounds of $\CZ$ gates is relatively small compared to $\CX$ gates and was neglected in our exact simulations (although it is included in our Pauli simulations). Thus the total amount of noise in the circuit is roughly equivalent in our exact simulations, to the circuit with parallelized check readout which would be used in practice, however due to the different gate order there will be some differences in how errors propagate between data and ancilla qubits.

\section{Kerr cat and purely dissipative cat gate fidelities}
\label{s:kerr_diss2}

\begin{figure}[!tbhp]
    \centering
   \includegraphics[width=\linewidth]{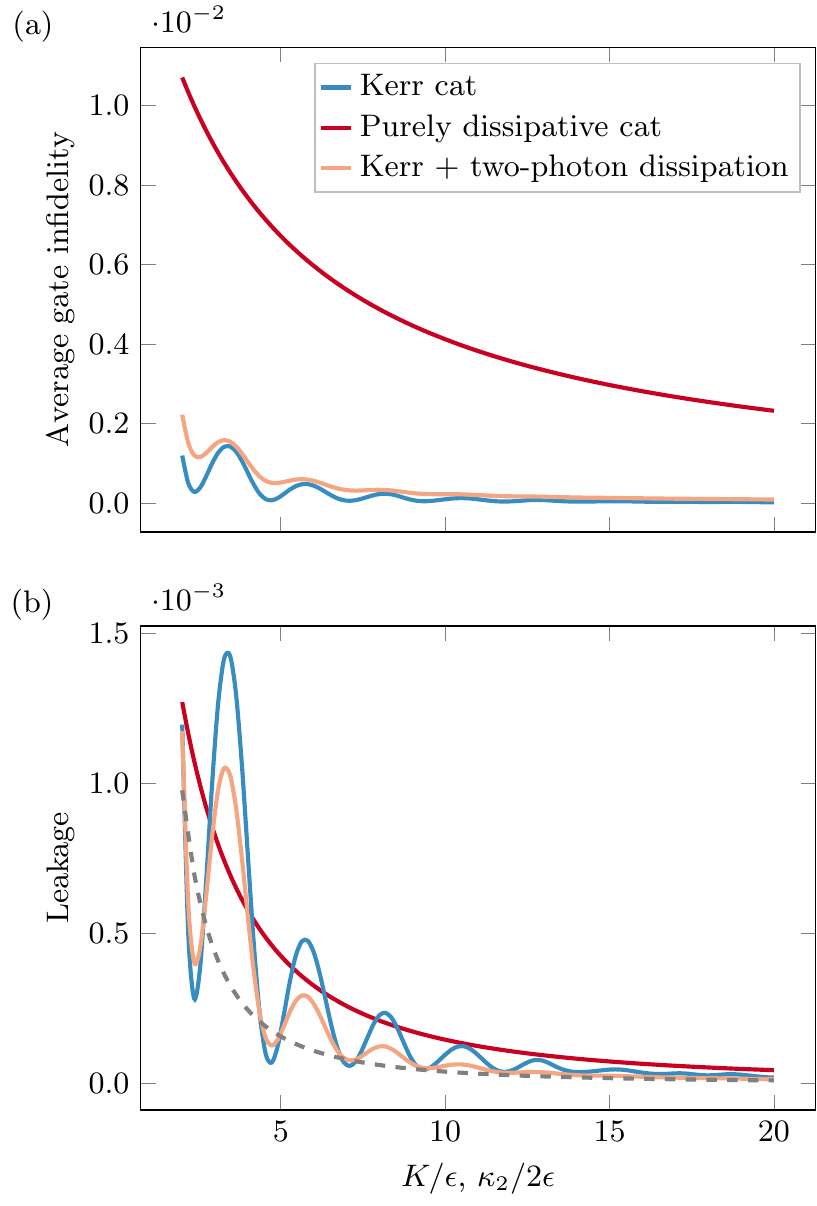}
    \caption{(a) $S$ gate infidelity and (b) leakage with Kerr cats (blue), purely dissipative cats (red), and Kerr cats with two-photon dissipation (orange), $\kappa_2=K/10$.
    In (b) the grey dashed line indicates the approximate analytic expression for average residual leakage $(\varepsilon/4K\alpha^2)^2,(\varepsilon/2\kappa_2\alpha^2)^2$.
    In these simulations we take $\alpha=2$ and all other sources of decoherence, such as single photon loss and thermal excitations, are neglected.
    Note that the oscillations in the leakage for Kerr cats results from off-resonant population transfer back and forth between the cat and leakage space due to the Hamiltonian dynamics.
    These oscillations disappear when a smoother time-dependent pulse $\varepsilon(t)$ is used for implementing the $S$-gate.
    Nonetheless, the average behaviour agrees well with the simple analytic estimate given by the grey line.  }
    \label{fig:S_comp}
\end{figure}

In this section, with the help of a simple numerical example, we outline an important difference between Kerr cats and dissipative cats which leads to higher-fidelity gate operations with Kerr-cat qubits. Consider the implementation of the $S$-gate in the Kerr-cat qubit. The required Hamiltonian is \begin{align}
    H&=H_\mathrm{cat}+\varepsilon(a^\dag+a)\nonumber\\
    H_\mathrm{cat}&=-Ka^{\dag 2}a^2+K\alpha^2(a^{\dag 2}+a^2)\nonumber\\
    \label{eq:sgate}
\end{align}
In order to understand the effect of the microwave drive $\varepsilon(a^\dag+a)$ let us look at how the photon annihilation and creation operators affect the computational cat states. 
Action of the photon annihilation operator causes transitions between the cat states, $a\puricatpm=\pm \alpha r^{\pm 1}\puricatmp$. Here $r=\sqrt{(1-e^{-2\alpha^2})/(1+e^{-2\alpha^2})}$. The creation operator on the other hand, not only causes transitions between the cat states, but it can also cause transitions outside the cat-subspace. In fact, in the limit of large $\alpha$, $a^\dag\puricatpm\sim \pm \alpha \puricatmp+\ket{\psi^{\mp}_1}$, where $\ket{\psi^{\mp}_1}$ are the first excited states with odd and even photon-number parity respectively. These states are separated from the cat subspace by the energy gap $|\omega_\mathrm{gap}|$. While, the applied drive is resonant with the cat-qubit frequency, it is off-resonant from the excited states by an amount $|\omega_\mathrm{gap}|$. Consequently, this drive can only cause small amount of population transfer to the leakage state. More importantly, the amount of leakage on average $\propto (\varepsilon/\omega_\mathrm{gap})^2$, decreases with $|\omega_\mathrm{gap}|$~\cite{Puri2020}. 

An $S$-gate with the purely dissipative cat requires the Hamiltonian $H=i(\kappa_2/2)\alpha^2(a^{\dag 2}-a^2)+\varepsilon(a^\dag+a)$ and the engineered dissipation $\kappa_2\mathcal{D}[a^2]\rho$. The gate time is $\pi/8\alpha\varepsilon$, which is the same as the $S$ gate time with Kerr cats. In this case, $a^\dag$ also causes transition to the first excited manifold $\puricatpm\rightarrow\ket{\psi^{\mp}_1}$. The excited state population quickly decays back to the cat-subspace at rate $\sim 2\kappa_2\alpha^2$. The quantity $\sim 2\kappa_2\alpha^2$ is also called the Lindbladian gap. The residual excited state population is $(\varepsilon/2\kappa_2\alpha^2)^2$. Because of the transition matrix elements $\puricatpmb a^2\ket{\psi^{\pm}_1}\sim 2\alpha$ and $\puricatpmb a^2\ket{\psi^{\mp}_1}=0$, the dissipation causes transitions within the even parity and odd parity subspaces respectively. On the other hand, the leakage causes transitions between the even and odd parity subspace. Hence,
cooling down of the excited state population leads to additional phase-flip errors. These additional phase-flip errors in the purely-dissipative cats are also termed as non-adiabatic errors in literature~\cite{Guillaud2019,Chamberland2020}. The amount of phase flips, $\propto \varepsilon/\kappa_2\alpha^3$, increases with $\varepsilon$~\cite{Chamberland2020}. 
Fortunately, the additional phase-flip errors are noticeably absent in Kerr cats. 
Consequently, gates with Kerr cats have higher fidelity than those with purely-dissipative cats.

We demonstrate the effect of additional phase-flip errors in purely dissipative cats by simulating the $S$ gate and comparing its fidelity with the fidelity of $S$ gate with Kerr cats. \Cref{fig:S_comp} shows this comparison. To keep the analysis simple we neglect other sources of decoherence in either of the two cat qubits. In the simulations we fix $\varepsilon$, $\alpha=2$ and hence the gate time $\pi/8\alpha\varepsilon$ is the same for the two cats. We choose $\kappa_2=2K$ so as to keep the energy and Lindbladian gaps (or equivalently the residual leakage) approximately the same in the two cats. \Cref{fig:S_comp} clearly shows that the $S$ gate infidelity can be significantly higher for Kerr cats than purely  dissipative  cats for the same amount of leakage. We also show the the infidelity of the $S$ gate in the hybrid approach with small two-photon  dissipation $\kappa_2=K/10$ included in addition to the Hamiltonian \cref{eq:sgate}. The result  shows that addition of small amount of two-photon dissipation during the Kerr cat $S$ gate does not degrade the gate fidelity in the parameter regime used in this paper.

\bibstyle{plain}
\bibliography{QECReferences}

%apsrev4-2.bst 2019-01-14 (MD) hand-edited version of apsrev4-1.bst
%Control: key (0)
%Control: author (8) initials jnrlst
%Control: editor formatted (1) identically to author
%Control: production of article title (0) allowed
%Control: page (0) single
%Control: year (1) truncated
%Control: production of eprint (0) enabled
\begin{thebibliography}{84}%
\makeatletter
\providecommand \@ifxundefined [1]{%
 \@ifx{#1\undefined}
}%
\providecommand \@ifnum [1]{%
 \ifnum #1\expandafter \@firstoftwo
 \else \expandafter \@secondoftwo
 \fi
}%
\providecommand \@ifx [1]{%
 \ifx #1\expandafter \@firstoftwo
 \else \expandafter \@secondoftwo
 \fi
}%
\providecommand \natexlab [1]{#1}%
\providecommand \enquote  [1]{``#1''}%
\providecommand \bibnamefont  [1]{#1}%
\providecommand \bibfnamefont [1]{#1}%
\providecommand \citenamefont [1]{#1}%
\providecommand \href@noop [0]{\@secondoftwo}%
\providecommand \href [0]{\begingroup \@sanitize@url \@href}%
\providecommand \@href[1]{\@@startlink{#1}\@@href}%
\providecommand \@@href[1]{\endgroup#1\@@endlink}%
\providecommand \@sanitize@url [0]{\catcode `\\12\catcode `\$12\catcode
  `\&12\catcode `\#12\catcode `\^12\catcode `\_12\catcode `\%12\relax}%
\providecommand \@@startlink[1]{}%
\providecommand \@@endlink[0]{}%
\providecommand \url  [0]{\begingroup\@sanitize@url \@url }%
\providecommand \@url [1]{\endgroup\@href {#1}{\urlprefix }}%
\providecommand \urlprefix  [0]{URL }%
\providecommand \Eprint [0]{\href }%
\providecommand \doibase [0]{https://doi.org/}%
\providecommand \selectlanguage [0]{\@gobble}%
\providecommand \bibinfo  [0]{\@secondoftwo}%
\providecommand \bibfield  [0]{\@secondoftwo}%
\providecommand \translation [1]{[#1]}%
\providecommand \BibitemOpen [0]{}%
\providecommand \bibitemStop [0]{}%
\providecommand \bibitemNoStop [0]{.\EOS\space}%
\providecommand \EOS [0]{\spacefactor3000\relax}%
\providecommand \BibitemShut  [1]{\csname bibitem#1\endcsname}%
\let\auto@bib@innerbib\@empty
%</preamble>
\bibitem [{\citenamefont {Reed}\ \emph {et~al.}(2012)\citenamefont {Reed},
  \citenamefont {DiCarlo}, \citenamefont {Nigg}, \citenamefont {Sun},
  \citenamefont {Frunzio}, \citenamefont {Girvin},\ and\ \citenamefont
  {Schoelkopf}}]{Reed12}%
  \BibitemOpen
  \bibfield  {author} {\bibinfo {author} {\bibfnamefont {M.~D.}\ \bibnamefont
  {Reed}}, \bibinfo {author} {\bibfnamefont {L.}~\bibnamefont {DiCarlo}},
  \bibinfo {author} {\bibfnamefont {S.~E.}\ \bibnamefont {Nigg}}, \bibinfo
  {author} {\bibfnamefont {L.}~\bibnamefont {Sun}}, \bibinfo {author}
  {\bibfnamefont {L.}~\bibnamefont {Frunzio}}, \bibinfo {author} {\bibfnamefont
  {S.~M.}\ \bibnamefont {Girvin}},\ and\ \bibinfo {author} {\bibfnamefont
  {R.~J.}\ \bibnamefont {Schoelkopf}},\ }\bibfield  {title} {\bibinfo {title}
  {Realization of three-qubit quantum error correction with superconducting
  circuits},\ }\href {https://doi.org/10.1038/nature10786} {\bibfield
  {journal} {\bibinfo  {journal} {Nature}\ }\textbf {\bibinfo {volume} {482}},\
  \bibinfo {pages} {382} (\bibinfo {year} {2012})},\ \Eprint
  {https://arxiv.org/abs/1109.4948} {arXiv:1109.4948 [quant-ph]} \BibitemShut
  {NoStop}%
\bibitem [{\citenamefont {Barends}\ \emph {et~al.}(2014)\citenamefont
  {Barends}, \citenamefont {Kelly}, \citenamefont {Megrant}, \citenamefont
  {Veitia}, \citenamefont {Sank}, \citenamefont {Jeffry}, \citenamefont
  {White}, \citenamefont {Mutus}, \citenamefont {Fowler}, \citenamefont
  {Campbell} \emph {et~al.}}]{Barends14}%
  \BibitemOpen
  \bibfield  {author} {\bibinfo {author} {\bibfnamefont {R.}~\bibnamefont
  {Barends}}, \bibinfo {author} {\bibfnamefont {J.}~\bibnamefont {Kelly}},
  \bibinfo {author} {\bibfnamefont {A.}~\bibnamefont {Megrant}}, \bibinfo
  {author} {\bibfnamefont {A.}~\bibnamefont {Veitia}}, \bibinfo {author}
  {\bibfnamefont {D.}~\bibnamefont {Sank}}, \bibinfo {author} {\bibfnamefont
  {E.}~\bibnamefont {Jeffry}}, \bibinfo {author} {\bibfnamefont {T.~C.}\
  \bibnamefont {White}}, \bibinfo {author} {\bibfnamefont {J.}~\bibnamefont
  {Mutus}}, \bibinfo {author} {\bibfnamefont {A.~G.}\ \bibnamefont {Fowler}},
  \bibinfo {author} {\bibfnamefont {B.}~\bibnamefont {Campbell}}, \emph
  {et~al.},\ }\bibfield  {title} {\bibinfo {title} {Superconducting quantum
  circuits at the surface code threshold for fault tolerance},\ }\href
  {https://doi.org/10.1038/nature13171} {\bibfield  {journal} {\bibinfo
  {journal} {Nature}\ }\textbf {\bibinfo {volume} {508}},\ \bibinfo {pages}
  {500} (\bibinfo {year} {2014})},\ \Eprint {https://arxiv.org/abs/1402.4848}
  {arXiv:1402.4848 [quant-ph]} \BibitemShut {NoStop}%
\bibitem [{\citenamefont {Nigg}\ \emph {et~al.}(2014)\citenamefont {Nigg},
  \citenamefont {Martinez}, \citenamefont {Schindler}, \citenamefont
  {Hennrich}, \citenamefont {Monz}, \citenamefont {Martin-Delagado},\ and\
  \citenamefont {Blatt}}]{Nigg14}%
  \BibitemOpen
  \bibfield  {author} {\bibinfo {author} {\bibfnamefont {D.}~\bibnamefont
  {Nigg}}, \bibinfo {author} {\bibfnamefont {M.~M. E.~A.}\ \bibnamefont
  {Martinez}}, \bibinfo {author} {\bibfnamefont {P.}~\bibnamefont {Schindler}},
  \bibinfo {author} {\bibfnamefont {M.}~\bibnamefont {Hennrich}}, \bibinfo
  {author} {\bibfnamefont {T.}~\bibnamefont {Monz}}, \bibinfo {author}
  {\bibfnamefont {M.~A.}\ \bibnamefont {Martin-Delagado}},\ and\ \bibinfo
  {author} {\bibfnamefont {R.}~\bibnamefont {Blatt}},\ }\bibfield  {title}
  {\bibinfo {title} {Experimental quantum computations on a topologically
  encoded qubit},\ }\href {https://doi.org/10.1126/science.1253742} {\bibfield
  {journal} {\bibinfo  {journal} {Science}\ }\textbf {\bibinfo {volume}
  {345}},\ \bibinfo {pages} {302} (\bibinfo {year} {2014})},\ \Eprint
  {https://arxiv.org/abs/1403.5426} {arXiv:1403.5426 [quant-ph]} \BibitemShut
  {NoStop}%
\bibitem [{\citenamefont {Kelly}\ \emph {et~al.}(2015)\citenamefont {Kelly},
  \citenamefont {Barends}, \citenamefont {Fowler}, \citenamefont {Megrant},
  \citenamefont {Jeffrey}, \citenamefont {White}, \citenamefont {Sank},
  \citenamefont {Mutus}, \citenamefont {Campbell}, \citenamefont {Chen} \emph
  {et~al.}}]{kelly2015state}%
  \BibitemOpen
  \bibfield  {author} {\bibinfo {author} {\bibfnamefont {J.}~\bibnamefont
  {Kelly}}, \bibinfo {author} {\bibfnamefont {R.}~\bibnamefont {Barends}},
  \bibinfo {author} {\bibfnamefont {A.~G.}\ \bibnamefont {Fowler}}, \bibinfo
  {author} {\bibfnamefont {A.}~\bibnamefont {Megrant}}, \bibinfo {author}
  {\bibfnamefont {E.}~\bibnamefont {Jeffrey}}, \bibinfo {author} {\bibfnamefont
  {T.~C.}\ \bibnamefont {White}}, \bibinfo {author} {\bibfnamefont
  {D.}~\bibnamefont {Sank}}, \bibinfo {author} {\bibfnamefont {J.~Y.}\
  \bibnamefont {Mutus}}, \bibinfo {author} {\bibfnamefont {B.}~\bibnamefont
  {Campbell}}, \bibinfo {author} {\bibfnamefont {Y.}~\bibnamefont {Chen}},
  \emph {et~al.},\ }\bibfield  {title} {\bibinfo {title} {State preservation by
  repetitive error detection in a superconducting quantum circuit},\ }\href
  {https://doi.org/10.1038/nature14270} {\bibfield  {journal} {\bibinfo
  {journal} {Nature}\ }\textbf {\bibinfo {volume} {519}},\ \bibinfo {pages}
  {66} (\bibinfo {year} {2015})},\ \Eprint {https://arxiv.org/abs/1411.7403}
  {arXiv:1411.7403 [quant-ph]} \BibitemShut {NoStop}%
\bibitem [{\citenamefont {C{\'o}rcoles}\ \emph {et~al.}(2015)\citenamefont
  {C{\'o}rcoles}, \citenamefont {Magesan}, \citenamefont {Srinivasan},
  \citenamefont {Cross}, \citenamefont {Steffen}, \citenamefont {Gambetta},\
  and\ \citenamefont {Chow}}]{Corcoles15}%
  \BibitemOpen
  \bibfield  {author} {\bibinfo {author} {\bibfnamefont {A.~D.}\ \bibnamefont
  {C{\'o}rcoles}}, \bibinfo {author} {\bibfnamefont {E.}~\bibnamefont
  {Magesan}}, \bibinfo {author} {\bibfnamefont {S.~J.}\ \bibnamefont
  {Srinivasan}}, \bibinfo {author} {\bibfnamefont {A.~W.}\ \bibnamefont
  {Cross}}, \bibinfo {author} {\bibfnamefont {M.}~\bibnamefont {Steffen}},
  \bibinfo {author} {\bibfnamefont {J.~M.}\ \bibnamefont {Gambetta}},\ and\
  \bibinfo {author} {\bibfnamefont {J.~M.}\ \bibnamefont {Chow}},\ }\bibfield
  {title} {\bibinfo {title} {Demonstration of a quantum error detection code
  using a square lattice of four superconducting qubits},\ }\href
  {https://doi.org/10.1038/ncomms7979} {\bibfield  {journal} {\bibinfo
  {journal} {Nat. Commun.}\ }\textbf {\bibinfo {volume} {6}},\ \bibinfo {pages}
  {6979} (\bibinfo {year} {2015})},\ \Eprint {https://arxiv.org/abs/1410.6419}
  {arXiv:1410.6419 [quant-ph]} \BibitemShut {NoStop}%
\bibitem [{\citenamefont {Takita}\ \emph {et~al.}(2016)\citenamefont {Takita},
  \citenamefont {C\'{o}rcoles}, \citenamefont {Magesan}, \citenamefont {Abdo},
  \citenamefont {Brink}, \citenamefont {Cross}, \citenamefont {Chow},\ and\
  \citenamefont {Gambetta}}]{Takita16}%
  \BibitemOpen
  \bibfield  {author} {\bibinfo {author} {\bibfnamefont {M.}~\bibnamefont
  {Takita}}, \bibinfo {author} {\bibfnamefont {A.~D.}\ \bibnamefont
  {C\'{o}rcoles}}, \bibinfo {author} {\bibfnamefont {E.}~\bibnamefont
  {Magesan}}, \bibinfo {author} {\bibfnamefont {B.}~\bibnamefont {Abdo}},
  \bibinfo {author} {\bibfnamefont {M.}~\bibnamefont {Brink}}, \bibinfo
  {author} {\bibfnamefont {A.~W.}\ \bibnamefont {Cross}}, \bibinfo {author}
  {\bibfnamefont {J.~M.}\ \bibnamefont {Chow}},\ and\ \bibinfo {author}
  {\bibfnamefont {J.~M.}\ \bibnamefont {Gambetta}},\ }\bibfield  {title}
  {\bibinfo {title} {Demonstration of weight-four parity measurements in the
  surface code architecture},\ }\href
  {https://doi.org/10.1103/PhysRevLett.117.210505} {\bibfield  {journal}
  {\bibinfo  {journal} {Phys. Rev. Lett.}\ }\textbf {\bibinfo {volume} {117}},\
  \bibinfo {pages} {210505} (\bibinfo {year} {2016})},\ \Eprint
  {https://arxiv.org/abs/1605.01351} {arXiv:1605.01351 [quant-ph]} \BibitemShut
  {NoStop}%
\bibitem [{\citenamefont {Ofek}\ \emph {et~al.}(2016)\citenamefont {Ofek},
  \citenamefont {Petrenko}, \citenamefont {Heeres}, \citenamefont {Reinhold},
  \citenamefont {Leghtas}, \citenamefont {Vlastakis}, \citenamefont {Liu},
  \citenamefont {Frunzio}, \citenamefont {Girvin}, \citenamefont {Jiang} \emph
  {et~al.}}]{Ofek16}%
  \BibitemOpen
  \bibfield  {author} {\bibinfo {author} {\bibfnamefont {N.}~\bibnamefont
  {Ofek}}, \bibinfo {author} {\bibfnamefont {A.}~\bibnamefont {Petrenko}},
  \bibinfo {author} {\bibfnamefont {R.}~\bibnamefont {Heeres}}, \bibinfo
  {author} {\bibfnamefont {P.}~\bibnamefont {Reinhold}}, \bibinfo {author}
  {\bibfnamefont {Z.}~\bibnamefont {Leghtas}}, \bibinfo {author} {\bibfnamefont
  {B.}~\bibnamefont {Vlastakis}}, \bibinfo {author} {\bibfnamefont
  {Y.}~\bibnamefont {Liu}}, \bibinfo {author} {\bibfnamefont {L.}~\bibnamefont
  {Frunzio}}, \bibinfo {author} {\bibfnamefont {S.~M.}\ \bibnamefont {Girvin}},
  \bibinfo {author} {\bibfnamefont {L.}~\bibnamefont {Jiang}}, \emph {et~al.},\
  }\bibfield  {title} {\bibinfo {title} {Extending the lifetime of a quantum
  bit with error correction in superconducting circuits},\ }\href
  {https://doi.org/10.1038/nature18949} {\bibfield  {journal} {\bibinfo
  {journal} {Nature}\ }\textbf {\bibinfo {volume} {536}},\ \bibinfo {pages}
  {441} (\bibinfo {year} {2016})},\ \Eprint {https://arxiv.org/abs/1602.04768}
  {arXiv:1602.04768 [quant-ph]} \BibitemShut {NoStop}%
\bibitem [{\citenamefont {Hu}\ \emph {et~al.}(2019)\citenamefont {Hu},
  \citenamefont {Ma}, \citenamefont {Cai}, \citenamefont {Mu}, \citenamefont
  {Xu}, \citenamefont {Wang}, \citenamefont {Wu}, \citenamefont {Wang},
  \citenamefont {Song}, \citenamefont {Zou} \emph {et~al.}}]{Hu:2019aa}%
  \BibitemOpen
  \bibfield  {author} {\bibinfo {author} {\bibfnamefont {L.}~\bibnamefont
  {Hu}}, \bibinfo {author} {\bibfnamefont {Y.}~\bibnamefont {Ma}}, \bibinfo
  {author} {\bibfnamefont {W.}~\bibnamefont {Cai}}, \bibinfo {author}
  {\bibfnamefont {X.}~\bibnamefont {Mu}}, \bibinfo {author} {\bibfnamefont
  {Y.}~\bibnamefont {Xu}}, \bibinfo {author} {\bibfnamefont {W.}~\bibnamefont
  {Wang}}, \bibinfo {author} {\bibfnamefont {Y.}~\bibnamefont {Wu}}, \bibinfo
  {author} {\bibfnamefont {H.}~\bibnamefont {Wang}}, \bibinfo {author}
  {\bibfnamefont {Y.~P.}\ \bibnamefont {Song}}, \bibinfo {author}
  {\bibfnamefont {C.~L.}\ \bibnamefont {Zou}}, \emph {et~al.},\ }\bibfield
  {title} {\bibinfo {title} {Quantum error correction and universal gate set
  operation on a binomial bosonic logical qubit},\ }\href
  {https://doi.org/10.1038/s41567-018-0414-3} {\bibfield  {journal} {\bibinfo
  {journal} {Nat. Phys.}\ }\textbf {\bibinfo {volume} {15}},\ \bibinfo {pages}
  {503} (\bibinfo {year} {2019})},\ \Eprint {https://arxiv.org/abs/1805.09072}
  {arXiv:1805.09072 [quant-ph]} \BibitemShut {NoStop}%
\bibitem [{\citenamefont {Fl{\"u}hmann}\ \emph {et~al.}(2019)\citenamefont
  {Fl{\"u}hmann}, \citenamefont {Nguyen}, \citenamefont {Marinelli},
  \citenamefont {Negnevitsky}, \citenamefont {Mehta},\ and\ \citenamefont
  {Home}}]{Fluhmann:2019aa}%
  \BibitemOpen
  \bibfield  {author} {\bibinfo {author} {\bibfnamefont {C.}~\bibnamefont
  {Fl{\"u}hmann}}, \bibinfo {author} {\bibfnamefont {T.~L.}\ \bibnamefont
  {Nguyen}}, \bibinfo {author} {\bibfnamefont {M.}~\bibnamefont {Marinelli}},
  \bibinfo {author} {\bibfnamefont {V.}~\bibnamefont {Negnevitsky}}, \bibinfo
  {author} {\bibfnamefont {K.}~\bibnamefont {Mehta}},\ and\ \bibinfo {author}
  {\bibfnamefont {J.~P.}\ \bibnamefont {Home}},\ }\bibfield  {title} {\bibinfo
  {title} {Encoding a qubit in a trapped-ion mechanical oscillator},\ }\href
  {https://doi.org/10.1038/s41586-019-0960-6} {\bibfield  {journal} {\bibinfo
  {journal} {Nature}\ }\textbf {\bibinfo {volume} {566}},\ \bibinfo {pages}
  {513} (\bibinfo {year} {2019})},\ \Eprint {https://arxiv.org/abs/1807.01033}
  {arXiv:1807.01033 [quant-ph]} \BibitemShut {NoStop}%
\bibitem [{\citenamefont {Campagne-Ibarcq}\ \emph {et~al.}(2020)\citenamefont
  {Campagne-Ibarcq}, \citenamefont {Eickbusch}, \citenamefont {Touzard},
  \citenamefont {Zalys-Geller}, \citenamefont {Frattini}, \citenamefont
  {Sivak}, \citenamefont {Reinhold}, \citenamefont {Puri}, \citenamefont
  {Shankar}, \citenamefont {Schoelkopf} \emph {et~al.}}]{Campagne2020}%
  \BibitemOpen
  \bibfield  {author} {\bibinfo {author} {\bibfnamefont {P.}~\bibnamefont
  {Campagne-Ibarcq}}, \bibinfo {author} {\bibfnamefont {A.}~\bibnamefont
  {Eickbusch}}, \bibinfo {author} {\bibfnamefont {S.}~\bibnamefont {Touzard}},
  \bibinfo {author} {\bibfnamefont {E.}~\bibnamefont {Zalys-Geller}}, \bibinfo
  {author} {\bibfnamefont {N.~E.}\ \bibnamefont {Frattini}}, \bibinfo {author}
  {\bibfnamefont {V.~V.}\ \bibnamefont {Sivak}}, \bibinfo {author}
  {\bibfnamefont {P.}~\bibnamefont {Reinhold}}, \bibinfo {author}
  {\bibfnamefont {S.}~\bibnamefont {Puri}}, \bibinfo {author} {\bibfnamefont
  {S.}~\bibnamefont {Shankar}}, \bibinfo {author} {\bibfnamefont {R.~J.}\
  \bibnamefont {Schoelkopf}}, \emph {et~al.},\ }\bibfield  {title} {\bibinfo
  {title} {Quantum error correction of a qubit encoded in grid states of an
  oscillator},\ }\href {https://doi.org/10.1038/s41586-020-2603-3} {\bibfield
  {journal} {\bibinfo  {journal} {Nature}\ }\textbf {\bibinfo {volume} {584}},\
  \bibinfo {pages} {368} (\bibinfo {year} {2020})},\ \Eprint
  {https://arxiv.org/abs/1907.12487} {arXiv:1907.12487 [quant-ph]} \BibitemShut
  {NoStop}%
\bibitem [{\citenamefont {Grimm}\ \emph {et~al.}(2020)\citenamefont {Grimm},
  \citenamefont {Frattini}, \citenamefont {Puri}, \citenamefont {Mundhada},
  \citenamefont {Touzard}, \citenamefont {Mirrahimi}, \citenamefont {Girvin},
  \citenamefont {Shankar},\ and\ \citenamefont {Devoret}}]{Grimm2020}%
  \BibitemOpen
  \bibfield  {author} {\bibinfo {author} {\bibfnamefont {A.}~\bibnamefont
  {Grimm}}, \bibinfo {author} {\bibfnamefont {N.~E.}\ \bibnamefont {Frattini}},
  \bibinfo {author} {\bibfnamefont {S.}~\bibnamefont {Puri}}, \bibinfo {author}
  {\bibfnamefont {S.~O.}\ \bibnamefont {Mundhada}}, \bibinfo {author}
  {\bibfnamefont {S.}~\bibnamefont {Touzard}}, \bibinfo {author} {\bibfnamefont
  {M.}~\bibnamefont {Mirrahimi}}, \bibinfo {author} {\bibfnamefont {S.~M.}\
  \bibnamefont {Girvin}}, \bibinfo {author} {\bibfnamefont {S.}~\bibnamefont
  {Shankar}},\ and\ \bibinfo {author} {\bibfnamefont {M.~H.}\ \bibnamefont
  {Devoret}},\ }\bibfield  {title} {\bibinfo {title} {Stabilization and
  operation of a {K}err-cat qubit},\ }\href
  {https://doi.org/10.1038/s41586-020-2587-z} {\bibfield  {journal} {\bibinfo
  {journal} {Nature}\ }\textbf {\bibinfo {volume} {584}},\ \bibinfo {pages}
  {205} (\bibinfo {year} {2020})},\ \Eprint {https://arxiv.org/abs/1907.12131}
  {arXiv:1907.12131 [quant-ph]} \BibitemShut {NoStop}%
\bibitem [{\citenamefont {Lescanne}\ \emph {et~al.}(2020)\citenamefont
  {Lescanne}, \citenamefont {Villiers}, \citenamefont {Peronnin}, \citenamefont
  {Sarlette}, \citenamefont {Delbecq}, \citenamefont {Huard}, \citenamefont
  {Kontos}, \citenamefont {Mirrahimi},\ and\ \citenamefont
  {Leghtas}}]{Lescanne2020}%
  \BibitemOpen
  \bibfield  {author} {\bibinfo {author} {\bibfnamefont {R.}~\bibnamefont
  {Lescanne}}, \bibinfo {author} {\bibfnamefont {M.}~\bibnamefont {Villiers}},
  \bibinfo {author} {\bibfnamefont {T.}~\bibnamefont {Peronnin}}, \bibinfo
  {author} {\bibfnamefont {A.}~\bibnamefont {Sarlette}}, \bibinfo {author}
  {\bibfnamefont {M.}~\bibnamefont {Delbecq}}, \bibinfo {author} {\bibfnamefont
  {B.}~\bibnamefont {Huard}}, \bibinfo {author} {\bibfnamefont
  {T.}~\bibnamefont {Kontos}}, \bibinfo {author} {\bibfnamefont
  {M.}~\bibnamefont {Mirrahimi}},\ and\ \bibinfo {author} {\bibfnamefont
  {Z.}~\bibnamefont {Leghtas}},\ }\bibfield  {title} {\bibinfo {title}
  {Exponential suppression of bit-flips in a qubit encoded in an oscillator},\
  }\href {https://doi.org/10.1038/s41567-020-0824-x} {\bibfield  {journal}
  {\bibinfo  {journal} {Nat. Phys.}\ }\textbf {\bibinfo {volume} {16}},\
  \bibinfo {pages} {509} (\bibinfo {year} {2020})},\ \Eprint
  {https://arxiv.org/abs/1907.11729} {arXiv:1907.11729 [quant-ph]} \BibitemShut
  {NoStop}%
\bibitem [{\citenamefont {Ma}\ \emph {et~al.}(2020)\citenamefont {Ma},
  \citenamefont {Xu}, \citenamefont {Mu}, \citenamefont {Cai}, \citenamefont
  {Hu}, \citenamefont {Wang}, \citenamefont {Pan}, \citenamefont {Wang},
  \citenamefont {Song}, \citenamefont {Zou},\ and\ \citenamefont
  {Sun}}]{Sun2020}%
  \BibitemOpen
  \bibfield  {author} {\bibinfo {author} {\bibfnamefont {Y.}~\bibnamefont
  {Ma}}, \bibinfo {author} {\bibfnamefont {Y.}~\bibnamefont {Xu}}, \bibinfo
  {author} {\bibfnamefont {X.}~\bibnamefont {Mu}}, \bibinfo {author}
  {\bibfnamefont {W.}~\bibnamefont {Cai}}, \bibinfo {author} {\bibfnamefont
  {L.}~\bibnamefont {Hu}}, \bibinfo {author} {\bibfnamefont {W.}~\bibnamefont
  {Wang}}, \bibinfo {author} {\bibfnamefont {X.}~\bibnamefont {Pan}}, \bibinfo
  {author} {\bibfnamefont {H.}~\bibnamefont {Wang}}, \bibinfo {author}
  {\bibfnamefont {Y.~P.}\ \bibnamefont {Song}}, \bibinfo {author}
  {\bibfnamefont {C.~L.}\ \bibnamefont {Zou}},\ and\ \bibinfo {author}
  {\bibfnamefont {L.}~\bibnamefont {Sun}},\ }\bibfield  {title} {\bibinfo
  {title} {Error-transparent operations on a logical qubit protected by quantum
  error correction},\ }\href {https://doi.org/10.1038/s41567-020-0893-x}
  {\bibfield  {journal} {\bibinfo  {journal} {Nat. Phys.}\ }\textbf {\bibinfo
  {volume} {16}},\ \bibinfo {pages} {827} (\bibinfo {year} {2020})},\ \Eprint
  {https://arxiv.org/abs/1909.06803} {arXiv:1909.06803 [quant-ph]} \BibitemShut
  {NoStop}%
\bibitem [{\citenamefont {Chen}\ \emph {et~al.}(2021)\citenamefont {Chen},
  \citenamefont {Satzinger}, \citenamefont {Atalaya}, \citenamefont {Korotkov},
  \citenamefont {Dunsworth}, \citenamefont {Sank}, \citenamefont {Quintana},
  \citenamefont {McEwen}, \citenamefont {Barends}, \citenamefont {Klimov} \emph
  {et~al.}}]{chen_exponential_2021}%
  \BibitemOpen
  \bibfield  {author} {\bibinfo {author} {\bibfnamefont {Z.}~\bibnamefont
  {Chen}}, \bibinfo {author} {\bibfnamefont {K.~J.}\ \bibnamefont {Satzinger}},
  \bibinfo {author} {\bibfnamefont {J.}~\bibnamefont {Atalaya}}, \bibinfo
  {author} {\bibfnamefont {A.~N.}\ \bibnamefont {Korotkov}}, \bibinfo {author}
  {\bibfnamefont {A.}~\bibnamefont {Dunsworth}}, \bibinfo {author}
  {\bibfnamefont {D.}~\bibnamefont {Sank}}, \bibinfo {author} {\bibfnamefont
  {C.}~\bibnamefont {Quintana}}, \bibinfo {author} {\bibfnamefont
  {M.}~\bibnamefont {McEwen}}, \bibinfo {author} {\bibfnamefont
  {R.}~\bibnamefont {Barends}}, \bibinfo {author} {\bibfnamefont {P.~V.}\
  \bibnamefont {Klimov}}, \emph {et~al.},\ }\bibfield  {title} {\bibinfo
  {title} {Exponential suppression of bit or phase errors with cyclic error
  correction},\ }\href {https://doi.org/10.1038/s41586-021-03588-y} {\bibfield
  {journal} {\bibinfo  {journal} {Nature}\ }\textbf {\bibinfo {volume} {595}},\
  \bibinfo {pages} {383} (\bibinfo {year} {2021})},\ \Eprint
  {https://arxiv.org/abs/2102.06132} {arXiv:2102.06132 [quant-ph]} \BibitemShut
  {NoStop}%
\bibitem [{\citenamefont {Shor}(1996)}]{Shor96}%
  \BibitemOpen
  \bibfield  {author} {\bibinfo {author} {\bibfnamefont {P.~W.}\ \bibnamefont
  {Shor}},\ }\bibfield  {title} {\bibinfo {title} {Fault-tolerant quantum
  computation},\ }in\ \href {https://doi.org/10.1109/SFCS.1996.548464} {\emph
  {\bibinfo {booktitle} {Proceedings of 37th Conference on Foundations of
  Computer Science}}},\ \bibinfo {series and number} {FOCS '96}\ (\bibinfo
  {publisher} {IEEE Computer Society},\ \bibinfo {address} {USA},\ \bibinfo
  {year} {1996})\ p.~\bibinfo {pages} {56},\ \Eprint
  {https://arxiv.org/abs/quant-ph/9605011} {arXiv:quant-ph/9605011}
  \BibitemShut {NoStop}%
\bibitem [{\citenamefont {Kitaev}(2003)}]{Kitaev03}%
  \BibitemOpen
  \bibfield  {author} {\bibinfo {author} {\bibfnamefont {A.}~\bibnamefont
  {Kitaev}},\ }\bibfield  {title} {\bibinfo {title} {Fault-tolerant quantum
  computation by anyons},\ }\href
  {https://doi.org/10.1016/S0003-4916(02)00018-0} {\bibfield  {journal}
  {\bibinfo  {journal} {Ann. Phys.}\ }\textbf {\bibinfo {volume} {303}},\
  \bibinfo {pages} {2} (\bibinfo {year} {2003})},\ \Eprint
  {https://arxiv.org/abs/quant-ph/9707021} {arXiv:quant-ph/9707021}
  \BibitemShut {NoStop}%
\bibitem [{\citenamefont {Dennis}\ \emph {et~al.}(2002)\citenamefont {Dennis},
  \citenamefont {Kitaev}, \citenamefont {Landahl},\ and\ \citenamefont
  {Preskill}}]{Dennis02}%
  \BibitemOpen
  \bibfield  {author} {\bibinfo {author} {\bibfnamefont {E.}~\bibnamefont
  {Dennis}}, \bibinfo {author} {\bibfnamefont {A.}~\bibnamefont {Kitaev}},
  \bibinfo {author} {\bibfnamefont {A.}~\bibnamefont {Landahl}},\ and\ \bibinfo
  {author} {\bibfnamefont {J.}~\bibnamefont {Preskill}},\ }\bibfield  {title}
  {\bibinfo {title} {Topological quantum memory},\ }\href
  {https://doi.org/10.1063/1.1499754} {\bibfield  {journal} {\bibinfo
  {journal} {J. Math. Phys. (N.Y.)}\ }\textbf {\bibinfo {volume} {43}},\
  \bibinfo {pages} {4452} (\bibinfo {year} {2002})},\ \Eprint
  {https://arxiv.org/abs/quant-ph/0110143} {arXiv:quant-ph/0110143}
  \BibitemShut {NoStop}%
\bibitem [{\citenamefont {Terhal}(2015)}]{Terhal15}%
  \BibitemOpen
  \bibfield  {author} {\bibinfo {author} {\bibfnamefont {B.~M.}\ \bibnamefont
  {Terhal}},\ }\bibfield  {title} {\bibinfo {title} {Quantum error correction
  for quantum memories},\ }\href {https://doi.org/10.1103/RevModPhys.87.307}
  {\bibfield  {journal} {\bibinfo  {journal} {Rev. Mod. Phys.}\ }\textbf
  {\bibinfo {volume} {87}},\ \bibinfo {pages} {307} (\bibinfo {year} {2015})},\
  \Eprint {https://arxiv.org/abs/1302.3428} {arXiv:1302.3428 [quant-ph]}
  \BibitemShut {NoStop}%
\bibitem [{\citenamefont {Brown}\ \emph {et~al.}(2016)\citenamefont {Brown},
  \citenamefont {Loss}, \citenamefont {Pachos}, \citenamefont {Self},\ and\
  \citenamefont {Wootton}}]{Brown16}%
  \BibitemOpen
  \bibfield  {author} {\bibinfo {author} {\bibfnamefont {B.~J.}\ \bibnamefont
  {Brown}}, \bibinfo {author} {\bibfnamefont {D.}~\bibnamefont {Loss}},
  \bibinfo {author} {\bibfnamefont {J.~K.}\ \bibnamefont {Pachos}}, \bibinfo
  {author} {\bibfnamefont {C.~N.}\ \bibnamefont {Self}},\ and\ \bibinfo
  {author} {\bibfnamefont {J.~R.}\ \bibnamefont {Wootton}},\ }\bibfield
  {title} {\bibinfo {title} {Quantum memories at finite temperature},\ }\href
  {https://doi.org/10.1103/RevModPhys.88.045005} {\bibfield  {journal}
  {\bibinfo  {journal} {Rev. Mod. Phys.}\ }\textbf {\bibinfo {volume} {88}},\
  \bibinfo {pages} {045005} (\bibinfo {year} {2016})},\ \Eprint
  {https://arxiv.org/abs/1411.6643} {arXiv:1411.6643 [quant-ph]} \BibitemShut
  {NoStop}%
\bibitem [{\citenamefont {Campbell}\ \emph {et~al.}(2017)\citenamefont
  {Campbell}, \citenamefont {Terhal},\ and\ \citenamefont
  {Vuillot}}]{Campbell17}%
  \BibitemOpen
  \bibfield  {author} {\bibinfo {author} {\bibfnamefont {E.~T.}\ \bibnamefont
  {Campbell}}, \bibinfo {author} {\bibfnamefont {B.~M.}\ \bibnamefont
  {Terhal}},\ and\ \bibinfo {author} {\bibfnamefont {C.}~\bibnamefont
  {Vuillot}},\ }\bibfield  {title} {\bibinfo {title} {Roads towards
  fault-tolerant universal quantum computation},\ }\href
  {https://doi.org/10.1038/nature23460} {\bibfield  {journal} {\bibinfo
  {journal} {Nature}\ }\textbf {\bibinfo {volume} {549}},\ \bibinfo {pages}
  {172} (\bibinfo {year} {2017})},\ \Eprint {https://arxiv.org/abs/1612.07330}
  {arXiv:1612.07330 [quant-ph]} \BibitemShut {NoStop}%
\bibitem [{\citenamefont {Aharonov}\ and\ \citenamefont
  {Ben-Or}(1997)}]{AharonovBen-Or97}%
  \BibitemOpen
  \bibfield  {author} {\bibinfo {author} {\bibfnamefont {D.}~\bibnamefont
  {Aharonov}}\ and\ \bibinfo {author} {\bibfnamefont {M.}~\bibnamefont
  {Ben-Or}},\ }\bibfield  {title} {\bibinfo {title} {Fault-tolerant quantum
  computation with constant error},\ }in\ \href
  {https://doi.org/10.1145/258533.258579} {\emph {\bibinfo {booktitle} {STOC
  '97 Proceedings of the Twenty-Ninth Annual ACM Symposium on Theory of
  Computing}}}\ (\bibinfo {year} {1997})\ p.\ \bibinfo {pages} {176},\ \Eprint
  {https://arxiv.org/abs/quant-ph/9611025} {arXiv:quant-ph/9611025}
  \BibitemShut {NoStop}%
\bibitem [{\citenamefont {Chuang}\ \emph {et~al.}(1997)\citenamefont {Chuang},
  \citenamefont {Leung},\ and\ \citenamefont {Yamamoto}}]{Chuang97}%
  \BibitemOpen
  \bibfield  {author} {\bibinfo {author} {\bibfnamefont {I.~L.}\ \bibnamefont
  {Chuang}}, \bibinfo {author} {\bibfnamefont {D.~W.}\ \bibnamefont {Leung}},\
  and\ \bibinfo {author} {\bibfnamefont {Y.}~\bibnamefont {Yamamoto}},\
  }\bibfield  {title} {\bibinfo {title} {Bosonic quantum codes for amplitude
  damping},\ }\href {https://doi.org/10.1103/PhysRevA.56.1114} {\bibfield
  {journal} {\bibinfo  {journal} {Phys. Rev. A}\ }\textbf {\bibinfo {volume}
  {56}},\ \bibinfo {pages} {1114} (\bibinfo {year} {1997})},\ \Eprint
  {https://arxiv.org/abs/quant-ph/9610043} {arXiv:quant-ph/9610043}
  \BibitemShut {NoStop}%
\bibitem [{\citenamefont {Cochrane}\ \emph {et~al.}(1999)\citenamefont
  {Cochrane}, \citenamefont {Milburn},\ and\ \citenamefont
  {Munro}}]{cochrane99}%
  \BibitemOpen
  \bibfield  {author} {\bibinfo {author} {\bibfnamefont {P.~T.}\ \bibnamefont
  {Cochrane}}, \bibinfo {author} {\bibfnamefont {G.~J.}\ \bibnamefont
  {Milburn}},\ and\ \bibinfo {author} {\bibfnamefont {W.~J.}\ \bibnamefont
  {Munro}},\ }\bibfield  {title} {\bibinfo {title} {Macroscopically distinct
  quantum-superposition states as a bosonic code for amplitude damping},\
  }\href {https://doi.org/10.1103/PhysRevA.59.2631} {\bibfield  {journal}
  {\bibinfo  {journal} {Phys. Rev. A}\ }\textbf {\bibinfo {volume} {59}},\
  \bibinfo {pages} {2631} (\bibinfo {year} {1999})},\ \Eprint
  {https://arxiv.org/abs/quant-ph/9809037} {arXiv:quant-ph/9809037}
  \BibitemShut {NoStop}%
\bibitem [{\citenamefont {Gottesman}\ \emph {et~al.}(2001)\citenamefont
  {Gottesman}, \citenamefont {Kitaev},\ and\ \citenamefont
  {Preskill}}]{gottesman01}%
  \BibitemOpen
  \bibfield  {author} {\bibinfo {author} {\bibfnamefont {D.}~\bibnamefont
  {Gottesman}}, \bibinfo {author} {\bibfnamefont {A.}~\bibnamefont {Kitaev}},\
  and\ \bibinfo {author} {\bibfnamefont {J.}~\bibnamefont {Preskill}},\
  }\bibfield  {title} {\bibinfo {title} {Encoding a qubit in an oscillator},\
  }\href {https://doi.org/10.1103/PhysRevA.64.012310} {\bibfield  {journal}
  {\bibinfo  {journal} {Phys. Rev. A}\ }\textbf {\bibinfo {volume} {64}},\
  \bibinfo {pages} {012310} (\bibinfo {year} {2001})},\ \Eprint
  {https://arxiv.org/abs/quant-ph/0008040} {arXiv:quant-ph/0008040}
  \BibitemShut {NoStop}%
\bibitem [{\citenamefont {Mirrahimi}\ \emph {et~al.}(2014)\citenamefont
  {Mirrahimi}, \citenamefont {Leghtas}, \citenamefont {Albert}, \citenamefont
  {Touzard}, \citenamefont {Schoelkopf}, \citenamefont {Jiang},\ and\
  \citenamefont {Devoret}}]{mirrahimi2014dynamically}%
  \BibitemOpen
  \bibfield  {author} {\bibinfo {author} {\bibfnamefont {M.}~\bibnamefont
  {Mirrahimi}}, \bibinfo {author} {\bibfnamefont {Z.}~\bibnamefont {Leghtas}},
  \bibinfo {author} {\bibfnamefont {V.~V.}\ \bibnamefont {Albert}}, \bibinfo
  {author} {\bibfnamefont {S.}~\bibnamefont {Touzard}}, \bibinfo {author}
  {\bibfnamefont {R.~J.}\ \bibnamefont {Schoelkopf}}, \bibinfo {author}
  {\bibfnamefont {L.}~\bibnamefont {Jiang}},\ and\ \bibinfo {author}
  {\bibfnamefont {M.~H.}\ \bibnamefont {Devoret}},\ }\bibfield  {title}
  {\bibinfo {title} {Dynamically protected cat-qubits: a new paradigm for
  universal quantum computation},\ }\href
  {https://doi.org/10.1088/1367-2630/16/4/045014} {\bibfield  {journal}
  {\bibinfo  {journal} {New J. Phys.}\ }\textbf {\bibinfo {volume} {16}},\
  \bibinfo {pages} {045014} (\bibinfo {year} {2014})},\ \Eprint
  {https://arxiv.org/abs/1312.2017} {arXiv:1312.2017 [quant-ph]} \BibitemShut
  {NoStop}%
\bibitem [{\citenamefont {Gertler}\ \emph {et~al.}(2021)\citenamefont
  {Gertler}, \citenamefont {Baker}, \citenamefont {Li}, \citenamefont {Shirol},
  \citenamefont {Koch},\ and\ \citenamefont {Wang}}]{gertler2021protecting}%
  \BibitemOpen
  \bibfield  {author} {\bibinfo {author} {\bibfnamefont {J.~M.}\ \bibnamefont
  {Gertler}}, \bibinfo {author} {\bibfnamefont {B.}~\bibnamefont {Baker}},
  \bibinfo {author} {\bibfnamefont {J.}~\bibnamefont {Li}}, \bibinfo {author}
  {\bibfnamefont {S.}~\bibnamefont {Shirol}}, \bibinfo {author} {\bibfnamefont
  {J.}~\bibnamefont {Koch}},\ and\ \bibinfo {author} {\bibfnamefont
  {C.}~\bibnamefont {Wang}},\ }\bibfield  {title} {\bibinfo {title} {Protecting
  a bosonic qubit with autonomous quantum error correction},\ }\href
  {https://doi.org/10.1038/s41586-021-03257-0} {\bibfield  {journal} {\bibinfo
  {journal} {Nature}\ }\textbf {\bibinfo {volume} {590}},\ \bibinfo {pages}
  {243} (\bibinfo {year} {2021})},\ \Eprint {https://arxiv.org/abs/2004.09322}
  {arXiv:2004.09322 [quant-ph]} \BibitemShut {NoStop}%
\bibitem [{\citenamefont {Vuillot}\ \emph {et~al.}(2019)\citenamefont
  {Vuillot}, \citenamefont {Asasi}, \citenamefont {Wang}, \citenamefont
  {Pryadko},\ and\ \citenamefont {Terhal}}]{Vuillot2018}%
  \BibitemOpen
  \bibfield  {author} {\bibinfo {author} {\bibfnamefont {C.}~\bibnamefont
  {Vuillot}}, \bibinfo {author} {\bibfnamefont {H.}~\bibnamefont {Asasi}},
  \bibinfo {author} {\bibfnamefont {Y.}~\bibnamefont {Wang}}, \bibinfo {author}
  {\bibfnamefont {L.~P.}\ \bibnamefont {Pryadko}},\ and\ \bibinfo {author}
  {\bibfnamefont {B.~M.}\ \bibnamefont {Terhal}},\ }\bibfield  {title}
  {\bibinfo {title} {Quantum error correction with the toric
  {G}ottesman-{K}itaev-{P}reskill code},\ }\href
  {https://doi.org/10.1103/PhysRevA.99.032344} {\bibfield  {journal} {\bibinfo
  {journal} {Phys. Rev. A}\ }\textbf {\bibinfo {volume} {99}},\ \bibinfo
  {pages} {032344} (\bibinfo {year} {2019})},\ \Eprint
  {https://arxiv.org/abs/1810.00047} {arXiv:1810.00047 [quant-ph]} \BibitemShut
  {NoStop}%
\bibitem [{\citenamefont {Noh}\ and\ \citenamefont
  {Chamberland}(2020)}]{Noh:2019aa}%
  \BibitemOpen
  \bibfield  {author} {\bibinfo {author} {\bibfnamefont {K.}~\bibnamefont
  {Noh}}\ and\ \bibinfo {author} {\bibfnamefont {C.}~\bibnamefont
  {Chamberland}},\ }\bibfield  {title} {\bibinfo {title} {Fault-tolerant
  bosonic quantum error correction with the
  surface--{G}ottesman-{K}itaev-{P}reskill code},\ }\href
  {https://doi.org/10.1103/PhysRevA.101.012316} {\bibfield  {journal} {\bibinfo
   {journal} {Phys. Rev. A}\ }\textbf {\bibinfo {volume} {101}},\ \bibinfo
  {pages} {012316} (\bibinfo {year} {2020})},\ \Eprint
  {https://arxiv.org/abs/1908.03579} {arXiv:1908.03579 [quant-ph]} \BibitemShut
  {NoStop}%
\bibitem [{\citenamefont {Guillaud}\ and\ \citenamefont
  {Mirrahimi}(2019)}]{Guillaud2019}%
  \BibitemOpen
  \bibfield  {author} {\bibinfo {author} {\bibfnamefont {J.}~\bibnamefont
  {Guillaud}}\ and\ \bibinfo {author} {\bibfnamefont {M.}~\bibnamefont
  {Mirrahimi}},\ }\bibfield  {title} {\bibinfo {title} {Repetition cat qubits
  for fault-tolerant quantum computation},\ }\href
  {https://doi.org/10.1103/PhysRevX.9.041053} {\bibfield  {journal} {\bibinfo
  {journal} {Phys. Rev. X}\ }\textbf {\bibinfo {volume} {9}},\ \bibinfo {pages}
  {041053} (\bibinfo {year} {2019})},\ \Eprint
  {https://arxiv.org/abs/1904.09474} {arXiv:1904.09474 [quant-ph]} \BibitemShut
  {NoStop}%
\bibitem [{\citenamefont {Grimsmo}\ \emph {et~al.}(2020)\citenamefont
  {Grimsmo}, \citenamefont {Combes},\ and\ \citenamefont
  {Baragiola}}]{Grimsmo2020}%
  \BibitemOpen
  \bibfield  {author} {\bibinfo {author} {\bibfnamefont {A.~L.}\ \bibnamefont
  {Grimsmo}}, \bibinfo {author} {\bibfnamefont {J.}~\bibnamefont {Combes}},\
  and\ \bibinfo {author} {\bibfnamefont {B.~Q.}\ \bibnamefont {Baragiola}},\
  }\bibfield  {title} {\bibinfo {title} {Quantum computing with
  rotation-symmetric bosonic codes},\ }\href
  {https://doi.org/10.1103/PhysRevX.10.011058} {\bibfield  {journal} {\bibinfo
  {journal} {Phys. Rev. X}\ }\textbf {\bibinfo {volume} {10}},\ \bibinfo
  {pages} {011058} (\bibinfo {year} {2020})},\ \Eprint
  {https://arxiv.org/abs/1901.08071} {arXiv:1901.08071 [quant-ph]} \BibitemShut
  {NoStop}%
\bibitem [{\citenamefont {Chamberland}\ \emph
  {et~al.}(2020{\natexlab{a}})\citenamefont {Chamberland}, \citenamefont {Noh},
  \citenamefont {Arrangoiz-Arriola}, \citenamefont {Campbell}, \citenamefont
  {Hann}, \citenamefont {Iverson}, \citenamefont {Putterman}, \citenamefont
  {Bohdanowicz}, \citenamefont {Flammia}, \citenamefont {Keller} \emph
  {et~al.}}]{chamberland2020building}%
  \BibitemOpen
  \bibfield  {author} {\bibinfo {author} {\bibfnamefont {C.}~\bibnamefont
  {Chamberland}}, \bibinfo {author} {\bibfnamefont {K.}~\bibnamefont {Noh}},
  \bibinfo {author} {\bibfnamefont {P.}~\bibnamefont {Arrangoiz-Arriola}},
  \bibinfo {author} {\bibfnamefont {E.~T.}\ \bibnamefont {Campbell}}, \bibinfo
  {author} {\bibfnamefont {C.~T.}\ \bibnamefont {Hann}}, \bibinfo {author}
  {\bibfnamefont {J.}~\bibnamefont {Iverson}}, \bibinfo {author} {\bibfnamefont
  {H.}~\bibnamefont {Putterman}}, \bibinfo {author} {\bibfnamefont {T.~C.}\
  \bibnamefont {Bohdanowicz}}, \bibinfo {author} {\bibfnamefont {S.~T.}\
  \bibnamefont {Flammia}}, \bibinfo {author} {\bibfnamefont {A.}~\bibnamefont
  {Keller}}, \emph {et~al.},\ }\href@noop {} {\bibinfo {title} {Building a
  fault-tolerant quantum computer using concatenated cat codes}} (\bibinfo
  {year} {2020}{\natexlab{a}}),\ \Eprint {https://arxiv.org/abs/2012.04108}
  {arXiv:2012.04108 [quant-ph]} \BibitemShut {NoStop}%
\bibitem [{\citenamefont {Noh}\ \emph {et~al.}(2021)\citenamefont {Noh},
  \citenamefont {Chamberland},\ and\ \citenamefont {Brandão}}]{noh2021low}%
  \BibitemOpen
  \bibfield  {author} {\bibinfo {author} {\bibfnamefont {K.}~\bibnamefont
  {Noh}}, \bibinfo {author} {\bibfnamefont {C.}~\bibnamefont {Chamberland}},\
  and\ \bibinfo {author} {\bibfnamefont {F.~G. S.~L.}\ \bibnamefont
  {Brandão}},\ }\href@noop {} {\bibinfo {title} {Low overhead fault-tolerant
  quantum error correction with the surface-{GKP} code}} (\bibinfo {year}
  {2021}),\ \Eprint {https://arxiv.org/abs/2103.06994} {arXiv:2103.06994
  [quant-ph]} \BibitemShut {NoStop}%
\bibitem [{\citenamefont {Terhal}\ \emph {et~al.}(2020)\citenamefont {Terhal},
  \citenamefont {Conrad},\ and\ \citenamefont {Vuillot}}]{Terhal2020}%
  \BibitemOpen
  \bibfield  {author} {\bibinfo {author} {\bibfnamefont {B.~M.}\ \bibnamefont
  {Terhal}}, \bibinfo {author} {\bibfnamefont {J.}~\bibnamefont {Conrad}},\
  and\ \bibinfo {author} {\bibfnamefont {C.}~\bibnamefont {Vuillot}},\
  }\bibfield  {title} {\bibinfo {title} {Towards scalable bosonic quantum error
  correction},\ }\href {https://doi.org/10.1088/2058-9565/ab98a5} {\bibfield
  {journal} {\bibinfo  {journal} {Quantum Sci. Technol.}\ }\textbf {\bibinfo
  {volume} {5}},\ \bibinfo {pages} {043001} (\bibinfo {year} {2020})},\ \Eprint
  {https://arxiv.org/abs/2002.11008} {arXiv:2002.11008 [quant-ph]} \BibitemShut
  {NoStop}%
\bibitem [{\citenamefont {Joshi}\ \emph {et~al.}(2021)\citenamefont {Joshi},
  \citenamefont {Noh},\ and\ \citenamefont {Gao}}]{joshi2021quantum}%
  \BibitemOpen
  \bibfield  {author} {\bibinfo {author} {\bibfnamefont {A.}~\bibnamefont
  {Joshi}}, \bibinfo {author} {\bibfnamefont {K.}~\bibnamefont {Noh}},\ and\
  \bibinfo {author} {\bibfnamefont {Y.~Y.}\ \bibnamefont {Gao}},\ }\bibfield
  {title} {\bibinfo {title} {Quantum information processing with bosonic qubits
  in circuit {QED}},\ }\href {https://doi.org/10.1088/2058-9565/abe989}
  {\bibfield  {journal} {\bibinfo  {journal} {Quantum Sci. Technol.}\ }\textbf
  {\bibinfo {volume} {6}},\ \bibinfo {pages} {033001} (\bibinfo {year}
  {2021})},\ \Eprint {https://arxiv.org/abs/2008.13471} {arXiv:2008.13471
  [quant-ph]} \BibitemShut {NoStop}%
\bibitem [{\citenamefont {Goto}(2016)}]{Goto2016}%
  \BibitemOpen
  \bibfield  {author} {\bibinfo {author} {\bibfnamefont {H.}~\bibnamefont
  {Goto}},\ }\bibfield  {title} {\bibinfo {title} {Universal quantum
  computation with a nonlinear oscillator network},\ }\href
  {https://doi.org/10.1103/PhysRevA.93.050301} {\bibfield  {journal} {\bibinfo
  {journal} {Phys. Rev. A}\ }\textbf {\bibinfo {volume} {93}},\ \bibinfo
  {pages} {050301} (\bibinfo {year} {2016})},\ \Eprint
  {https://arxiv.org/abs/1605.03250} {arXiv:1605.03250 [quant-ph]} \BibitemShut
  {NoStop}%
\bibitem [{\citenamefont {Puri}\ \emph {et~al.}(2017)\citenamefont {Puri},
  \citenamefont {Boutin},\ and\ \citenamefont {Blais}}]{Puri2017}%
  \BibitemOpen
  \bibfield  {author} {\bibinfo {author} {\bibfnamefont {S.}~\bibnamefont
  {Puri}}, \bibinfo {author} {\bibfnamefont {S.}~\bibnamefont {Boutin}},\ and\
  \bibinfo {author} {\bibfnamefont {A.}~\bibnamefont {Blais}},\ }\bibfield
  {title} {\bibinfo {title} {Engineering the quantum states of light in a
  kerr-nonlinear resonator by two-photon driving},\ }\href
  {https://doi.org/10.1038/s41534-017-0019-1} {\bibfield  {journal} {\bibinfo
  {journal} {npj Quantum Inf.}\ }\textbf {\bibinfo {volume} {3}},\ \bibinfo
  {pages} {18} (\bibinfo {year} {2017})},\ \Eprint
  {https://arxiv.org/abs/1605.09408} {arXiv:1605.09408 [quant-ph]} \BibitemShut
  {NoStop}%
\bibitem [{\citenamefont {Puri}\ \emph {et~al.}(2020)\citenamefont {Puri},
  \citenamefont {St-Jean}, \citenamefont {Gross}, \citenamefont {Grimm},
  \citenamefont {Frattini}, \citenamefont {Iyer}, \citenamefont {Krishna},
  \citenamefont {Touzard}, \citenamefont {Jiang}, \citenamefont {Blais} \emph
  {et~al.}}]{Puri2020}%
  \BibitemOpen
  \bibfield  {author} {\bibinfo {author} {\bibfnamefont {S.}~\bibnamefont
  {Puri}}, \bibinfo {author} {\bibfnamefont {L.}~\bibnamefont {St-Jean}},
  \bibinfo {author} {\bibfnamefont {J.~A.}\ \bibnamefont {Gross}}, \bibinfo
  {author} {\bibfnamefont {A.}~\bibnamefont {Grimm}}, \bibinfo {author}
  {\bibfnamefont {N.~E.}\ \bibnamefont {Frattini}}, \bibinfo {author}
  {\bibfnamefont {P.~S.}\ \bibnamefont {Iyer}}, \bibinfo {author}
  {\bibfnamefont {A.}~\bibnamefont {Krishna}}, \bibinfo {author} {\bibfnamefont
  {S.}~\bibnamefont {Touzard}}, \bibinfo {author} {\bibfnamefont
  {L.}~\bibnamefont {Jiang}}, \bibinfo {author} {\bibfnamefont
  {A.}~\bibnamefont {Blais}}, \emph {et~al.},\ }\bibfield  {title} {\bibinfo
  {title} {Bias-preserving gates with stabilized cat qubits},\ }\href
  {https://doi.org/10.1126/sciadv.aay5901} {\bibfield  {journal} {\bibinfo
  {journal} {Sci. Adv.}\ }\textbf {\bibinfo {volume} {6}},\ \bibinfo {pages}
  {eaay5901} (\bibinfo {year} {2020})},\ \Eprint
  {https://arxiv.org/abs/1905.00450} {arXiv:1905.00450 [quant-ph]} \BibitemShut
  {NoStop}%
\bibitem [{\citenamefont {Aliferis}\ and\ \citenamefont
  {Preskill}(2008)}]{aliferis_fault-tolerant_2008}%
  \BibitemOpen
  \bibfield  {author} {\bibinfo {author} {\bibfnamefont {P.}~\bibnamefont
  {Aliferis}}\ and\ \bibinfo {author} {\bibfnamefont {J.}~\bibnamefont
  {Preskill}},\ }\bibfield  {title} {\bibinfo {title} {Fault-tolerant quantum
  computation against biased noise},\ }\href
  {https://doi.org/10.1103/PhysRevA.78.052331} {\bibfield  {journal} {\bibinfo
  {journal} {Phys. Rev. A}\ }\textbf {\bibinfo {volume} {78}},\ \bibinfo
  {pages} {052331} (\bibinfo {year} {2008})},\ \Eprint
  {https://arxiv.org/abs/0710.1301} {arXiv:0710.1301 [quant-ph]} \BibitemShut
  {NoStop}%
\bibitem [{\citenamefont {Aliferis}\ \emph {et~al.}(2009)\citenamefont
  {Aliferis}, \citenamefont {Brito}, \citenamefont {DiVincenzo}, \citenamefont
  {Preskill}, \citenamefont {Steffen},\ and\ \citenamefont
  {Terhal}}]{aliferis_fault-tolerant_2009}%
  \BibitemOpen
  \bibfield  {author} {\bibinfo {author} {\bibfnamefont {P.}~\bibnamefont
  {Aliferis}}, \bibinfo {author} {\bibfnamefont {F.}~\bibnamefont {Brito}},
  \bibinfo {author} {\bibfnamefont {D.~P.}\ \bibnamefont {DiVincenzo}},
  \bibinfo {author} {\bibfnamefont {J.}~\bibnamefont {Preskill}}, \bibinfo
  {author} {\bibfnamefont {M.}~\bibnamefont {Steffen}},\ and\ \bibinfo {author}
  {\bibfnamefont {B.~M.}\ \bibnamefont {Terhal}},\ }\bibfield  {title}
  {\bibinfo {title} {Fault-tolerant computing with biased-noise superconducting
  qubits: a case study},\ }\href
  {https://doi.org/10.1088/1367-2630/11/1/013061} {\bibfield  {journal}
  {\bibinfo  {journal} {New J. Phys}\ }\textbf {\bibinfo {volume} {11}},\
  \bibinfo {pages} {013061} (\bibinfo {year} {2009})},\ \Eprint
  {https://arxiv.org/abs/0806.0383} {arXiv:0806.0383 [quant-ph]} \BibitemShut
  {NoStop}%
\bibitem [{\citenamefont {Stephens}\ \emph {et~al.}(2013)\citenamefont
  {Stephens}, \citenamefont {Munro},\ and\ \citenamefont
  {Nemoto}}]{Stephens13bias}%
  \BibitemOpen
  \bibfield  {author} {\bibinfo {author} {\bibfnamefont {A.~M.}\ \bibnamefont
  {Stephens}}, \bibinfo {author} {\bibfnamefont {W.~J.}\ \bibnamefont
  {Munro}},\ and\ \bibinfo {author} {\bibfnamefont {K.}~\bibnamefont
  {Nemoto}},\ }\bibfield  {title} {\bibinfo {title} {High-threshold topological
  quantum error correction against biased noise},\ }\href
  {https://doi.org/10.1103/PhysRevA.88.060301} {\bibfield  {journal} {\bibinfo
  {journal} {Phys. Rev. A}\ }\textbf {\bibinfo {volume} {88}},\ \bibinfo
  {pages} {060301} (\bibinfo {year} {2013})},\ \Eprint
  {https://arxiv.org/abs/1308.4776} {arXiv:1308.4776} \BibitemShut {NoStop}%
\bibitem [{\citenamefont {Tuckett}\ \emph {et~al.}(2018)\citenamefont
  {Tuckett}, \citenamefont {Bartlett},\ and\ \citenamefont
  {Flammia}}]{Tuckett18}%
  \BibitemOpen
  \bibfield  {author} {\bibinfo {author} {\bibfnamefont {D.~K.}\ \bibnamefont
  {Tuckett}}, \bibinfo {author} {\bibfnamefont {S.~D.}\ \bibnamefont
  {Bartlett}},\ and\ \bibinfo {author} {\bibfnamefont {S.~T.}\ \bibnamefont
  {Flammia}},\ }\bibfield  {title} {\bibinfo {title} {Ultrahigh error threshold
  for surface codes with biased noise},\ }\href
  {https://doi.org/10.1103/PhysRevLett.120.050505} {\bibfield  {journal}
  {\bibinfo  {journal} {Phys. Rev. Lett.}\ }\textbf {\bibinfo {volume} {120}},\
  \bibinfo {pages} {050505} (\bibinfo {year} {2018})},\ \Eprint
  {https://arxiv.org/abs/1708.08474} {arXiv:1708.08474 [quant-ph]} \BibitemShut
  {NoStop}%
\bibitem [{\citenamefont {Tuckett}\ \emph {et~al.}(2019)\citenamefont
  {Tuckett}, \citenamefont {Darmawan}, \citenamefont {Chubb}, \citenamefont
  {Bravyi}, \citenamefont {Bartlett},\ and\ \citenamefont
  {Flammia}}]{Tuckett19}%
  \BibitemOpen
  \bibfield  {author} {\bibinfo {author} {\bibfnamefont {D.~K.}\ \bibnamefont
  {Tuckett}}, \bibinfo {author} {\bibfnamefont {A.~S.}\ \bibnamefont
  {Darmawan}}, \bibinfo {author} {\bibfnamefont {C.~T.}\ \bibnamefont {Chubb}},
  \bibinfo {author} {\bibfnamefont {S.}~\bibnamefont {Bravyi}}, \bibinfo
  {author} {\bibfnamefont {S.~D.}\ \bibnamefont {Bartlett}},\ and\ \bibinfo
  {author} {\bibfnamefont {S.~T.}\ \bibnamefont {Flammia}},\ }\bibfield
  {title} {\bibinfo {title} {Tailoring surface codes for highly biased noise},\
  }\href {https://doi.org/10.1103/PhysRevX.9.041031} {\bibfield  {journal}
  {\bibinfo  {journal} {Phys. Rev. X}\ }\textbf {\bibinfo {volume} {9}},\
  \bibinfo {pages} {041031} (\bibinfo {year} {2019})},\ \Eprint
  {https://arxiv.org/abs/1812.08186} {arXiv:1812.08186 [quant-ph]} \BibitemShut
  {NoStop}%
\bibitem [{\citenamefont {Xu}\ \emph {et~al.}(2019)\citenamefont {Xu},
  \citenamefont {Zhao}, \citenamefont {Yuan},\ and\ \citenamefont
  {Benjamin}}]{Xu19}%
  \BibitemOpen
  \bibfield  {author} {\bibinfo {author} {\bibfnamefont {X.}~\bibnamefont
  {Xu}}, \bibinfo {author} {\bibfnamefont {Q.}~\bibnamefont {Zhao}}, \bibinfo
  {author} {\bibfnamefont {X.}~\bibnamefont {Yuan}},\ and\ \bibinfo {author}
  {\bibfnamefont {S.~C.}\ \bibnamefont {Benjamin}},\ }\bibfield  {title}
  {\bibinfo {title} {High-threshold code for modular hardware with asymmetric
  noise},\ }\href {https://doi.org/10.1103/PhysRevApplied.12.064006} {\bibfield
   {journal} {\bibinfo  {journal} {Phys. Rev. Applied}\ }\textbf {\bibinfo
  {volume} {12}},\ \bibinfo {pages} {064006} (\bibinfo {year} {2019})},\
  \Eprint {https://arxiv.org/abs/1812.01505} {arXiv:1812.01505} \BibitemShut
  {NoStop}%
\bibitem [{\citenamefont {Li}\ \emph {et~al.}(2019)\citenamefont {Li},
  \citenamefont {Miller}, \citenamefont {Newman}, \citenamefont {Wu},\ and\
  \citenamefont {Brown}}]{Li2019}%
  \BibitemOpen
  \bibfield  {author} {\bibinfo {author} {\bibfnamefont {M.}~\bibnamefont
  {Li}}, \bibinfo {author} {\bibfnamefont {D.}~\bibnamefont {Miller}}, \bibinfo
  {author} {\bibfnamefont {M.}~\bibnamefont {Newman}}, \bibinfo {author}
  {\bibfnamefont {Y.}~\bibnamefont {Wu}},\ and\ \bibinfo {author}
  {\bibfnamefont {K.~R.}\ \bibnamefont {Brown}},\ }\bibfield  {title} {\bibinfo
  {title} {{2D} compass codes},\ }\href
  {https://doi.org/10.1103/PhysRevX.9.021041} {\bibfield  {journal} {\bibinfo
  {journal} {Phys. Rev. X}\ }\textbf {\bibinfo {volume} {9}},\ \bibinfo {pages}
  {021041} (\bibinfo {year} {2019})},\ \Eprint
  {https://arxiv.org/abs/1809.01193} {arXiv:1809.01193 [quant-ph]} \BibitemShut
  {NoStop}%
\bibitem [{\citenamefont {Tuckett}\ \emph {et~al.}(2020)\citenamefont
  {Tuckett}, \citenamefont {Bartlett}, \citenamefont {Flammia},\ and\
  \citenamefont {Brown}}]{Tuckett20}%
  \BibitemOpen
  \bibfield  {author} {\bibinfo {author} {\bibfnamefont {D.~K.}\ \bibnamefont
  {Tuckett}}, \bibinfo {author} {\bibfnamefont {S.~D.}\ \bibnamefont
  {Bartlett}}, \bibinfo {author} {\bibfnamefont {S.~T.}\ \bibnamefont
  {Flammia}},\ and\ \bibinfo {author} {\bibfnamefont {B.~J.}\ \bibnamefont
  {Brown}},\ }\bibfield  {title} {\bibinfo {title} {Fault-tolerant thresholds
  for the surface code in excess of $5\%$ under biased noise},\ }\href
  {https://doi.org/10.1103/PhysRevLett.124.130501} {\bibfield  {journal}
  {\bibinfo  {journal} {Phys. Rev. Lett.}\ }\textbf {\bibinfo {volume} {124}},\
  \bibinfo {pages} {130501} (\bibinfo {year} {2020})},\ \Eprint
  {https://arxiv.org/abs/1907.02554} {arXiv:1907.02554 [quant-ph]} \BibitemShut
  {NoStop}%
\bibitem [{\citenamefont {Huang}\ \emph
  {et~al.}(2020{\natexlab{a}})\citenamefont {Huang}, \citenamefont {Newman},\
  and\ \citenamefont {Brown}}]{Huang20}%
  \BibitemOpen
  \bibfield  {author} {\bibinfo {author} {\bibfnamefont {S.}~\bibnamefont
  {Huang}}, \bibinfo {author} {\bibfnamefont {M.}~\bibnamefont {Newman}},\ and\
  \bibinfo {author} {\bibfnamefont {K.~R.}\ \bibnamefont {Brown}},\ }\bibfield
  {title} {\bibinfo {title} {Fault-tolerant weighted union-find decoding on the
  toric code},\ }\href {https://doi.org/10.1103/PhysRevA.102.012419} {\bibfield
   {journal} {\bibinfo  {journal} {Phys. Rev. A}\ }\textbf {\bibinfo {volume}
  {102}},\ \bibinfo {pages} {012419} (\bibinfo {year} {2020}{\natexlab{a}})},\
  \Eprint {https://arxiv.org/abs/2004.04693} {arXiv:2004.04693 [quant-ph]}
  \BibitemShut {NoStop}%
\bibitem [{\citenamefont {Guillaud}\ and\ \citenamefont
  {Mirrahimi}(2021)}]{guillaud2021error}%
  \BibitemOpen
  \bibfield  {author} {\bibinfo {author} {\bibfnamefont {J.}~\bibnamefont
  {Guillaud}}\ and\ \bibinfo {author} {\bibfnamefont {M.}~\bibnamefont
  {Mirrahimi}},\ }\bibfield  {title} {\bibinfo {title} {Error rates and
  resource overheads of repetition cat qubits},\ }\href
  {https://doi.org/10.1103/PhysRevA.103.042413} {\bibfield  {journal} {\bibinfo
   {journal} {Phys. Rev. A}\ }\textbf {\bibinfo {volume} {103}},\ \bibinfo
  {pages} {042413} (\bibinfo {year} {2021})},\ \Eprint
  {https://arxiv.org/abs/2009.10756} {arXiv:2009.10756 [quant-ph]} \BibitemShut
  {NoStop}%
\bibitem [{\citenamefont {H\"anggli}\ \emph {et~al.}(2020)\citenamefont
  {H\"anggli}, \citenamefont {Heinze},\ and\ \citenamefont
  {K\"onig}}]{hanggli_enhanced_2020}%
  \BibitemOpen
  \bibfield  {author} {\bibinfo {author} {\bibfnamefont {L.}~\bibnamefont
  {H\"anggli}}, \bibinfo {author} {\bibfnamefont {M.}~\bibnamefont {Heinze}},\
  and\ \bibinfo {author} {\bibfnamefont {R.}~\bibnamefont {K\"onig}},\
  }\bibfield  {title} {\bibinfo {title} {Enhanced noise resilience of the
  surface--{G}ottesman-{K}itaev-{P}reskill code via designed bias},\ }\href
  {https://doi.org/10.1103/PhysRevA.102.052408} {\bibfield  {journal} {\bibinfo
   {journal} {Phys. Rev. A}\ }\textbf {\bibinfo {volume} {102}},\ \bibinfo
  {pages} {052408} (\bibinfo {year} {2020})},\ \Eprint
  {https://arxiv.org/abs/2004.00541} {arXiv:2004.00541 [quant-ph]} \BibitemShut
  {NoStop}%
\bibitem [{\citenamefont {Higgott}\ and\ \citenamefont
  {Breuckmann}(2020)}]{higgott_subsystem_2020}%
  \BibitemOpen
  \bibfield  {author} {\bibinfo {author} {\bibfnamefont {O.}~\bibnamefont
  {Higgott}}\ and\ \bibinfo {author} {\bibfnamefont {N.~P.}\ \bibnamefont
  {Breuckmann}},\ }\href@noop {} {\bibinfo {title} {Subsystem codes with high
  thresholds by gauge fixing and reduced qubit overhead}} (\bibinfo {year}
  {2020}),\ \Eprint {https://arxiv.org/abs/2010.09626} {arXiv:2010.09626
  [quant-ph]} \BibitemShut {NoStop}%
\bibitem [{\citenamefont {Wen}(2003)}]{Wen03}%
  \BibitemOpen
  \bibfield  {author} {\bibinfo {author} {\bibfnamefont {X.-G.}\ \bibnamefont
  {Wen}},\ }\bibfield  {title} {\bibinfo {title} {Quantum orders in an exact
  soluble model},\ }\href {https://doi.org/10.1103/physrevlett.90.016803}
  {\bibfield  {journal} {\bibinfo  {journal} {Phys. Rev. Lett.}\ }\textbf
  {\bibinfo {volume} {90}},\ \bibinfo {pages} {016803} (\bibinfo {year}
  {2003})},\ \Eprint {https://arxiv.org/abs/quant-ph/0205004}
  {arXiv:quant-ph/0205004} \BibitemShut {NoStop}%
\bibitem [{\citenamefont {\surname{Bonilla Ataides}}\ \emph
  {et~al.}(2021)\citenamefont {\surname{Bonilla Ataides}}, \citenamefont
  {Tuckett}, \citenamefont {Bartlett}, \citenamefont {Flammia},\ and\
  \citenamefont {Brown}}]{BonillaAtaides20}%
  \BibitemOpen
  \bibfield  {author} {\bibinfo {author} {\bibfnamefont {J.~P.}\ \bibnamefont
  {\surname{Bonilla Ataides}}}, \bibinfo {author} {\bibfnamefont {D.~K.}\
  \bibnamefont {Tuckett}}, \bibinfo {author} {\bibfnamefont {S.~D.}\
  \bibnamefont {Bartlett}}, \bibinfo {author} {\bibfnamefont {S.~T.}\
  \bibnamefont {Flammia}},\ and\ \bibinfo {author} {\bibfnamefont {B.~J.}\
  \bibnamefont {Brown}},\ }\bibfield  {title} {\bibinfo {title} {The {XZZX}
  surface code},\ }\href {https://doi.org/10.1038/s41467-021-22274-1}
  {\bibfield  {journal} {\bibinfo  {journal} {Nat. Commun.}\ }\textbf {\bibinfo
  {volume} {11}},\ \bibinfo {pages} {2172} (\bibinfo {year} {2021})},\ \Eprint
  {https://arxiv.org/abs/2009.07851} {arXiv:2009.07851 [quant-ph]} \BibitemShut
  {NoStop}%
\bibitem [{\citenamefont {Bravyi}\ and\ \citenamefont
  {Kitaev}(1998)}]{Bravyi1998}%
  \BibitemOpen
  \bibfield  {author} {\bibinfo {author} {\bibfnamefont {S.~B.}\ \bibnamefont
  {Bravyi}}\ and\ \bibinfo {author} {\bibfnamefont {A.~Y.}\ \bibnamefont
  {Kitaev}},\ }\href@noop {} {\bibinfo {title} {Quantum codes on a lattice with
  boundary}} (\bibinfo {year} {1998}),\ \Eprint
  {https://arxiv.org/abs/quant-ph/9811052} {arXiv:quant-ph/9811052 [quant-ph]}
  \BibitemShut {NoStop}%
\bibitem [{\citenamefont {Edmonds}(1965)}]{Edmonds65}%
  \BibitemOpen
  \bibfield  {author} {\bibinfo {author} {\bibfnamefont {J.}~\bibnamefont
  {Edmonds}},\ }\bibfield  {title} {\bibinfo {title} {Paths, trees and
  flowers},\ }\href {https://doi.org/10.4153/CJM-1965-045-4} {\bibfield
  {journal} {\bibinfo  {journal} {Can. J. Math.}\ }\textbf {\bibinfo {volume}
  {17}},\ \bibinfo {pages} {449} (\bibinfo {year} {1965})}\BibitemShut
  {NoStop}%
\bibitem [{\citenamefont {Kolmogorov}(2009)}]{Kolmogorov09}%
  \BibitemOpen
  \bibfield  {author} {\bibinfo {author} {\bibfnamefont {V.}~\bibnamefont
  {Kolmogorov}},\ }\bibfield  {title} {\bibinfo {title} {Blossom {V}: A new
  implementation of a minimum cost perfect matching algorithm},\ }\href
  {https://doi.org/10.1007/s12532-009-0002-8} {\bibfield  {journal} {\bibinfo
  {journal} {Math. Prog. Comput.}\ }\textbf {\bibinfo {volume} {1}},\ \bibinfo
  {pages} {43} (\bibinfo {year} {2009})}\BibitemShut {NoStop}%
\bibitem [{\citenamefont {Brown}\ and\ \citenamefont
  {Williamson}(2020)}]{Brown20}%
  \BibitemOpen
  \bibfield  {author} {\bibinfo {author} {\bibfnamefont {B.~J.}\ \bibnamefont
  {Brown}}\ and\ \bibinfo {author} {\bibfnamefont {D.~J.}\ \bibnamefont
  {Williamson}},\ }\bibfield  {title} {\bibinfo {title} {Parallelized quantum
  error correction with fracton topological codes},\ }\href
  {https://doi.org/10.1103/PhysRevResearch.2.013303} {\bibfield  {journal}
  {\bibinfo  {journal} {Phys. Rev. Research}\ }\textbf {\bibinfo {volume}
  {2}},\ \bibinfo {pages} {013303} (\bibinfo {year} {2020})},\ \Eprint
  {https://arxiv.org/abs/1901.08061} {arXiv:1901.08061 [quant-ph]} \BibitemShut
  {NoStop}%
\bibitem [{\citenamefont {Jurcevic}\ \emph {et~al.}(2021)\citenamefont
  {Jurcevic}, \citenamefont {Javadi-Abhari}, \citenamefont {Bishop},
  \citenamefont {Lauer}, \citenamefont {Borgorin}, \citenamefont {Brink},
  \citenamefont {Capelluto}, \citenamefont {Gunluk}, \citenamefont {Itoko},
  \citenamefont {Kanazawa} \emph {et~al.}}]{jurcevic2021demonstration}%
  \BibitemOpen
  \bibfield  {author} {\bibinfo {author} {\bibfnamefont {P.}~\bibnamefont
  {Jurcevic}}, \bibinfo {author} {\bibfnamefont {A.}~\bibnamefont
  {Javadi-Abhari}}, \bibinfo {author} {\bibfnamefont {L.~S.}\ \bibnamefont
  {Bishop}}, \bibinfo {author} {\bibfnamefont {I.}~\bibnamefont {Lauer}},
  \bibinfo {author} {\bibfnamefont {D.}~\bibnamefont {Borgorin}}, \bibinfo
  {author} {\bibfnamefont {M.}~\bibnamefont {Brink}}, \bibinfo {author}
  {\bibfnamefont {L.}~\bibnamefont {Capelluto}}, \bibinfo {author}
  {\bibfnamefont {O.}~\bibnamefont {Gunluk}}, \bibinfo {author} {\bibfnamefont
  {T.}~\bibnamefont {Itoko}}, \bibinfo {author} {\bibfnamefont
  {N.}~\bibnamefont {Kanazawa}}, \emph {et~al.},\ }\bibfield  {title} {\bibinfo
  {title} {Demonstration of quantum volume 64 on a superconducting quantum
  computing system},\ }\href {https://doi.org/10.1088/2058-9565/abe519}
  {\bibfield  {journal} {\bibinfo  {journal} {Quantum Sci. Technol.}\ }\textbf
  {\bibinfo {volume} {6}},\ \bibinfo {pages} {025020} (\bibinfo {year}
  {2021})},\ \Eprint {https://arxiv.org/abs/2008.08571} {arXiv:2008.08571
  [quant-ph]} \BibitemShut {NoStop}%
\bibitem [{\citenamefont {Kjaergaard}\ \emph {et~al.}(2020)\citenamefont
  {Kjaergaard}, \citenamefont {Schwartz}, \citenamefont {Braum{\"u}ller},
  \citenamefont {Krantz}, \citenamefont {Wang}, \citenamefont {Gustavsson},\
  and\ \citenamefont {Oliver}}]{kjaergaard2020superconducting}%
  \BibitemOpen
  \bibfield  {author} {\bibinfo {author} {\bibfnamefont {M.}~\bibnamefont
  {Kjaergaard}}, \bibinfo {author} {\bibfnamefont {M.~E.}\ \bibnamefont
  {Schwartz}}, \bibinfo {author} {\bibfnamefont {J.}~\bibnamefont
  {Braum{\"u}ller}}, \bibinfo {author} {\bibfnamefont {P.}~\bibnamefont
  {Krantz}}, \bibinfo {author} {\bibfnamefont {J.~I.-J.}\ \bibnamefont {Wang}},
  \bibinfo {author} {\bibfnamefont {S.}~\bibnamefont {Gustavsson}},\ and\
  \bibinfo {author} {\bibfnamefont {W.~D.}\ \bibnamefont {Oliver}},\ }\bibfield
   {title} {\bibinfo {title} {Superconducting qubits: Current state of play},\
  }\href {https://doi.org/10.1146/annurev-conmatphys-031119-050605} {\bibfield
  {journal} {\bibinfo  {journal} {Annu. Rev. Condens. Matter Phys.}\ }\textbf
  {\bibinfo {volume} {11}},\ \bibinfo {pages} {369} (\bibinfo {year} {2020})},\
  \Eprint {https://arxiv.org/abs/1905.13641} {arXiv:1905.13641 [quant-ph]}
  \BibitemShut {NoStop}%
\bibitem [{\citenamefont {Nussinov}\ and\ \citenamefont
  {Ortiz}(2009)}]{Nussinov09}%
  \BibitemOpen
  \bibfield  {author} {\bibinfo {author} {\bibfnamefont {Z.}~\bibnamefont
  {Nussinov}}\ and\ \bibinfo {author} {\bibfnamefont {G.}~\bibnamefont
  {Ortiz}},\ }\bibfield  {title} {\bibinfo {title} {A symmetry principle for
  topological quantum order},\ }\href
  {https://doi.org/10.1016/j.aop.2008.11.002} {\bibfield  {journal} {\bibinfo
  {journal} {Ann. Phys.}\ }\textbf {\bibinfo {volume} {324}},\ \bibinfo {pages}
  {977} (\bibinfo {year} {2009})},\ \Eprint
  {https://arxiv.org/abs/cond-mat/0702377} {arXiv:cond-mat/0702377
  [cond-mat.str-el]} \BibitemShut {NoStop}%
\bibitem [{\citenamefont {Brown}\ \emph {et~al.}(2011)\citenamefont {Brown},
  \citenamefont {Son}, \citenamefont {Kraus}, \citenamefont {Fazio},\ and\
  \citenamefont {Vedral}}]{Brown11}%
  \BibitemOpen
  \bibfield  {author} {\bibinfo {author} {\bibfnamefont {B.~J.}\ \bibnamefont
  {Brown}}, \bibinfo {author} {\bibfnamefont {W.}~\bibnamefont {Son}}, \bibinfo
  {author} {\bibfnamefont {C.~V.}\ \bibnamefont {Kraus}}, \bibinfo {author}
  {\bibfnamefont {R.}~\bibnamefont {Fazio}},\ and\ \bibinfo {author}
  {\bibfnamefont {V.}~\bibnamefont {Vedral}},\ }\bibfield  {title} {\bibinfo
  {title} {Generating topological order from a two-dimensional cluster state
  using a duality mapping},\ }\href
  {https://doi.org/10.1088/1367-2630/13/6/065010} {\bibfield  {journal}
  {\bibinfo  {journal} {New J. Phys.}\ }\textbf {\bibinfo {volume} {13}},\
  \bibinfo {pages} {065010} (\bibinfo {year} {2011})},\ \Eprint
  {https://arxiv.org/abs/1105.2111} {arXiv:1105.2111 [quant-ph]} \BibitemShut
  {NoStop}%
\bibitem [{\citenamefont {Fowler}\ \emph {et~al.}(2012)\citenamefont {Fowler},
  \citenamefont {Mariantoni}, \citenamefont {Martinis},\ and\ \citenamefont
  {Cleland}}]{Fowler12a}%
  \BibitemOpen
  \bibfield  {author} {\bibinfo {author} {\bibfnamefont {A.~G.}\ \bibnamefont
  {Fowler}}, \bibinfo {author} {\bibfnamefont {M.}~\bibnamefont {Mariantoni}},
  \bibinfo {author} {\bibfnamefont {J.~M.}\ \bibnamefont {Martinis}},\ and\
  \bibinfo {author} {\bibfnamefont {A.~N.}\ \bibnamefont {Cleland}},\
  }\bibfield  {title} {\bibinfo {title} {Surface codes: Towards practical
  large-scale quantum computation},\ }\href
  {https://doi.org/10.1103/PhysRevA.86.032324} {\bibfield  {journal} {\bibinfo
  {journal} {Phys. Rev. A}\ }\textbf {\bibinfo {volume} {86}},\ \bibinfo
  {pages} {032324} (\bibinfo {year} {2012})},\ \Eprint
  {https://arxiv.org/abs/1208.0928} {arXiv:1208.0928 [quant-ph]} \BibitemShut
  {NoStop}%
\bibitem [{\citenamefont {Tomita}\ and\ \citenamefont
  {Svore}(2014)}]{tomita_low-distance_2014}%
  \BibitemOpen
  \bibfield  {author} {\bibinfo {author} {\bibfnamefont {Y.}~\bibnamefont
  {Tomita}}\ and\ \bibinfo {author} {\bibfnamefont {K.~M.}\ \bibnamefont
  {Svore}},\ }\bibfield  {title} {\bibinfo {title} {Low-distance surface codes
  under realistic quantum noise},\ }\href
  {https://doi.org/10.1103/PhysRevA.90.062320} {\bibfield  {journal} {\bibinfo
  {journal} {Phys. Rev. A}\ }\textbf {\bibinfo {volume} {90}},\ \bibinfo
  {pages} {062320} (\bibinfo {year} {2014})},\ \Eprint
  {https://arxiv.org/abs/1404.3747} {arXiv:1404.3747 [quant-ph]} \BibitemShut
  {NoStop}%
\bibitem [{\citenamefont {Sete}\ \emph {et~al.}(2021)\citenamefont {Sete},
  \citenamefont {Didier}, \citenamefont {Chen}, \citenamefont {Kulshreshtha},
  \citenamefont {Manenti},\ and\ \citenamefont {Poletto}}]{sete2021parametric}%
  \BibitemOpen
  \bibfield  {author} {\bibinfo {author} {\bibfnamefont {E.~A.}\ \bibnamefont
  {Sete}}, \bibinfo {author} {\bibfnamefont {N.}~\bibnamefont {Didier}},
  \bibinfo {author} {\bibfnamefont {A.~Q.}\ \bibnamefont {Chen}}, \bibinfo
  {author} {\bibfnamefont {S.}~\bibnamefont {Kulshreshtha}}, \bibinfo {author}
  {\bibfnamefont {R.}~\bibnamefont {Manenti}},\ and\ \bibinfo {author}
  {\bibfnamefont {S.}~\bibnamefont {Poletto}},\ }\href@noop {} {\bibinfo
  {title} {Parametric-resonance entangling gates with a tunable coupler}}
  (\bibinfo {year} {2021}),\ \Eprint {https://arxiv.org/abs/2104.03511}
  {arXiv:2104.03511 [quant-ph]} \BibitemShut {NoStop}%
\bibitem [{\citenamefont {Leghtas}\ \emph {et~al.}(2015)\citenamefont
  {Leghtas}, \citenamefont {Touzard}, \citenamefont {Pop}, \citenamefont {Kou},
  \citenamefont {Vlastakis}, \citenamefont {Petrenko}, \citenamefont {Sliwa},
  \citenamefont {Narla}, \citenamefont {Shankar}, \citenamefont {Hatridge}
  \emph {et~al.}}]{leghtas2015confining}%
  \BibitemOpen
  \bibfield  {author} {\bibinfo {author} {\bibfnamefont {Z.}~\bibnamefont
  {Leghtas}}, \bibinfo {author} {\bibfnamefont {S.}~\bibnamefont {Touzard}},
  \bibinfo {author} {\bibfnamefont {I.~M.}\ \bibnamefont {Pop}}, \bibinfo
  {author} {\bibfnamefont {A.}~\bibnamefont {Kou}}, \bibinfo {author}
  {\bibfnamefont {B.}~\bibnamefont {Vlastakis}}, \bibinfo {author}
  {\bibfnamefont {A.}~\bibnamefont {Petrenko}}, \bibinfo {author}
  {\bibfnamefont {K.~M.}\ \bibnamefont {Sliwa}}, \bibinfo {author}
  {\bibfnamefont {A.}~\bibnamefont {Narla}}, \bibinfo {author} {\bibfnamefont
  {S.}~\bibnamefont {Shankar}}, \bibinfo {author} {\bibfnamefont {M.~J.}\
  \bibnamefont {Hatridge}}, \emph {et~al.},\ }\bibfield  {title} {\bibinfo
  {title} {Confining the state of light to a quantum manifold by engineered
  two-photon loss},\ }\href {https://doi.org/10.1126/science.aaa2085}
  {\bibfield  {journal} {\bibinfo  {journal} {Science}\ }\textbf {\bibinfo
  {volume} {347}},\ \bibinfo {pages} {853} (\bibinfo {year} {2015})},\ \Eprint
  {https://arxiv.org/abs/1412.4633} {arXiv:1412.4633 [quant-ph]} \BibitemShut
  {NoStop}%
\bibitem [{\citenamefont {Touzard}\ \emph {et~al.}(2018)\citenamefont
  {Touzard}, \citenamefont {Grimm}, \citenamefont {Leghtas}, \citenamefont
  {Mundhada}, \citenamefont {Reinhold}, \citenamefont {Axline}, \citenamefont
  {Reagor}, \citenamefont {Chou}, \citenamefont {Blumoff}, \citenamefont
  {Sliwa} \emph {et~al.}}]{touzard2018coherent}%
  \BibitemOpen
  \bibfield  {author} {\bibinfo {author} {\bibfnamefont {S.}~\bibnamefont
  {Touzard}}, \bibinfo {author} {\bibfnamefont {A.}~\bibnamefont {Grimm}},
  \bibinfo {author} {\bibfnamefont {Z.}~\bibnamefont {Leghtas}}, \bibinfo
  {author} {\bibfnamefont {S.~O.}\ \bibnamefont {Mundhada}}, \bibinfo {author}
  {\bibfnamefont {P.}~\bibnamefont {Reinhold}}, \bibinfo {author}
  {\bibfnamefont {C.}~\bibnamefont {Axline}}, \bibinfo {author} {\bibfnamefont
  {M.}~\bibnamefont {Reagor}}, \bibinfo {author} {\bibfnamefont
  {K.}~\bibnamefont {Chou}}, \bibinfo {author} {\bibfnamefont {J.}~\bibnamefont
  {Blumoff}}, \bibinfo {author} {\bibfnamefont {K.~M.}\ \bibnamefont {Sliwa}},
  \emph {et~al.},\ }\bibfield  {title} {\bibinfo {title} {Coherent oscillations
  inside a quantum manifold stabilized by dissipation},\ }\href
  {https://doi.org/10.1103/PhysRevX.8.021005} {\bibfield  {journal} {\bibinfo
  {journal} {Phys. Rev. X}\ }\textbf {\bibinfo {volume} {8}},\ \bibinfo {pages}
  {021005} (\bibinfo {year} {2018})},\ \Eprint
  {https://arxiv.org/abs/1705.02401} {arXiv:1705.02401 [quant-ph]} \BibitemShut
  {NoStop}%
\bibitem [{\citenamefont {Woeginger}(2008)}]{woeginger2008}%
  \BibitemOpen
  \bibfield  {author} {\bibinfo {author} {\bibfnamefont {G.~J.}\ \bibnamefont
  {Woeginger}},\ }\bibfield  {title} {\bibinfo {title} {Open problems around
  exact algorithms},\ }\href
  {https://doi.org/https://doi.org/10.1016/j.dam.2007.03.023} {\bibfield
  {journal} {\bibinfo  {journal} {Discret. Appl. Math.}\ }\textbf {\bibinfo
  {volume} {156}},\ \bibinfo {pages} {397} (\bibinfo {year}
  {2008})}\BibitemShut {NoStop}%
\bibitem [{\citenamefont {Gambetta}\ \emph {et~al.}(2006)\citenamefont
  {Gambetta}, \citenamefont {Blais}, \citenamefont {Schuster}, \citenamefont
  {Wallraff}, \citenamefont {Frunzio}, \citenamefont {Majer}, \citenamefont
  {Devoret}, \citenamefont {Girvin},\ and\ \citenamefont
  {Schoelkopf}}]{gambetta2006qubit}%
  \BibitemOpen
  \bibfield  {author} {\bibinfo {author} {\bibfnamefont {J.}~\bibnamefont
  {Gambetta}}, \bibinfo {author} {\bibfnamefont {A.}~\bibnamefont {Blais}},
  \bibinfo {author} {\bibfnamefont {D.~I.}\ \bibnamefont {Schuster}}, \bibinfo
  {author} {\bibfnamefont {A.}~\bibnamefont {Wallraff}}, \bibinfo {author}
  {\bibfnamefont {L.}~\bibnamefont {Frunzio}}, \bibinfo {author} {\bibfnamefont
  {J.}~\bibnamefont {Majer}}, \bibinfo {author} {\bibfnamefont {M.~H.}\
  \bibnamefont {Devoret}}, \bibinfo {author} {\bibfnamefont {S.~M.}\
  \bibnamefont {Girvin}},\ and\ \bibinfo {author} {\bibfnamefont {R.~J.}\
  \bibnamefont {Schoelkopf}},\ }\bibfield  {title} {\bibinfo {title}
  {Qubit-photon interactions in a cavity: Measurement-induced dephasing and
  number splitting},\ }\href {https://doi.org/10.1103/PhysRevA.74.042318}
  {\bibfield  {journal} {\bibinfo  {journal} {Phys. Rev. A}\ }\textbf {\bibinfo
  {volume} {74}},\ \bibinfo {pages} {042318} (\bibinfo {year} {2006})},\
  \Eprint {https://arxiv.org/abs/cond-mat/0602322} {arXiv:cond-mat/0602322
  [cond-mat.mes-hall]} \BibitemShut {NoStop}%
\bibitem [{\citenamefont {Puri}\ \emph {et~al.}(2019)\citenamefont {Puri},
  \citenamefont {Grimm}, \citenamefont {Campagne-Ibarcq}, \citenamefont
  {Eickbusch}, \citenamefont {Noh}, \citenamefont {Roberts}, \citenamefont
  {Jiang}, \citenamefont {Mirrahimi}, \citenamefont {Devoret},\ and\
  \citenamefont {Girvin}}]{puri2019stabilized}%
  \BibitemOpen
  \bibfield  {author} {\bibinfo {author} {\bibfnamefont {S.}~\bibnamefont
  {Puri}}, \bibinfo {author} {\bibfnamefont {A.}~\bibnamefont {Grimm}},
  \bibinfo {author} {\bibfnamefont {P.}~\bibnamefont {Campagne-Ibarcq}},
  \bibinfo {author} {\bibfnamefont {A.}~\bibnamefont {Eickbusch}}, \bibinfo
  {author} {\bibfnamefont {K.}~\bibnamefont {Noh}}, \bibinfo {author}
  {\bibfnamefont {G.}~\bibnamefont {Roberts}}, \bibinfo {author} {\bibfnamefont
  {L.}~\bibnamefont {Jiang}}, \bibinfo {author} {\bibfnamefont
  {M.}~\bibnamefont {Mirrahimi}}, \bibinfo {author} {\bibfnamefont {M.~H.}\
  \bibnamefont {Devoret}},\ and\ \bibinfo {author} {\bibfnamefont {S.~M.}\
  \bibnamefont {Girvin}},\ }\bibfield  {title} {\bibinfo {title} {Stabilized
  cat in a driven nonlinear cavity: A fault-tolerant error syndrome detector},\
  }\href {https://doi.org/10.1103/PhysRevX.9.041009} {\bibfield  {journal}
  {\bibinfo  {journal} {Phys. Rev. X}\ }\textbf {\bibinfo {volume} {9}},\
  \bibinfo {pages} {041009} (\bibinfo {year} {2019})},\ \Eprint
  {https://arxiv.org/abs/1807.09334} {arXiv:1807.09334 [quant-ph]} \BibitemShut
  {NoStop}%
\bibitem [{\citenamefont {Bombin}\ and\ \citenamefont
  {Martin-Delgado}(2007)}]{bombin_optimal_2007}%
  \BibitemOpen
  \bibfield  {author} {\bibinfo {author} {\bibfnamefont {H.}~\bibnamefont
  {Bombin}}\ and\ \bibinfo {author} {\bibfnamefont {M.~A.}\ \bibnamefont
  {Martin-Delgado}},\ }\bibfield  {title} {\bibinfo {title} {Optimal resources
  for topological two-dimensional stabilizer codes: {Comparative} study},\
  }\href {https://doi.org/10.1103/PhysRevA.76.012305} {\bibfield  {journal}
  {\bibinfo  {journal} {Phys. Rev. A}\ }\textbf {\bibinfo {volume} {76}},\
  \bibinfo {pages} {012305} (\bibinfo {year} {2007})},\ \Eprint
  {https://arxiv.org/abs/quant-ph/0703272} {arXiv:quant-ph/0703272}
  \BibitemShut {NoStop}%
\bibitem [{\citenamefont {Touzard}\ \emph {et~al.}(2019)\citenamefont
  {Touzard}, \citenamefont {Kou}, \citenamefont {Frattini}, \citenamefont
  {Sivak}, \citenamefont {Puri}, \citenamefont {Grimm}, \citenamefont
  {Frunzio}, \citenamefont {Shankar},\ and\ \citenamefont
  {Devoret}}]{touzard2019gated}%
  \BibitemOpen
  \bibfield  {author} {\bibinfo {author} {\bibfnamefont {S.}~\bibnamefont
  {Touzard}}, \bibinfo {author} {\bibfnamefont {A.}~\bibnamefont {Kou}},
  \bibinfo {author} {\bibfnamefont {N.}~\bibnamefont {Frattini}}, \bibinfo
  {author} {\bibfnamefont {V.}~\bibnamefont {Sivak}}, \bibinfo {author}
  {\bibfnamefont {S.}~\bibnamefont {Puri}}, \bibinfo {author} {\bibfnamefont
  {A.}~\bibnamefont {Grimm}}, \bibinfo {author} {\bibfnamefont
  {L.}~\bibnamefont {Frunzio}}, \bibinfo {author} {\bibfnamefont
  {S.}~\bibnamefont {Shankar}},\ and\ \bibinfo {author} {\bibfnamefont
  {M.}~\bibnamefont {Devoret}},\ }\bibfield  {title} {\bibinfo {title} {Gated
  conditional displacement readout of superconducting qubits},\ }\href
  {https://doi.org/10.1103/PhysRevLett.122.080502} {\bibfield  {journal}
  {\bibinfo  {journal} {Phys. Rev. Lett.}\ }\textbf {\bibinfo {volume} {122}},\
  \bibinfo {pages} {080502} (\bibinfo {year} {2019})},\ \Eprint
  {https://arxiv.org/abs/1809.06964} {arXiv:1809.06964 [quant-ph]} \BibitemShut
  {NoStop}%
\bibitem [{\citenamefont {Didier}\ \emph {et~al.}(2015)\citenamefont {Didier},
  \citenamefont {Bourassa},\ and\ \citenamefont {Blais}}]{didier2015fast}%
  \BibitemOpen
  \bibfield  {author} {\bibinfo {author} {\bibfnamefont {N.}~\bibnamefont
  {Didier}}, \bibinfo {author} {\bibfnamefont {J.}~\bibnamefont {Bourassa}},\
  and\ \bibinfo {author} {\bibfnamefont {A.}~\bibnamefont {Blais}},\ }\bibfield
   {title} {\bibinfo {title} {Fast quantum nondemolition readout by parametric
  modulation of longitudinal qubit-oscillator interaction},\ }\href
  {https://doi.org/10.1103/PhysRevLett.115.203601} {\bibfield  {journal}
  {\bibinfo  {journal} {Phys. Rev. Lett.}\ }\textbf {\bibinfo {volume} {115}},\
  \bibinfo {pages} {203601} (\bibinfo {year} {2015})},\ \Eprint
  {https://arxiv.org/abs/1504.04002} {arXiv:1504.04002 [quant-ph]} \BibitemShut
  {NoStop}%
\bibitem [{\citenamefont {Werninghaus}\ \emph {et~al.}(2021)\citenamefont
  {Werninghaus}, \citenamefont {Egger}, \citenamefont {Roy}, \citenamefont
  {Machnes}, \citenamefont {Wilhelm},\ and\ \citenamefont
  {Filipp}}]{werninghaus2021leakage}%
  \BibitemOpen
  \bibfield  {author} {\bibinfo {author} {\bibfnamefont {M.}~\bibnamefont
  {Werninghaus}}, \bibinfo {author} {\bibfnamefont {D.~J.}\ \bibnamefont
  {Egger}}, \bibinfo {author} {\bibfnamefont {F.}~\bibnamefont {Roy}}, \bibinfo
  {author} {\bibfnamefont {S.}~\bibnamefont {Machnes}}, \bibinfo {author}
  {\bibfnamefont {F.~K.}\ \bibnamefont {Wilhelm}},\ and\ \bibinfo {author}
  {\bibfnamefont {S.}~\bibnamefont {Filipp}},\ }\bibfield  {title} {\bibinfo
  {title} {Leakage reduction in fast superconducting qubit gates via optimal
  control},\ }\href {https://doi.org/10.1038/s41534-020-00346-2} {\bibfield
  {journal} {\bibinfo  {journal} {npj Quantum Inf.}\ }\textbf {\bibinfo
  {volume} {7}},\ \bibinfo {pages} {14} (\bibinfo {year} {2021})},\ \Eprint
  {https://arxiv.org/abs/2003.05952} {arXiv:2003.05952 [quant-ph]} \BibitemShut
  {NoStop}%
\bibitem [{\citenamefont {Chen}\ \emph {et~al.}(2016)\citenamefont {Chen},
  \citenamefont {Kelly}, \citenamefont {Quintana}, \citenamefont {Barends},
  \citenamefont {Campbell}, \citenamefont {Chen}, \citenamefont {Chiaro},
  \citenamefont {Dunsworth}, \citenamefont {Fowler}, \citenamefont {Lucero}
  \emph {et~al.}}]{chen2016measuring}%
  \BibitemOpen
  \bibfield  {author} {\bibinfo {author} {\bibfnamefont {Z.}~\bibnamefont
  {Chen}}, \bibinfo {author} {\bibfnamefont {J.}~\bibnamefont {Kelly}},
  \bibinfo {author} {\bibfnamefont {C.}~\bibnamefont {Quintana}}, \bibinfo
  {author} {\bibfnamefont {R.}~\bibnamefont {Barends}}, \bibinfo {author}
  {\bibfnamefont {B.}~\bibnamefont {Campbell}}, \bibinfo {author}
  {\bibfnamefont {Y.}~\bibnamefont {Chen}}, \bibinfo {author} {\bibfnamefont
  {B.}~\bibnamefont {Chiaro}}, \bibinfo {author} {\bibfnamefont
  {A.}~\bibnamefont {Dunsworth}}, \bibinfo {author} {\bibfnamefont {A.~G.}\
  \bibnamefont {Fowler}}, \bibinfo {author} {\bibfnamefont {E.}~\bibnamefont
  {Lucero}}, \emph {et~al.},\ }\bibfield  {title} {\bibinfo {title} {Measuring
  and suppressing quantum state leakage in a superconducting qubit},\ }\href
  {https://doi.org/10.1103/PhysRevLett.116.020501} {\bibfield  {journal}
  {\bibinfo  {journal} {Phys. Rev. Lett.}\ }\textbf {\bibinfo {volume} {116}},\
  \bibinfo {pages} {020501} (\bibinfo {year} {2016})},\ \Eprint
  {https://arxiv.org/abs/1509.05470} {arXiv:1509.05470 [quant-ph]} \BibitemShut
  {NoStop}%
\bibitem [{\citenamefont {McEwen}\ \emph {et~al.}(2021)\citenamefont {McEwen},
  \citenamefont {Kafri}, \citenamefont {Chen}, \citenamefont {Atalaya},
  \citenamefont {Satzinger}, \citenamefont {Quintana}, \citenamefont {Klimov},
  \citenamefont {Sank}, \citenamefont {Gidney}, \citenamefont {Fowler} \emph
  {et~al.}}]{mcewen2021removing}%
  \BibitemOpen
  \bibfield  {author} {\bibinfo {author} {\bibfnamefont {M.}~\bibnamefont
  {McEwen}}, \bibinfo {author} {\bibfnamefont {D.}~\bibnamefont {Kafri}},
  \bibinfo {author} {\bibfnamefont {Z.}~\bibnamefont {Chen}}, \bibinfo {author}
  {\bibfnamefont {J.}~\bibnamefont {Atalaya}}, \bibinfo {author} {\bibfnamefont
  {K.}~\bibnamefont {Satzinger}}, \bibinfo {author} {\bibfnamefont
  {C.}~\bibnamefont {Quintana}}, \bibinfo {author} {\bibfnamefont
  {P.}~\bibnamefont {Klimov}}, \bibinfo {author} {\bibfnamefont
  {D.}~\bibnamefont {Sank}}, \bibinfo {author} {\bibfnamefont {C.}~\bibnamefont
  {Gidney}}, \bibinfo {author} {\bibfnamefont {A.}~\bibnamefont {Fowler}},
  \emph {et~al.},\ }\bibfield  {title} {\bibinfo {title} {Removing
  leakage-induced correlated errors in superconducting quantum error
  correction},\ }\href {https://doi.org/10.1038/s41467-021-21982-y} {\bibfield
  {journal} {\bibinfo  {journal} {Nat. Commun.}\ }\textbf {\bibinfo {volume}
  {12}},\ \bibinfo {pages} {1761} (\bibinfo {year} {2021})},\ \Eprint
  {https://arxiv.org/abs/arXiv:2102.06131} {arXiv:arXiv:2102.06131 [quant-ph]}
  \BibitemShut {NoStop}%
\bibitem [{\citenamefont {Aliferis}\ and\ \citenamefont
  {Terhal}(2007)}]{aliferis2005fault}%
  \BibitemOpen
  \bibfield  {author} {\bibinfo {author} {\bibfnamefont {P.}~\bibnamefont
  {Aliferis}}\ and\ \bibinfo {author} {\bibfnamefont {B.~M.}\ \bibnamefont
  {Terhal}},\ }\bibfield  {title} {\bibinfo {title} {Fault-tolerant quantum
  computation for local leakage faults},\ }\href
  {https://doi.org/10.26421/QIC7.1-2-9} {\bibfield  {journal} {\bibinfo
  {journal} {Quant. Inf. Comput.}\ }\textbf {\bibinfo {volume} {7}},\ \bibinfo
  {pages} {139} (\bibinfo {year} {2007})},\ \Eprint
  {https://arxiv.org/abs/quant-ph/0511065} {arXiv:quant-ph/0511065}
  \BibitemShut {NoStop}%
\bibitem [{\citenamefont {Suchara}\ \emph {et~al.}(2015)\citenamefont
  {Suchara}, \citenamefont {Cross},\ and\ \citenamefont
  {Gambetta}}]{suchara2015leakage}%
  \BibitemOpen
  \bibfield  {author} {\bibinfo {author} {\bibfnamefont {M.}~\bibnamefont
  {Suchara}}, \bibinfo {author} {\bibfnamefont {A.~W.}\ \bibnamefont {Cross}},\
  and\ \bibinfo {author} {\bibfnamefont {J.~M.}\ \bibnamefont {Gambetta}},\
  }\bibfield  {title} {\bibinfo {title} {Leakage suppression in the toric
  code},\ }\href {https://doi.org/10.26421/QIC15.11-12-8} {\bibfield  {journal}
  {\bibinfo  {journal} {Quant. Inf. Comput.}\ }\textbf {\bibinfo {volume}
  {15}},\ \bibinfo {pages} {997} (\bibinfo {year} {2015})},\ \Eprint
  {https://arxiv.org/abs/1410.8562} {arXiv:1410.8562 [quant-ph]} \BibitemShut
  {NoStop}%
\bibitem [{\citenamefont {Place}\ \emph {et~al.}(2021)\citenamefont {Place},
  \citenamefont {Rodgers}, \citenamefont {Mundada}, \citenamefont {Smitham},
  \citenamefont {Fitzpatrick}, \citenamefont {Leng}, \citenamefont {Premkumar},
  \citenamefont {Bryon}, \citenamefont {Vrajitoarea}, \citenamefont {Sussman}
  \emph {et~al.}}]{place2021new}%
  \BibitemOpen
  \bibfield  {author} {\bibinfo {author} {\bibfnamefont {A.~P.}\ \bibnamefont
  {Place}}, \bibinfo {author} {\bibfnamefont {L.~V.}\ \bibnamefont {Rodgers}},
  \bibinfo {author} {\bibfnamefont {P.}~\bibnamefont {Mundada}}, \bibinfo
  {author} {\bibfnamefont {B.~M.}\ \bibnamefont {Smitham}}, \bibinfo {author}
  {\bibfnamefont {M.}~\bibnamefont {Fitzpatrick}}, \bibinfo {author}
  {\bibfnamefont {Z.}~\bibnamefont {Leng}}, \bibinfo {author} {\bibfnamefont
  {A.}~\bibnamefont {Premkumar}}, \bibinfo {author} {\bibfnamefont
  {J.}~\bibnamefont {Bryon}}, \bibinfo {author} {\bibfnamefont
  {A.}~\bibnamefont {Vrajitoarea}}, \bibinfo {author} {\bibfnamefont
  {S.}~\bibnamefont {Sussman}}, \emph {et~al.},\ }\bibfield  {title} {\bibinfo
  {title} {New material platform for superconducting transmon qubits with
  coherence times exceeding 0.3 milliseconds},\ }\href
  {https://doi.org/10.1038/s41467-021-22030-5} {\bibfield  {journal} {\bibinfo
  {journal} {Nat. Commun.}\ }\textbf {\bibinfo {volume} {12}},\ \bibinfo
  {pages} {1779} (\bibinfo {year} {2021})},\ \Eprint
  {https://arxiv.org/abs/2003.00024} {arXiv:2003.00024 [quant-ph]} \BibitemShut
  {NoStop}%
\bibitem [{\citenamefont {Motzoi}\ \emph {et~al.}(2009)\citenamefont {Motzoi},
  \citenamefont {Gambetta}, \citenamefont {Rebentrost},\ and\ \citenamefont
  {Wilhelm}}]{Motzoi2009}%
  \BibitemOpen
  \bibfield  {author} {\bibinfo {author} {\bibfnamefont {F.}~\bibnamefont
  {Motzoi}}, \bibinfo {author} {\bibfnamefont {J.~M.}\ \bibnamefont
  {Gambetta}}, \bibinfo {author} {\bibfnamefont {P.}~\bibnamefont
  {Rebentrost}},\ and\ \bibinfo {author} {\bibfnamefont {F.~K.}\ \bibnamefont
  {Wilhelm}},\ }\bibfield  {title} {\bibinfo {title} {Simple pulses for
  elimination of leakage in weakly nonlinear qubits},\ }\href
  {https://doi.org/10.1103/PhysRevLett.103.110501} {\bibfield  {journal}
  {\bibinfo  {journal} {Phys. Rev. Lett.}\ }\textbf {\bibinfo {volume} {103}},\
  \bibinfo {pages} {110501} (\bibinfo {year} {2009})},\ \Eprint
  {https://arxiv.org/abs/0901.0534} {arXiv:0901.0534 [cond-mat.mes-hall]}
  \BibitemShut {NoStop}%
\bibitem [{\citenamefont {Beverland}\ \emph {et~al.}(2019)\citenamefont
  {Beverland}, \citenamefont {Brown}, \citenamefont {Kastoryano},\ and\
  \citenamefont {Marolleau}}]{Beverland18}%
  \BibitemOpen
  \bibfield  {author} {\bibinfo {author} {\bibfnamefont {M.~E.}\ \bibnamefont
  {Beverland}}, \bibinfo {author} {\bibfnamefont {B.~J.}\ \bibnamefont
  {Brown}}, \bibinfo {author} {\bibfnamefont {M.~J.}\ \bibnamefont
  {Kastoryano}},\ and\ \bibinfo {author} {\bibfnamefont {Q.}~\bibnamefont
  {Marolleau}},\ }\bibfield  {title} {\bibinfo {title} {The role of entropy in
  topological quantum error correction},\ }\href
  {https://doi.org/10.1088/1742-5468/ab25de} {\bibfield  {journal} {\bibinfo
  {journal} {J. Stat. Mech.}\ }\textbf {\bibinfo {volume} {2019}},\ \bibinfo
  {pages} {073404} (\bibinfo {year} {2019})},\ \Eprint
  {https://arxiv.org/abs/1812.05117} {arXiv:1812.05117 [quant-ph]} \BibitemShut
  {NoStop}%
\bibitem [{\citenamefont {Harper}\ \emph {et~al.}(2020)\citenamefont {Harper},
  \citenamefont {Flammia},\ and\ \citenamefont {Wallman}}]{Harper2020}%
  \BibitemOpen
  \bibfield  {author} {\bibinfo {author} {\bibfnamefont {R.}~\bibnamefont
  {Harper}}, \bibinfo {author} {\bibfnamefont {S.~T.}\ \bibnamefont
  {Flammia}},\ and\ \bibinfo {author} {\bibfnamefont {J.~J.}\ \bibnamefont
  {Wallman}},\ }\bibfield  {title} {\bibinfo {title} {Efficient learning of
  quantum noise},\ }\href {https://doi.org/10.1038/s41567-020-0992-8}
  {\bibfield  {journal} {\bibinfo  {journal} {Nat. Phys.}\ }\textbf {\bibinfo
  {volume} {16}},\ \bibinfo {pages} {1184} (\bibinfo {year} {2020})},\ \Eprint
  {https://arxiv.org/abs/1907.13022} {arXiv:1907.13022 [quant-ph]} \BibitemShut
  {NoStop}%
\bibitem [{\citenamefont {Arute}\ \emph {et~al.}(2019)\citenamefont {Arute},
  \citenamefont {Arya}, \citenamefont {Babbush}, \citenamefont {Bacon},
  \citenamefont {Bardin}, \citenamefont {Barends}, \citenamefont {Biswas},
  \citenamefont {Boixo}, \citenamefont {Brandao}, \citenamefont {Buell} \emph
  {et~al.}}]{arute2019quantum}%
  \BibitemOpen
  \bibfield  {author} {\bibinfo {author} {\bibfnamefont {F.}~\bibnamefont
  {Arute}}, \bibinfo {author} {\bibfnamefont {K.}~\bibnamefont {Arya}},
  \bibinfo {author} {\bibfnamefont {R.}~\bibnamefont {Babbush}}, \bibinfo
  {author} {\bibfnamefont {D.}~\bibnamefont {Bacon}}, \bibinfo {author}
  {\bibfnamefont {J.~C.}\ \bibnamefont {Bardin}}, \bibinfo {author}
  {\bibfnamefont {R.}~\bibnamefont {Barends}}, \bibinfo {author} {\bibfnamefont
  {R.}~\bibnamefont {Biswas}}, \bibinfo {author} {\bibfnamefont
  {S.}~\bibnamefont {Boixo}}, \bibinfo {author} {\bibfnamefont {F.~G.}\
  \bibnamefont {Brandao}}, \bibinfo {author} {\bibfnamefont {D.~A.}\
  \bibnamefont {Buell}}, \emph {et~al.},\ }\bibfield  {title} {\bibinfo {title}
  {Quantum supremacy using a programmable superconducting processor},\ }\href
  {https://doi.org/10.1038/s41586-019-1666-5} {\bibfield  {journal} {\bibinfo
  {journal} {Nature}\ }\textbf {\bibinfo {volume} {574}},\ \bibinfo {pages}
  {505} (\bibinfo {year} {2019})}\BibitemShut {NoStop}%
\bibitem [{\citenamefont {Darmawan}\ and\ \citenamefont
  {Poulin}(2017)}]{darmawan17}%
  \BibitemOpen
  \bibfield  {author} {\bibinfo {author} {\bibfnamefont {A.~S.}\ \bibnamefont
  {Darmawan}}\ and\ \bibinfo {author} {\bibfnamefont {D.}~\bibnamefont
  {Poulin}},\ }\bibfield  {title} {\bibinfo {title} {Tensor-{Network}
  {Simulations} of the {Surface} {Code} under {Realistic} {Noise}},\ }\href
  {https://doi.org/10.1103/PhysRevLett.119.040502} {\bibfield  {journal}
  {\bibinfo  {journal} {Phys. Rev. Lett.}\ }\textbf {\bibinfo {volume} {119}},\
  \bibinfo {pages} {040502} (\bibinfo {year} {2017})},\ \Eprint
  {https://arxiv.org/abs/1607.06460} {arXiv:1607.06460 [quant-ph]} \BibitemShut
  {NoStop}%
\bibitem [{\citenamefont {Darmawan}\ and\ \citenamefont
  {Poulin}(2018)}]{Darmawan18}%
  \BibitemOpen
  \bibfield  {author} {\bibinfo {author} {\bibfnamefont {A.~S.}\ \bibnamefont
  {Darmawan}}\ and\ \bibinfo {author} {\bibfnamefont {D.}~\bibnamefont
  {Poulin}},\ }\bibfield  {title} {\bibinfo {title} {Linear-time general
  decoding algorithm for the surface code},\ }\href
  {https://doi.org/10.1103/PhysRevE.97.051302} {\bibfield  {journal} {\bibinfo
  {journal} {Phys. Rev. E}\ }\textbf {\bibinfo {volume} {97}},\ \bibinfo
  {pages} {051302} (\bibinfo {year} {2018})},\ \Eprint
  {https://arxiv.org/abs/1801.01879} {arXiv:1801.01879 [quant-ph]} \BibitemShut
  {NoStop}%
\bibitem [{\citenamefont {Huang}\ \emph
  {et~al.}(2020{\natexlab{b}})\citenamefont {Huang}, \citenamefont {Ni},
  \citenamefont {Zhang}, \citenamefont {Newman}, \citenamefont {Ding},
  \citenamefont {Gao}, \citenamefont {Wang}, \citenamefont {Zhao},
  \citenamefont {Wu}, \citenamefont {Zhang} \emph
  {et~al.}}]{huang_alibaba_2020}%
  \BibitemOpen
  \bibfield  {author} {\bibinfo {author} {\bibfnamefont {C.}~\bibnamefont
  {Huang}}, \bibinfo {author} {\bibfnamefont {X.}~\bibnamefont {Ni}}, \bibinfo
  {author} {\bibfnamefont {F.}~\bibnamefont {Zhang}}, \bibinfo {author}
  {\bibfnamefont {M.}~\bibnamefont {Newman}}, \bibinfo {author} {\bibfnamefont
  {D.}~\bibnamefont {Ding}}, \bibinfo {author} {\bibfnamefont {X.}~\bibnamefont
  {Gao}}, \bibinfo {author} {\bibfnamefont {T.}~\bibnamefont {Wang}}, \bibinfo
  {author} {\bibfnamefont {H.-H.}\ \bibnamefont {Zhao}}, \bibinfo {author}
  {\bibfnamefont {F.}~\bibnamefont {Wu}}, \bibinfo {author} {\bibfnamefont
  {G.}~\bibnamefont {Zhang}}, \emph {et~al.},\ }\href@noop {} {\bibinfo {title}
  {Alibaba cloud quantum development platform: Surface code simulations with
  crosstalk}} (\bibinfo {year} {2020}{\natexlab{b}}),\ \Eprint
  {https://arxiv.org/abs/2002.08918} {arXiv:2002.08918 [quant-ph]} \BibitemShut
  {NoStop}%
\bibitem [{\citenamefont {Chamberland}\ \emph
  {et~al.}(2020{\natexlab{b}})\citenamefont {Chamberland}, \citenamefont
  {Kubica}, \citenamefont {Yoder},\ and\ \citenamefont
  {Zhu}}]{Chamberland2020}%
  \BibitemOpen
  \bibfield  {author} {\bibinfo {author} {\bibfnamefont {C.}~\bibnamefont
  {Chamberland}}, \bibinfo {author} {\bibfnamefont {A.}~\bibnamefont {Kubica}},
  \bibinfo {author} {\bibfnamefont {T.~J.}\ \bibnamefont {Yoder}},\ and\
  \bibinfo {author} {\bibfnamefont {G.}~\bibnamefont {Zhu}},\ }\bibfield
  {title} {\bibinfo {title} {Triangular color codes on trivalent graphs with
  flag qubits},\ }\href {https://doi.org/10.1088/1367-2630/ab68fd} {\bibfield
  {journal} {\bibinfo  {journal} {New J. Phys.}\ }\textbf {\bibinfo {volume}
  {22}},\ \bibinfo {pages} {023019} (\bibinfo {year} {2020}{\natexlab{b}})},\
  \Eprint {https://arxiv.org/abs/1911.00355} {arXiv:1911.00355 [quant-ph]}
  \BibitemShut {NoStop}%
\end{thebibliography}%
\end{document}